\RequirePackage{fix-cm}
\documentclass[smallextended,natbib]{svjour3}       % onecolumn (second format)
\smartqed  % flush right qed marks, e.g. at end of proof
\usepackage{graphicx}
\usepackage{booktabs} % For formal tables
\usepackage[utf8]{inputenc}
\usepackage{tikz}
\usepackage{todonotes}
\usepackage{hyperref}
\usepackage{balance}
\usepackage{enumerate}
\usepackage{enumitem}
\usepackage{listings}
\usepackage{subcaption}
\usepackage{multirow}
\usepackage{adjustbox}
\usepackage[super]{nth}
\usepackage{hhline}

\usepackage{color}
\usepackage{xcolor}
\usepackage{colortbl}
\definecolor{ckeywordcolor}{RGB}{127,0,85}
\definecolor{cstringcolor}{RGB}{42,0,255}
\definecolor{ccommentcolor}{RGB}{63,127,95}
\definecolor{Lightgray}{gray}{0.9}
\definecolor{Gray}{gray}{0.7}

\definecolor{lightgray}{rgb}{.9,.9,.9}
\definecolor{red}{rgb}{1,0,0}

\newcommand{\cse}{{\textit{Chair of Software Engineering}}}
\newcommand{\fin}{{\textit{Faculty of Computer Science}}}
\newcommand{\ovgu}{{\textit{Otto-von-Guericke University Magdeburg}}}

\newcommand*\circled[1]{\tikz[baseline=(char.base)]{
		\node[shape=circle,draw,inner sep=0.75pt] (char) {\textbf{#1}};}}

\graphicspath{{./}}

\begin{document}

\title{Guided Pattern Mining for API Misuse Detection by Change-Based Code Analysis%\thanks{Grants or other notes
}

\titlerunning{Guided Pattern Mining for API Misuse Detection}        % if too long for running head

\author{Sebastian Nielebock         \and
	Robert Heumüller \and
	Kevin Michael Schott \and
	Frank Ortmeier %etc.
}


\institute{Sebastian Nielebock\at \email{sebastian.nielebock@ovgu.de}
	\and Robert Heum\"uller\at \email{robert.heumueller@ovgu.de}
	\and Kevin Michael Schott\at \email{kevin.schott@ovgu.de}
	\and Frank Ortmeier\at \email{frank.ortmeier@ovgu.de}\\
	\at	
	\ovgu, \fin,\\ \cse
}

\date{Received: \today / Accepted: date}

% commands for research question to simplify reuse
\newcommand{\RQone}{Does the changed-based code analysis sufficiently reduce the number of code snippets to efficiently perform API usage pattern mining}
\newcommand{\RQtwo}{Which filtering strategy yields to the highest relative frequency of fixing patterns in the retrieved source files?}
\newcommand{\RQthree}{Does an existing API usage pattern miner increases the number of detected fixing patterns by means of the selected filtering strategy?}

\maketitle

\begin{abstract}
	Lack of experience, inadequate documentation, and sub-optimal API design frequently cause developers to make mistakes when re-using third-party implementations. 
Such \emph{API misuses} can result in unintended behavior, performance losses, or software crashes.
Therefore, current research aims to automatically detect such misuses by comparing the way a developer used an API to previously inferred patterns of the correct API usage.
While research has made significant progress, these techniques have not yet been adopted in practice. 
In part, this is due to the lack of a process capable of seamlessly integrating with software development processes. Particularly, existing approaches do not consider how to collect relevant source code samples from which to infer patterns. In fact, an inadequate collection can cause API usage pattern miners to infer irrelevant patterns which leads to false alarms instead of finding true API misuses.

In this paper, we target this problem (a) by providing a method that increases the likelihood of finding relevant and true-positive patterns concerning a given set of code changes and agnostic to a concrete static, intra-procedural mining technique and (b) by introducing a concept for just-in-time API misuse detection which analyzes changes at the time of commit. 

Particularly, we introduce different, lightweight code search and filtering strategies and evaluate them on two real-world API misuse datasets to determine their usefulness in finding relevant intra-procedural API usage patterns.

Our main results are (1) commit-based search with subsequent filtering effectively decreases the amount of code to be analyzed, (2) in particular method-level filtering is superior to file-level filtering, (3) project-internal and project-external code search find solutions for different types of misuses and thus are complementary, (4) incorporating prior knowledge of the misused API into the search has a negligible effect.
\end{abstract}

\section{Motivation}
\label{sec:motivation}

Application Programming Interfaces (APIs) enable programmers to reuse existing functionality from libraries.
However, since programmers are not always familiar with the particularities of a certain library, they can misuse its API.
Generally spoken, these API misuses denote usages that were non-intended by the developers of the library and eventually lead to negative behavior in the client code, for example, performance losses or software crashes.

The reasons why programmers introduce API misuses are manyfold and include unknown knowledge on the API usage domain, missing documentation, inconsistent or complex API design (e.g.,  confusing method names), or unknown internal dependencies of the API implementation~\cite{Robillard2011, Zibran2011, Hou2011, Nadi2016, Oliveira2018}.
In a study on real bug fixes~\cite{zhong2015empirical}, the authors showed that half of their analyzed fixes involved at least one API-specific change to resolve the bug. 
Even worse, API misuses appear in different shapes.
A study of 90 API misuses, therefore, introduced a first Misuse Classification scheme (MuC ~\cite{Amann2018}. The MuC distinguishes between missing and redundant API elements such as method calls, conditions, iterations, and exception handling.
In addition to this classification, other typical classes of misuses are incorrect ordering of method calls or incorrect usage of parameters~\cite{robillard2013automated, FrolinS.Ocariza2013}.

Therefore, recent research strives to invent automated methods for detecting API misuses.
Due to increasing computational power, data mining techniques have become prominent in finding so-called API specifications~\cite{robillard2013automated}.
An API specification is a formal model describing properties of the correct usage of APIs and thus is used as an oracle to detect API misuses as code parts violating this specification. 
On the highest level, API specifications have so far been distinguished into two categories, namely, dynamic invariants together with pre- and post-conditions as well as usage patterns~\cite{Ammons2002}.
Dynamic invariants reason about program state and how it changes with regard to API usages~\cite{Ernst2001, Ernst2007}. They vary in their representation form regular expressions over finite-state automata~\cite{Ammons2002,yang2004automatically,gabel2008javert,pradel2009automatic}, up to temporal specifications~\cite{wasylkowski2007detecting,wasylkowski2011mining}.
Similarly, various representations of API usage patterns have been proposed in prior research such as API method call pairs~\cite{Weimer2005}, association rules~\cite{livshits2005dynamine,li2005pr}, API method call sequences~\cite{Thummalapenta2007}, trees~\cite{Allamanis2014}, or graphs~\cite{Nguyen2009,Amann2018a}. Some approaches use Bayesian inference to learn the correct usage of APIs from code examples as probability distribution~\cite{Allamanis2014,murali2017bayesian}. Zhou et al. infer specifications from the API documentation to detect inconsistencies with the respective specifications from the code~\cite{Zhou2017}. In essence, they represent structural or temporal constraints between the elements of an API.

In this paper, we focus on API usage patterns that are inferred from existing source code through data mining. Moreover, we only consider intra-procedural patterns, namely, those that only occur within single method declarations. Thus, our results do not refer to inter-procedural patterns which are scattered among multiple method declarations. The general procedure of mining usage patterns and detecting API misuses comprises the following five steps:

\begin{enumerate}
	\item Collect a representative set of source code for mining
	\item Transform this code set into an intermediate representation (e.g., execution traces~\citep{yang2006perracotta}, syntax trees~\citep{Allamanis2014}, API usage graphs~\citep{Amann2018a})
	\item Conduct a frequent pattern mining approach (e.g., association rule mining~\citep{li2005pr}, sequence mining~\citep{zhong2009mapo}, subgraph mining~\citep{Amann2018a}) on this representation
	\item Filter generated patterns based on suitable ranking metrics (e.g., support, confidence, or others~\citep{Le2015})
	\item Compare the usage of the API with those of the highest-ranked patterns and report violations as misuses~\citep{Amann2018a}
\end{enumerate}

Research on API misuse detection and API usage pattern mining, in particular, has focused on reducing the high number of false positives, i.e., patterns originating from random co-occurrences of code elements, (e.g., method calls). These false positives can cause false alarms during API misuse detection which impedes the practical application of such automated detectors. Particularly, the approaches mainly improve the last four steps, namely, the intermediate representation, the frequent pattern mining approach, the filtering and ranking strategy, as well as the violation detection ~\cite{li2005pr,livshits2005dynamine,Thummalapenta2007,wasylkowski2007detecting,gabel2008javert,Nguyen2009,pradel2009automatic,zhong2009mapo,wasylkowski2011mining,Allamanis2014,murali2017bayesian,Amann2018a}. 

However, little effort (e.g., by Le~Goues and Weimer~\citep{Goues2012}) was put into the initial step of collecting code before mining, even though the quality of the input data has a significant impact on the results of any data mining algorithm.
False or noisy data results in bad --or at least unpredictable-- results, as seen with classifiers~\citep{Agrawal2018}.
Moreover, existing research gives only little insight into how code collection as well as subsequent pattern mining and misuse detection could conceptually be included in a software development cycle.

For that purpose, this paper concentrates on the investigation of strategies for collecting code samples before the mining step. Our envisioned strategies aim to improve the true positive rate agnostic of the concrete static, intra-procedural mining tool. In particular, such a strategy should select source code that contains a high density of relevant patterns concerning a particular API misuse. We refer to such patterns as \emph{fixing patterns}, meaning patterns that can detect and fix an API misuse. Thus, by increasing the \emph{relative frequency} of fixing patterns in the data set, we increase the likelihood that support-based miners will find the fixing patterns.
Note that we do not directly target the goal of decreasing the false positive rate since this was mainly part of previous API misuse detectors. However, our investigated search strategy gives an additional lightweight pre-processing step that improves the performance (i.e., the true positive rate) by reducing the number of code samples to mine from.

Moreover, we introduce a concept for how this strategy can be incorporated into a standard software development cycle. We exemplarily show that, due to the lightweight design of our search process, it requires only a little additional effort compared to later mining and filtering steps.

Our collection strategy is based on the analysis of code changes, i.e., commits in a version control system. The idea is to incrementally analyze only the small subsets of code that are affected by a change and to use this very specific context for a focused API misuse detection.
To this end, we search for source code files with similar but correct API usages regarding the changed code and further filter these using different strategies.
We compared these strategies by using known, real-word API misuses and their respected fixes from the two benchmarks MUBench~\citep{Amann2016} and the AU500~\cite{Kang2021}. In particular, we determine how frequently the known (or similar) fixes are found in the filtered sets and use this information to identify the best strategy (i.e., that with the highest relative frequency). 

Afterward, we check whether the results of an existing support-based API pattern mining approach are actually improved by comparing the performance with and without the filtering strategy.

This way, we answer the following three research questions (RQs):

\begin{enumerate}[label={RQ\textsubscript{\arabic*}}, leftmargin=*]
	
	\item \RQone \label{rq-1}
	\item \RQtwo \label{rq-2}
	\item \RQthree \label{rq-3}
	
\end{enumerate}

For replicability, we provide our data sets and evaluation scripts as a replication package\footnote{\label{fn:replpkg}available at \url{http://doi.org/10.5281/zenodo.5091822}}. This package can also be seen as a first prototype of the search and mining process. For example, this could be installed on a standard continuous integration (CI) server similarly as, e.g., automated testing approaches. Currently, a developer who commits her code changes to such a CI system would receive API usage patterns similar to the API usages in their change. Potential subsequent steps, such as misuse detectors and misuse fixing approaches, could also be added to the CI system. This way, misuse detection and correction are directly applicable for software developers.

Our process and the analyzed filtering strategies are introduced in the following Section~\ref{sec:approach}. Then, we present our results of the evaluation of the three research questions (Section~\ref{sec:evaluation}). Afterward, we discuss potential threats to validity (Section~\ref{sec:threats}) of our results as well as differences and similarities to related work (Section~\ref{sec:related}). Finally, we conclude our results and present future work (Section~\ref{sec:conclusion}).

\section{Process and Strategies of a Change-Based API Misuse Detection}
\label{sec:approach}

Within this section, we first present our envisioned process of an API misuse detection which leverages code search and filter strategies to improve the input data for API usage pattern mining. This section describes the concrete use case for which the collection strategies and subsequent mining approaches are designed. Thus, the misuse detection is not part of the contribution of this paper. Second, we discuss the notion of different search and filter strategies.

\subsection{A Vision of a Change-Based API Misuse Detection and Correction}
\label{ssec:approach-vision}
\begin{figure*}
	\includegraphics[width=\textwidth]{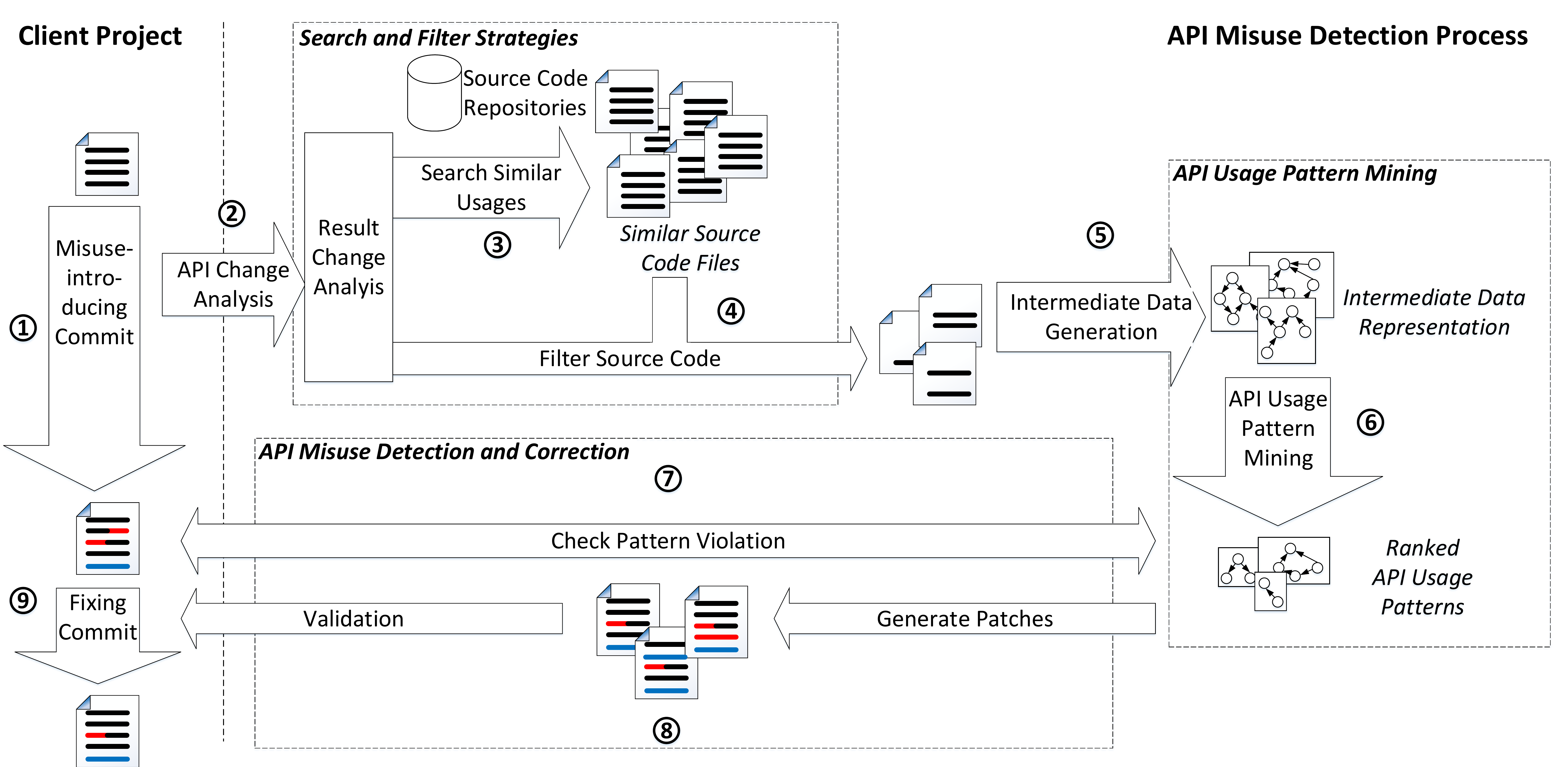}
	\caption{Envisioned API Misuse Detection and Correction Process}
	\label{fig:misuse-detection}
\end{figure*}

\autoref{fig:misuse-detection} depicts our envisioned API misuse detection process. It describes how our analyzed search and filter strategies, the API usage pattern mining, as well as the final detection and correction, can be seamlessly integrated into an ordinary continuous integration (CI) process. 

We consider a developer who commits changes in her client project into a version control system. Assuming that some of these changes could contain an API misuse (\circled{1}), our process conducts an \emph{API change analysis} (\circled{2}) based on the changeset of that commit. The goal of this step is to locate the changed methods and to extract the affected API usages and their context from each method. Based on this information, step \circled{3} searches for each changed method other source code samples (either from this or from foreign repositories) with similar API usages. These, in turn, are further filtered (\circled{4}).
The steps \circled{3} and \circled{4} describe the search and filter strategies, which we present in detail in the upcoming section and evaluate in Section~\ref{sec:evaluation}

The filtered code is then transformed (\circled{5}) into an intermediate data representation (e.g., method call sequences, abstract syntax trees, API usage graphs) to be further processed by an API usage pattern miner.
This miner conducts a frequent pattern mining approach (\circled{6}) and generates a list of ranked API usage patterns (e.g., ordered by support).

Then, the process checks whether the changed version of the client code violates one or multiple of the highest-ranked API usage patterns (\circled{7}). In case of violations, one can generate patch candidates based on the violated usage patterns (\circled{8}), for example, by using the difference between pattern and code as a patch. After validating the misuse and selecting a fitting patch, the developer can apply the fix (\circled{9}).

We envision these steps to be set up in an ordinary CI process. Thus, whenever developers commit code changes they get instant feedback whether or not the changes introduced an API misuse. In case of a misuse, commits can be automatically requested to be revised, for instance, by the patches suggested to correct the misuse.

Note that this approach assumes that every commit contains a  potential API misuse. Therefore, the API change analysis needs to reduce the number of methods to be analyzed so that the number of subsequent mining runs remains low as well. Moreover, to reduce the effort for a single mining run, the search and filter strategies effectively reduce the number of similar source files without harming the true-positive rate of the miner.
However, in case the subsequent pattern mining requires huge computation time, one may limit this analysis to specific testing branches in the CI system, or require a developer to deliberately trigger it.

This process shares some similarities with the work by Saied et al.~\cite{Saied2020}, which integrates API misuse detection as an interactive element in the coding task. However, our process does not require storing a set of previously inferred patterns but conducts online pattern inference during analysis from an evolving code base. Since here, the preprocessing and mining require some amount of time (approx. 5-10 minutes) it is currently neither intended nor recommended for interactive usage.

\subsection{Search and Filtering Strategies}
\label{ssec:approach-stratgies}
\begin{figure}
	\includegraphics[width=\textwidth]{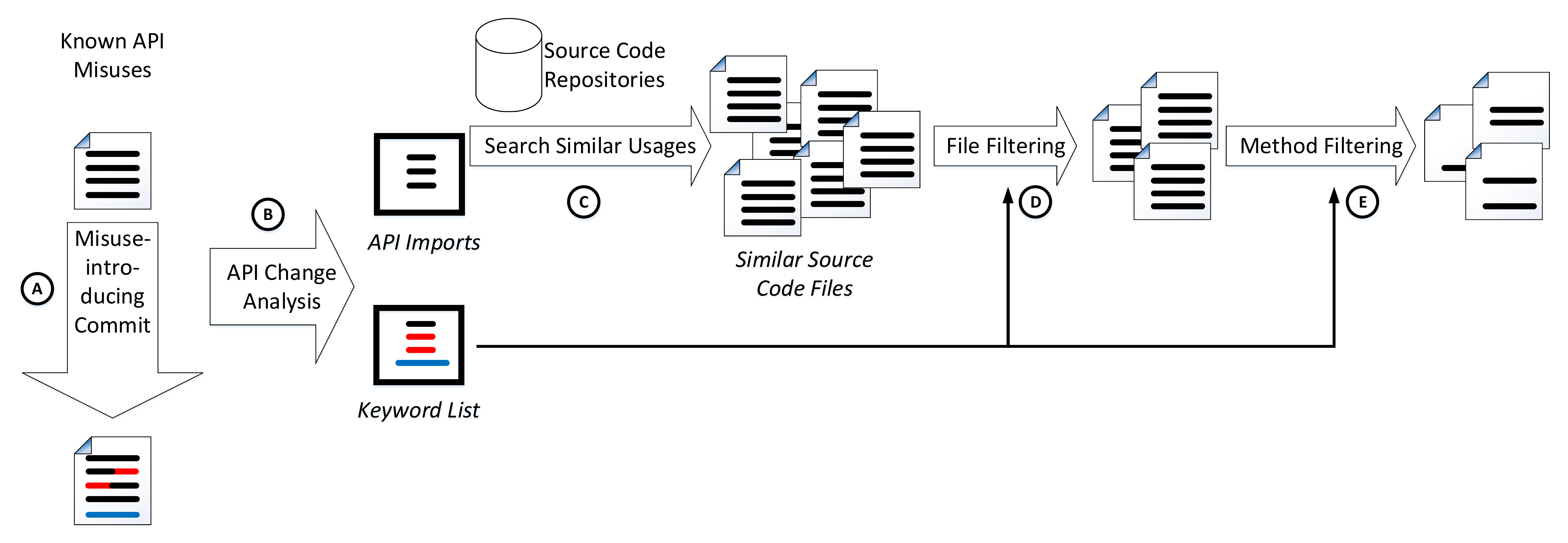}
	\caption{Process Steps of the Search and Filtering Strategies}
	\label{fig:filtering}
\end{figure}

To improve the input data of the mining step, we propose different strategies for searching and filtering source code to increase the ratio of relevant source code in terms of finding fixing patterns.
In this section, we describe the main steps and the intuition behind these strategies. Note that our evaluation (cf. Section~\ref{sec:evaluation}) is based on Java as a programming language. Therefore, some technical details refer to particularities of that language and may need to be adapted for use with other programming languages. 

In the general process (cf. \autoref{fig:filtering}), first, we extract the set of changed methods (\circled{A}) from a single commit. Then, for each changed method, we conduct a separate API change analysis (\circled{B}). This analysis extracts a set of relevant API import statements, i.e., those that import a type of a third-party library used in the analyzed method. It further extracts a set of \emph{keywords} describing the context of the API usage, i.e., within the analyzed method. Particularly, these keywords are the set of included class names from third-party libraries, as well as all method names used within the analyzed method.
Using the set of API import statements, we conducted a code search for files that also import these types (\circled{C}). Afterward, we filtered the files (\circled{D}) and the method declarations (\circled{E}).
In the following paragraphs, we describe each step in detail.

\paragraph{Commit}

The commit step (\circled{A}) extracts the set of changed methods from a commit that may have introduced API misuses. Note that during our evaluation, we specifically investigate misuse-introducing commits of known API misuses. We obtained these commits from the information given in the analyzed benchmark. Details are described in Section~\ref{sec:evaluation}.

We inferred the changed methods using the version control system (i.e., \texttt{git diff}) and extracted the set of changed source files with the respective changed lines. Then we parsed these changes and located those method declarations that were at least partially affected by these changes. These methods constitute the scope for the following analyses. We cannot restrict these only to the changed lines, because it does not necessarily contain all information required to detect misuses. For example, based on a previous method call order, an additional method call may introduce a misuse, e.g., an invalid double initialization of an object. Moreover, the effort to analyze the method scope is still manageable. This is important as a complex and long-lasting analysis would impede the development process.

\paragraph{API Change Analysis}

For each method detected in the previous step, we initiate a separate API change analysis and a subsequent search and filtering process. In the API change analysis (\circled{B}), we want to discover which API elements from outside of the current project's scope were changed by the commit.

Note that we only considered third-party libraries for two reasons. First, for project-internal API elements, i.e., types and methods that are declared within the analyzed project, it is very unlikely to find usage patterns in external code. Since we are comparing project-internal and project-external (i.e., in foreign projects) code search strategies this comparison would be fairly biased. Second, usages of the \texttt{java.lang} APIs are far too common and introduce too much noise into the filter process.

Intuitively, the discovered API elements correspond to the keywords that a developer would use when searching for similar code on the web.
To identify useful API elements, we reviewed the real-world misuses from the MUBench benchmark~\cite{Amann2016}. Based on these code samples, we identified common patterns of code features that describe the API usage and its context. In addition, Zhong et al. provided some insights into which code features indicate API usage~\cite{zhong2009mapo}.

\begin{figure}
	\begin{subfigure}{.8\textwidth}
\begin{lstlisting}[label=src:sample-extraction]
package my.own.pkg.subpkg;

import a.b.AClass;
import a.b.BClass;
import a.b.CClass;
import x.y.ZClass;
import x.v.*;
import my.own.pkg.QClass;

public class Foo extends AClass {
    @Override
    protected ZClass doSomething(BClass bObj) {
        super.doSomething(bObj);
        QClass myQObj = callThisMethod(RClass.rMethodCall());
        return myQObj.mergeWithZClass(bObj);
    }
}
\end{lstlisting}
	\end{subfigure}\hfill
	\begin{subfigure}{.19\textwidth}
		\includegraphics[width=\textwidth]{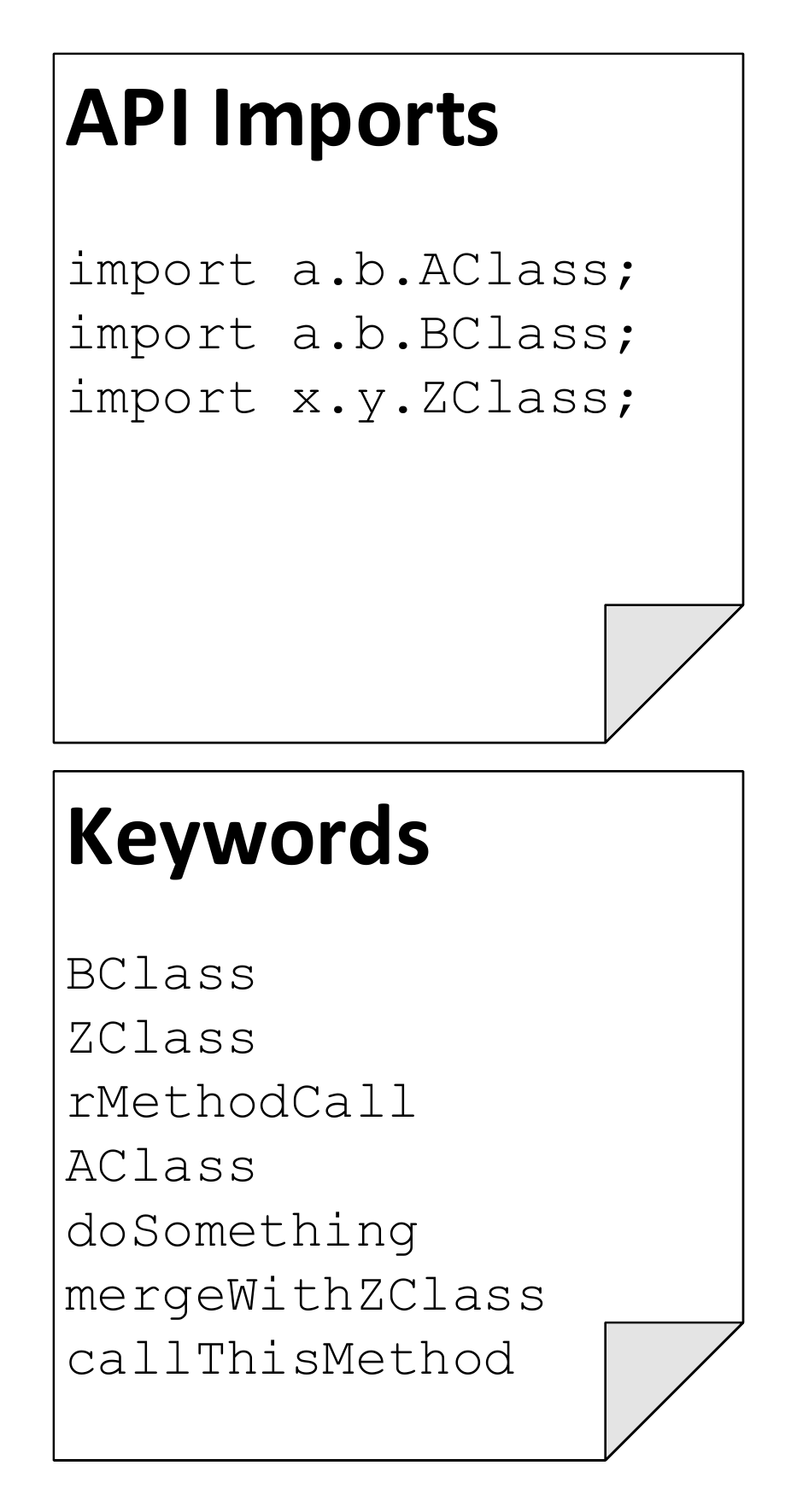}	
	\end{subfigure}
	\caption{Example of a keyword extraction for the \texttt{doSomething}-method. Left: Source Code from which keywords are extracted. Right: Set of extracted API \texttt{import} statements and keywords}
	\label{fig:keywords-extraction}
\end{figure}

We exemplify the identified code features describing the relevant API elements in \autoref{fig:keywords-extraction}. 

The used data types of third-party libraries are important indicators for API usage. In general, we consider a data type as a relevant API element if it is used in the analyzed method and originates from a third-party-library, i.e. is explicitly imported via an \texttt{import} statement and does not originate from the project itself. The usage of a data type in a method means one of the following five alternatives:

\begin{itemize}
	\item the type is applied as parameter type of that method	
	\item the type is applied as a return type of that method
	\item the type is applied as \texttt{throw}n type in the method's declaration 
	\item the type is explicitly mentioned in an expression inside of the method's body
	\item the type is inherited from the class of that method and the method overrides the method declaration, explicitly denoted by an \texttt{@Override} annotation
\end{itemize}

Considering the method \texttt{doSomething} in our example, this covers the data types \texttt{AClass}, \texttt{BClass}, and \texttt{ZClass}. Since type \texttt{CClass} is not used in the method, it is not extracted.

We consider data-types as imported when there exists an explicit \texttt{import}-statement that ends with that type name. For example in \autoref{fig:keywords-extraction}, the type \texttt{BClass} used in line 12 is imported via the \texttt{import}-statement in line 4 (\texttt{import a.b.BClass;}). On the other hand, assume that the type \texttt{RClass} used in line 14 is imported via the wildcard \texttt{import}-statement in line 7 (\texttt{import x.v.*;}). Since this type name is not explicitly mentioned, it is not considered as relevant.

Moreover, we checked whether the used types were related to a project-internal or a third-party library. This was done by checking whether the \texttt{import}-statements have the same prefix (i.e., qualifiers of that data type) as the \texttt{package}-statement in that class declaration. Particularly, we checked whether the first three qualifiers are identical to those of a particular import (e.g., \texttt{my.own.pkg} in \autoref{fig:keywords-extraction}). The rationale is that regarding the naming convention of packages in Java\footnote{\url{https://docs.oracle.com/javase/tutorial/java/package/namingpkgs.html}}, the first three qualifiers usually identify the package. In case that only one or two qualifiers are used in the package, we only check whether these are prefixes of the respective \texttt{import}-statement. If no package name is given, which did not occur in our evaluation data set, we ignore all data types.
For example, the \texttt{import}-statement of type \texttt{QClass} in line 8 has the same prefix, i.e. \texttt{my.own.pkg}, as the respective \texttt{package}-statement. Thus, it is not extracted as a relevant API element. 

All \texttt{import}-statements whose data types were found to be relevant API elements are added to the \emph{API imports set}. The respective type names are added to the \emph{keywords set}.

Note that besides the data types used in the method declaration and its body, we also add the names of inherited types, if the method under investigation is overridden (indicated by the \texttt{@Override} annotation). In such a case, we also add the method name of the overridden method declaration, which is usually not considered, to the keywords set. This is reasonable since some framework APIs (e.g., Eclipse, Android) are frequently accessed by inheritance and thus could be misused. 

Additionally, we extended the keyword list with the names of all methods that were called in the investigated method. This included calls to project-internal and \texttt{java.lang} methods. The reason is that we only consider source code changes and we cannot completely resolve all data types from this partial code. Thus, sometimes it was not possible to decide whether a method belonged to an internal, \texttt{java.lang}, or a third-party-library type. 

\paragraph{Searching Source Files ($search_{loc}$ and $search_{imp}$)}

The goal of the code search is to find code that is similar to the investigated method in terms of its API usages. In our cascaded search process, the first step (\circled{C}) applies the \emph{API import}-set as a set of keywords to find files that imported the relevant API types that were identified in the previous step.

Here, we used the two different search strategies $search_{loc}$ and $search_{imp}$. 
First, we distinguished between where we searched for similar code, namely \emph{internal} code, i.e. from the same project, or \emph{external} code, i.e. originating from a foreign project ($search_{loc}$).
Second, we varied which \emph{API imports} were used. In the first version, we applied \emph{all extracted import statements}. In the second one, we only used the \emph{misused import statements}, i.e., statements that imported misused APIs ($search_{imp}$).
We discuss the rationale for both strategies in the following.

Regarding $search_{loc}$, Amann discussed that other API usages can be found either \emph{internally}, i.e. in the same project, or \emph{externally}, i.e. in other pro\-jects~\citep{Amann2018a}. The internal search can find correct API usages or already fixed API misuses in other locations of the same project. This is, for example, indicated by the plastic surgery hypothesis from automatic program repair~\citep{LeGoues2019}. On the other hand, the external code search is likely to provide more and diverse data, and thus increases the likelihood to find similar source code. 

We applied $search_{imp}$ using different sets of extracted import statements, namely, \emph{all} vs. \emph{misused} imports. We assume that searching with misused imports will increase the true positive rate. However, it also requires a preliminary analysis to extract them since, in practice, we usually don't know the misused API. Therefore, we check whether such a preliminary analysis would significantly increase the likelihood to find fixing patterns or not. Note that in our evaluations, we already know the misused API from the information in our data set (cf. Section~\ref{ssec:data}) and therefore we did not implement such an approach.

\paragraph{File Filtering ($filter_{file}$)}

After searching, our process filters files regarding the keyword set (\circled{D}) that is further denoted as $filter_{file}$. We do not expect that all keywords have to be used in a similar file because some keywords could belong to a co-applied API or method names of internal APIs. 
Therefore, we introduce a measurement - so-called \emph{satisfaction ratio} ($sr$) for files to estimate to which degree these contain the keywords. The satisfaction ratio describes the proportion of keywords found in a source file $srcFile$ with respect to the keyword set $kwSet$. It is defined as follows:

\[sr(srcFile,kwSet) = \frac{|\{kw \in kwSet\textrm{ if }srcFile\textrm{ contains }kw\}|}{|kwSet|}\]

A satisfaction ratio of $0$ does not require \emph{any} keywords to be present, while a satisfaction ratio of $1$ requires \emph{all} keywords to be present. We do not believe that either extreme is useful to increase the relative frequency of fixing patterns. While the first strategy does not require the file to be similar at all, the second strategy would yield too few results. Thus, in our evaluation, we consider a range of values between $0$ and $1$ for the satisfaction ratio.

Note that this step could be technically integrated into the previous code search. However, in this paper, we want to determine the effect of each single filtering step and therefore these steps are separated.

\paragraph{Method Filtering ($filter_{method}$)}

In the second, so-called $filter_{method}$ strategy (\circled{E}), we extracted those methods from the source files that contained at least one of the keywords in the keyword set. This is a more fine-grained approach. Moreover, the envisioned API misuse detection considers the method scope, and therefore reducing the number of methods while keeping related ones can increase the relative frequency of the patterns. 

We filtered the methods by parsing each file and generating the token sets for all methods. After removing syntax elements and Java keywords, we check if at least one keyword is contained in the remaining set.

Note that we did not apply several different satisfaction ratios as in the previous step. This decision was based on three reasons. First, this would further increase the number of strategy configurations to be analyzed. As shown in Section~\ref{ssec:results_rq2}, the current number of configurations to be tested per misuse is 40 ($2\cdot2\cdot5\cdot2$) eventually leading with our 37 misuses from the MUBench dataset to 1,480 different configurations. By adding an equal fragmentation with the five different $sr$-values and assuming that we do not have to run method filtering with higher $sr$ on method level than on file level (e.g., a file filtered out with a $sr=0.5$ certainly does not contain a method with $sr=0.75$) the number of configuration would increase to 60 per misuse ($2\cdot2\cdot(\sum_{i=1}^{5}i)$). Thus we would have to analyze at most 2,220 configurations. Second, we can assume that the optimal $sr$ on the method level is usually lower than the one for the file level since method declarations contain fewer keywords. Therefore, the range to be analyzed must be lower, for instance, the interval $[0,0.25]$. However, then we need to test even more configurations since then we cannot exclude certain configuration combinations. Third, we deemed the effect to be non-significant compared to the effort of analyzing further configurations (i.e., generating AUGs and testing the pattern containment). Nevertheless, we computed the average $sr$ for the filtered method (i.e., those that contain at least one keyword). These values can then be used to further improve the method filtering.

After introducing the filtering steps, in the upcoming section, we discuss how we evaluate our process and in particular the different strategies for searching and filtering source files. We then interpret the results of the evaluation for our three research questions.
\section{Evaluation}
\label{sec:evaluation}

We primarily evaluated our approach by means of the MUBench benchmark ~\citep{Amann2016}\footnote{\url{https://github.com/stg-tud/MUBench}}. This benchmark represents a set of real, validated API misuses from open-source projects. 
Since we found not all misuses to be suitable for our evaluation, we selected a subset for our analyses. We describe this data acquisition in Section~\ref{ssec:data}. For RQ3, we additionally incorporated the AU500 dataset\footnote{AU500 is part of the ALP software artifact at \url{https://github.com/ALP-active-miner/ALP}} by Kang et al. consisting of 500 API usages manually labeled correct and incorrect API usages~\cite{Kang2021}. Moreover, we explain how we obtained the API misuse-introducing commits and the respective \emph{API Usage Graphs}, the intermediate data representation, based on the implementation by Amann~\citep{Amann2018a}\footnote{\url{https://github.com/stg-tud/MUDetect}}.
Afterward, we evaluate our three research questions. For each question, we first, describe our methodology, second, present the results of our evaluation, and third, summarize the main results and implications. 

\subsection{Data Acquisition and Processing}
\label{ssec:data}

\paragraph{Selecting API Misuses}

We considered an initial set of 245 API misuses from Java projects obtained on February, \nth{5} 2019 from the MUBench benchmark\footnote{commit \texttt{b8124077} from \url{https://github.com/stg-tud/MUBench.git}}~\citep{Amann2016}. For each misuse in this benchmark, the authors provide a file describing the meta information of the misuse, which essentially includes the version control system, the misused API, and the fixing commit (i.e., that commit that fixed the misuse). From these misuses, we selected those 103 that originated from projects using the git-version control system and for which the benchmark provides a fixing commit. The rationale for git is that it is one of the most frequently applied version control systems\footnote{\url{https://www.openhub.net/repositories/compare}}. Since we need the fixing commit to identify the misuse-introducing commit, this is also a mandatory requirement.

\noindent Then, we further removed 66 misuses due to the following reasons:

\begin{itemize}
	\item misuses are essentially duplicates, i.e., the misuse of the serialization of an object in a testing context in the jodatime-project - we only kept one version of that misuse (36)
	\item misuses of a \texttt{java.lang}-API, which is not covered by our method (18) 
	\item misuses of an internal, i.e., project-related, API, from which we do not expect to find correct usages in external projects (8)
	\item non-distinguishable misuse, i.e., same API misuse in the same method in the same class - we only kept one version of each misuse (2)
	\item misuse is a false parameter, which cannot be represented by the used intermediate representation (2)
\end{itemize}

\noindent Thus, we kept 37 misuses for our analysis.

The AU500 consists of 500 manually labeled API usages from 16 open-source projects~\cite{Kang2021}. These projects are disjunct from those of the MUBench dataset, and thus, this dataset is well suited for an independent validation. 385 entries of this dataset are labeled as correct while the other 115 are marked as misuses. All entries have the meta-information on the git-repository, the commit hash of that version, and the source file as well as the method containing the API usage. We use all those usages for the analysis of RQ3.

\paragraph{Detecting API Misuse-introducing Commits}

In MU\-Bench its creators already identified the fixing commit, i.e. the commit in which the misuse was corrected. However, we are also interested in the \emph{API misuse-introducing commit}, i.e., the commit that made the changes that lead to the misuse. 
For that purpose, we checked out the fixed version and identified those lines of code that had to be changed to fix the misuse via the command \texttt{git diff}. We then obtained the previous version of the fixing commit and run the command \texttt{git blame}\footnote{\url{https://git-scm.com/docs/git-blame}} to identify in which commit these lines were added to the repository. We denote this commit as the misuse-introducing commit. Note that in the case of multiple different commits, i.e., among differently added lines, we chose the latest commit, since this indicates the point in time when the misuse was `complete'. This essentially is a git-adapted version of the SZZ-algorithm~\citep{Sliwerski2005}, which was designed for usage with the CVS version control system.
With these remaining 37 misuse-introducing commits, we evaluated the change-based code analysis.

In the AU500 dataset, not all usages are misuses. This is why we do not determine the misuse-introducing commits for this dataset. In contrast, we extracted the keywords and import statements only from the single revision of that API usage. We identified this revision by the respective commit hash, source file, and method name. Particularly, we checked out the respective commit and analyzed the complete method declaration as if it was added all at once using the API Change Analysis as described in Section~\ref{ssec:approach-stratgies}.

\paragraph{Collecting Similar Source Files}

We also need different sets of source files for the \emph{internal} and \emph{external} code search (cf. Section~\ref{ssec:approach-stratgies}).
For the \emph{internal search} we used all source files from the same project and revision of the misuse-introducing commit, excluding the file that contains the misuse.

We conducted the \emph{external search} by means of the \emph{Searchcode}  engine\footnote{\url{https://searchcode.com/}}. Searchcode accesses well-known code repositories such as \emph{GitHub}, \emph{BitBucket}, \emph{Google Code}, and \emph{GitLab}. Compared to other code search engines such as Boa~\citep{Dyer2013} and GHTorrent~\citep{Gousios2013}, this engine provides access to individual source files without having to download the whole project. Due to internal restrictions at the time of our analysis, Searchcode returns at most 1,000 source files, which are ordered by relevance. 
We contacted Searchcode's developer to clarify the definition of relevance and were informed that it is estimated based on the proximity of detected keywords in a file. Thus, a file containing the keywords as ``foo bar'' is ranked higher than a file having ``foo'' and ``bar'' separated in different parts of the file. 
We ran two search sessions for each misuse to collect source files. In the first one, we only searched with the misused API import statement(s). In the second session, we used all extracted import statements. We downloaded both sets via Searchcode's REST API\footnote{\url{https://searchcode.com/api/}} between February \nth{7} and February \nth{8} 2019 and eventually merged both sets, yielding up to 2,000 source files. Due to an error, we repeated the analysis for the \texttt{logblock-logblock-2\_15} misuse on December \nth{12} 2019.
For the analysis of the AU500 dataset, we downloaded the source files on June \nth{15} 2021.
We prevented source files from the same project from being found by excluding all files with the same prefix in the \texttt{package}-statement as used in that original project.
Moreover, we excluded all source files whose generation of the API Usage Graphs occupied too much memory and caused the generation script to crash on our evaluation system. In particular, we had to exclude externally found source files for 13 misuses (i.e., nine misuses with a single excluded file and four misuses with two up to nine excluded files) for the MUBench dataset. 

Note that we kept both code file sets (i.e., internal and external without exclusions) consistent for all subsequent search and filtering steps. Therefore, the potential bias introduced by the search algorithm of Searchcode is also consistent across all search and filter strategies.

\paragraph{API Usage Graphs}

For the analysis of \ref{rq-2} and \ref{rq-3}, we utilized an intermediate source code representation, namely the API usage graph (AUG) introduced by Amann~\citep{Amann2018a}. This directed, labeled multigraph is a static code representation, which depicts data- as well as control flow properties. In this respect, it is specifically tailored to representing the API usages of a single method and is therefore ideal for our analyses. Moreover, this data structure enables us to also reuse the corresponding  API usage pattern miner, which was also introduced by Amann. 

\begin{figure}
	\begin{subfigure}{.54\textwidth}
\begin{lstlisting}[label=src:sample-aug,basicstyle=\fontsize{6}{7}\selectfont\ttfamily]
package my.own.pkg;

import a.b.AClass;
import a.b.BClass;
import a.b.CClass;

public class SampleClass {
 
 AClass aObj = new AClass();
 
 protected Integer myFancyMethod(CClass cObj) {
  BClass bObj = new BClass();
  return bObj.doSomething(aObj, cObj);
 }
}
\end{lstlisting}
	\end{subfigure}\hfill
	\begin{subfigure}{.45\textwidth}
		\includegraphics[trim=1.5cm 1.5cm 1.5cm 1.5cm, width=\textwidth]{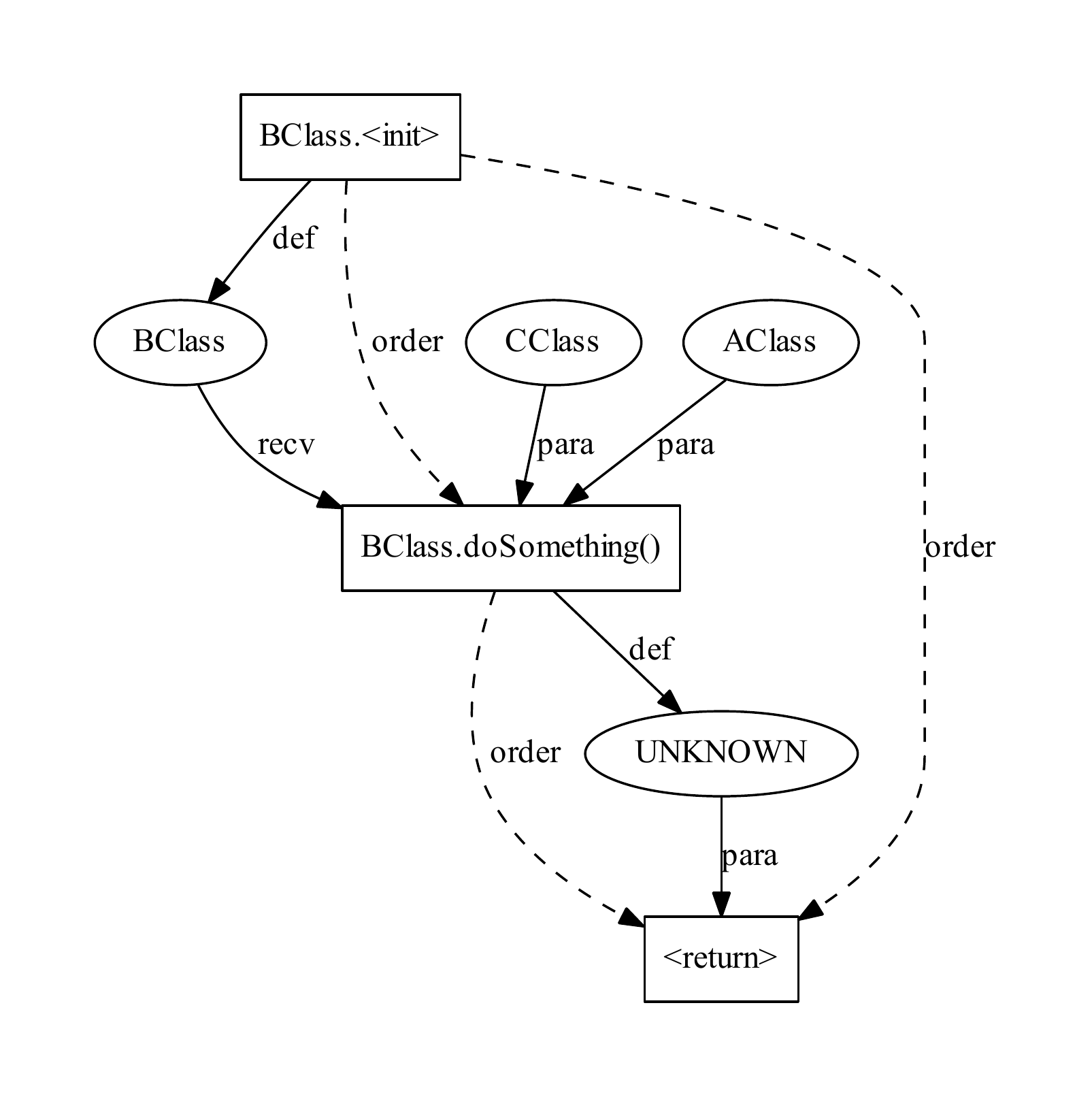}	
	\end{subfigure}
	\caption{Example of an API Usage Graph of the \texttt{myFancyMethod}-method in the \texttt{SampleClass} on the left hand-side}
	\label{fig:aug_sample}
\end{figure}

\autoref{fig:aug_sample} shows a small example of an AUG. The AUG consists of different types of nodes and edges. Rectangles denote \emph{action} nodes, e.g. method calls, while ellipses represent \emph{data} nodes, e.g. object instances. In addition to simple method calls, e.g. \texttt{doSomething}, there also exist special actions, e.g. \texttt{<init>} for constructor calls or \texttt{<return>} for return-statements. Data nodes usually represent object instances labeled with their respective type. If a type cannot be inferred statically from the code it is labeled as `\texttt{UNKNOWN}'. In the example, the return type of the \texttt{doSomething} method is \texttt{UNKNOWN} since the declaration of the method is missing and therefore the type resolution could not decide whether it is of type \texttt{Integer} or one of its subclasses. Besides having different types of nodes, AUGs also feature different kinds of edges, i.e., control flow edges (dashed) and data flow edges (solid). Control flow edges describe structural properties of the code, e.g. the order of actions (\texttt{order}-edge), while data flow edges describe how information in the form of objects is processed through the code. This includes instance definition (\texttt{def}-edge), calling methods on instances (\texttt{recv}-edge), or using instances as parameters to other methods (\texttt{para}-edge).

For further details on the AUG and the miner please refer to the work of Sven Amann~\citep{Amann2018a}.

\subsection{Commit Size Analysis (\ref{rq-1})}
\label{ssec:results_rq1}
\paragraph{Methodology}

First, our approach uses commits to reduce the amount of source code that is analyzed regarding an API misuse.
In this first experiment, we analyze the typical size of commits that contain API misuses, i.e., their number of changed methods, relevant external API imports, and their number of extracted keywords.
Since API usage pattern mining may require a lot of computational power and memory consumption, these values can be crucial for the overall performance.
For example, if a commit changes many methods, we also have to conduct the same number of searching and mining tasks.
Therefore, a small number of analyzed methods per commit is desirable.

We investigated these characteristics using the 37 misuse-introducing commits obtained from MUBench in the previous step. For each commit, we conducted the API change analysis by first determining the number of changed methods, and second, extracting for each changed method the set of API import statements and keywords as presented in Section~\ref{ssec:approach-stratgies}. For parsing source code, we used Eclipse's JDT parser\footnote{\url{https://www.eclipse.org/jdt/}}.

We locally stored each changed method. To avoid name clashes for methods (e.g., caused by overloading), we added a numerical ID to each method. For every method, we then stored the sets of API imports and keywords. 
We analyzed all collected information via python scripts and provide both the data and the evaluation scripts in our replication package\textsuperscript{\ref{fn:replpkg}}. 

\paragraph{Results}

In \autoref{tab:misuseIntro}, we show detailed information on each misuse and its respective misuse-introducing commit. In addition, it contains the number of all methods in the project (Column A), the number of methods changed in the misuse introducing commit (Column C), and the subset of those methods that were part of an external API (Column E). Note that for Column A, we obtained the total number of methods by parsing only unique source files identified by their md5-hash value. Thus, two methods originating from two identical source files are only counted once.
Moreover, some misuses were introduced in the same commit. Therefore, we analyze the degree of method reduction only for the \emph{32 unique commits}.

\begin{table}
\begin{adjustbox}{max width=\textwidth}
\begin{tabular}{ll|p{4cm}l|rrr|rr}
	\toprule
	\textbf{\#} & \textbf{misuse} & \textbf{repository (subdomain} & \textbf{MIC} & \multicolumn{3}{|c}{\textbf{\#methods}}&  \multicolumn{2}{|c}{\textbf{reduction (\%)}} \\	
	 & & \textbf{at https://github.com)} & & \textbf{A} & \textbf{C} & \textbf{E} & \textbf{A2C} & \textbf{C2E} \\ 
\midrule
1  &         alibaba-druid\_1 &                            /alibaba/druid.git &    de13143e0 &          16095 &                 12 &                  8 &                            99.9 &                            33.3 \\
2  &         alibaba-druid\_2 &                            /alibaba/druid.git &    de13143e0 &          16095 &                 12 &                  8 &                            99.9 &                            33.3 \\
3  &    android-rcs-rcsjta\_1 &                       /android-rcs/rcsjta.git &      b3445d9 &          10817 &               2275 &                642 &                            79.0 &                            71.8 \\
4  &            androiduil\_1 &  /nostra13/Android-Universal-Image-Loader.git &      9d77de9 &            737 &                279 &                131 &                            62.1 &                            53.0 \\
5  &        apache-gora\_56\_1 &                              /apache/gora.git &      e4db20a &           1565 &                141 &                 57 &                            91.0 &                            59.6 \\
6  &        apache-gora\_56\_2 &                              /apache/gora.git &    bbad5d213 &           1424 &                 39 &                 35 &                            97.3 &                            10.3 \\
7  &                bcel\_101 &                      /apache/commons-bcel.git &      d532ec1 &           3475 &               2517 &                269 &                            27.6 &                            89.3 \\
8  &           calligraphy\_1 &                    /chrisjenx/\-Calligraphy.git &     1a2d0f5d &             32 &                 10 &                  8 &                            68.8 &                            20.0 \\
9  &           calligraphy\_2 &                    /chrisjenx/\-Calligraphy.git &     1a2d0f5d &             32 &                 10 &                  8 &                            68.8 &                            20.0 \\
10 &               closure\_2 &                  /google/closure-compiler.git &   e5d3e5e012 &          11135 &                 28 &                 14 &                            99.7 &                            50.0 \\
11 &            jodatime\_269 &                        /emopers/joda-time.git &    08a925a31 &           4429 &                 54 &                 10 &                            98.8 &                            81.5 \\
12 &            jodatime\_339 &                        /emopers/joda-time.git &    9b01b9e8b &           9054 &                 21 &                 11 &                            99.8 &                            47.6 \\
13 &            jodatime\_361 &                        /emopers/joda-time.git &    7fe68f297 &           2556 &               2451 &                519 &                             4.1 &                            78.8 \\
14 &            jodatime\_362 &                        /emopers/joda-time.git &    7fe68f297 &           2556 &               2451 &                519 &                             4.1 &                            78.8 \\
15 &            jodatime\_363 &                        /emopers/joda-time.git &    7fe68f297 &           2556 &               2451 &                519 &                             4.1 &                            78.8 \\
16 &             lnreadera\_1 &            /calvinaquino/\-LNReader-Android.git &      a514f35 &           3329 &                 81 &                 72 &                            97.6 &                            11.1 \\
17 &             lnreadera\_2 &            /calvinaquino/\-LNReader-Android.git &     a514f35d &           3329 &                 81 &                 72 &                            97.6 &                            11.1 \\
18 &  logblock-logblock-2\_15 &                       /emopers/LogBlock-2.git &      5ea1b0b &             70 &                  5 &                  4 &                            92.9 &                            20.0 \\
19 &                mqtt\_389 &                   /emopers/\-paho.mqtt.java.git &     77aa39b9 &            670 &                608 &                115 &                             9.3 &                            81.1 \\
20 &                mqtt\_390 &                   /emopers/\-paho.mqtt.java.git &     f60b3721 &            990 &                 59 &                 23 &                            94.0 &                            61.0 \\
21 &             onosendai\_1 &                           /haku/Onosendai.git &      Cf2de97 &           1618 &                  3 &                  2 &                            99.8 &                            33.3 \\
22 &               openiab\_1 &                            /onepf/OpenIAB.git &      00e5612 &            957 &                173 &                100 &                            81.9 &                            42.2 \\
23 &  screen-notifications\_1 &              /lkorth/screen-notifications.git &     fa75a61f &             48 &                 21 &                 19 &                            56.2 &                             9.5 \\
24 &         tbuktu-ntru\_473 &                             /emopers/ntru.git &      4a095cc &            399 &                 13 &                  5 &                            96.7 &                            61.5 \\
25 &         tbuktu-ntru\_474 &                             /emopers/ntru.git &      e4f8688 &            187 &                  8 &                  4 &                            95.7 &                            50.0 \\
26 &         tbuktu-ntru\_475 &                             /emopers/ntru.git &      8cb6471 &            521 &                 41 &                 17 &                            92.1 &                            58.5 \\
27 &               testng\_16 &                            /cbeust/testng.git &    234c85874 &           5557 &                 21 &                 19 &                            99.6 &                             9.5 \\
28 &               testng\_17 &                            /cbeust/testng.git &    b68cf6de8 &           5479 &                 18 &                 17 &                            99.7 &                             5.6 \\
29 &               testng\_21 &                            /cbeust/testng.git &    24341340b &           5432 &                 14 &                 11 &                            99.7 &                            21.4 \\
30 &               testng\_22 &                            /cbeust/testng.git &     79cd443f &           4395 &                  4 &                  2 &                            99.9 &                            50.0 \\
31 &      thebluealliancea\_1 &       /Adam8234/the-blue-alliance-android.git &      be7b752 &           1168 &                 10 &                 10 &                            99.1 &                             0.0 \\
32 &  thomas-s-b-visualee\_29 &                         /emopers/visualee.git &      14e3f03 &            152 &                 76 &                 33 &                            50.0 &                            56.6 \\
33 &  thomas-s-b-visualee\_30 &                         /emopers/visualee.git &     14e3f03b &            152 &                 76 &                 33 &                            50.0 &                            56.6 \\
34 &  thomas-s-b-visualee\_32 &                         /emopers/visualee.git &      d4dc0ba &            250 &                  1 &                  1 &                            99.6 &                             0.0 \\
35 &           tucanmobile\_1 &                         /Tyde/\-TuCanMobile.git &      805f770 &             62 &                 11 &                  9 &                            82.3 &                            18.2 \\
36 &             ushahidia\_1 &                /ushahidi/\-Ushahidi\_Android.git &      db2b310 &           4405 &                 63 &                 40 &                            98.6 &                            36.5 \\
37 &            wordpressa\_1 &       /wordpress-mobile/\-WordPress-Android.git &  88368deadbe &           5453 &                 70 &                 39 &                            98.7 &                            44.3 \\
\bottomrule
\multicolumn{9}{c}{{\small MIC: Misuse-introducing commit; A: All methods; C: Changed methods in the MIC}}\\
\multicolumn{9}{c}{{\small E: All methods from C that contain at least one external (third-party) API}}\\
\multicolumn{9}{c}{{\small A2C: Reduction from all to changed methods}}\\
\multicolumn{9}{c}{{\small C2E Reduction from changed to changed methods that contain at least one external (third-party) API}}\\
\end{tabular}
\end{adjustbox}
\caption{Misuses with the characteristics of their misuse-introducing commits}
\label{tab:misuseIntro}
\end{table}

\begin{figure}
	\includegraphics[width=\textwidth]{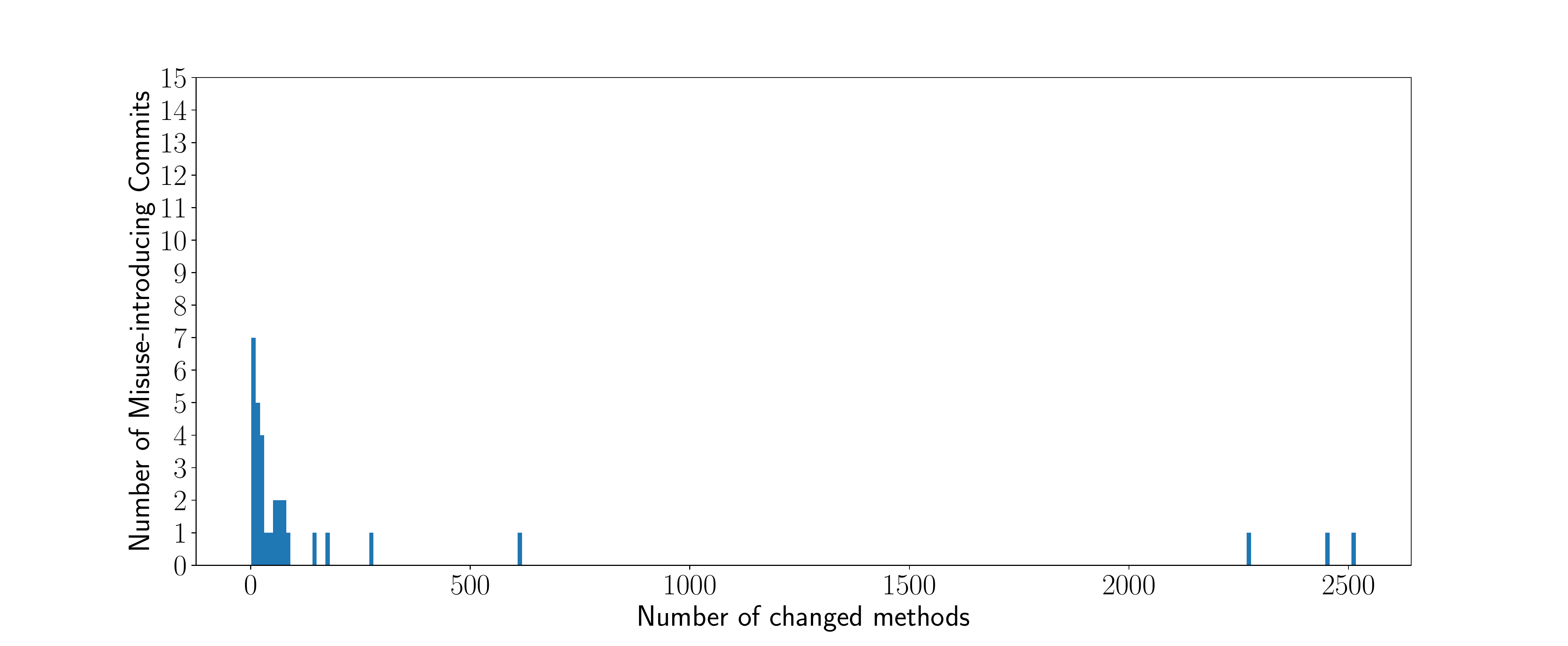}
	\caption{Distribution of Misuse-introducing commits among the number of changed methods (bin size of ten)}
	\label{fig:numAllMethod}
\end{figure}

\begin{figure}
	\includegraphics[width=\textwidth]{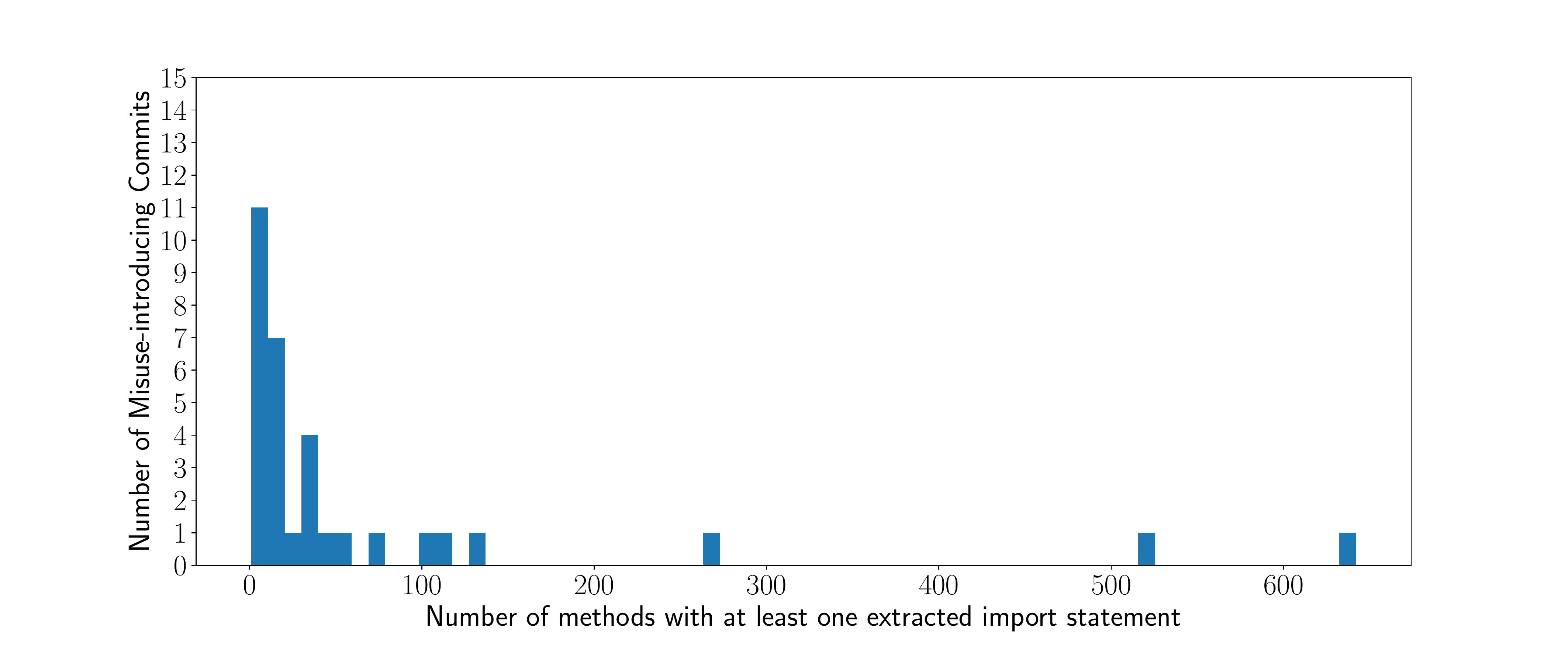}
	\caption{Distribution of Misuse-introducing commits among the number of changed methods with at least one extracted external API (bin size of ten)}
	\label{fig:numMethodOneImp}
\end{figure}

First, we considered all methods that were affected by the commit, regardless of whether this change edited an external API or not. \autoref{fig:numAllMethod} plots the distribution of misuse-introducing commits for increasing numbers of changed methods. The majority of 25 misuses modified less than 100 methods, while nine outliers had up to 2,517 changed methods (i.e., misuse \texttt{bcel\_101}). When considering only those methods for which we found imports of third-party libraries, the huge numbers shrink drastically as denoted by \autoref{fig:numMethodOneImp}. Since we only consider API misuses of third-party libraries, we do not need to investigate methods for which we cannot infer an import statement. Then, 26 misuses changed less than 100 methods, with 18 of them having less than 20 changed methods. Still, there exist six extreme outlier commits with 100 or more changed methods.

Our results show that we can effectively reduce the number of methods for later API misuse detection by considering only changed methods. Particularly, we compared the reduction against the number of methods that existed in the commit right after the misuse-introducing commit as a starting point. Considering all 32 unique commits we have an average reduction of $81.9\%$ (median $96.2\%$).
Further, we decreased the number of methods by checking whether these contained an API from a third-party library. By that means, we further reduced the number of methods on average by $15.6\%$ (median $11.6\%$). An interesting observation is that in cases in which the change-based approach could not reduce many methods (i.e., \texttt{jodaetime\_361}-\texttt{jodaetime\_363} and \texttt{mqtt\_389}) this step could effectively do so.
The mean number of changed methods is $\approx  287.6$ (median $33.5$), while after the removal of methods without a change to a third-party library this number is reduced to $\approx 71.2$ (median $18$). In total, we could reduce the number of methods on average by $86.4\%$ (median $96.8\%$).

In a second step, we investigated the number of extracted import statements and the number of keywords for those methods that referenced at least one external API (i.e., have at least one extracted import statement). These values are interesting since they indicate how many APIs potentially have to be analyzed. Moreover, having a huge number of keywords would also reduce the number of files satisfying the \emph{satisfaction ratio} in the \emph{file filtering} step. On the other hand, it increases the chance of including more methods in the \emph{method filtering} step, since more methods may match at least one of these words.

\begin{figure}
	\includegraphics[width=\textwidth]{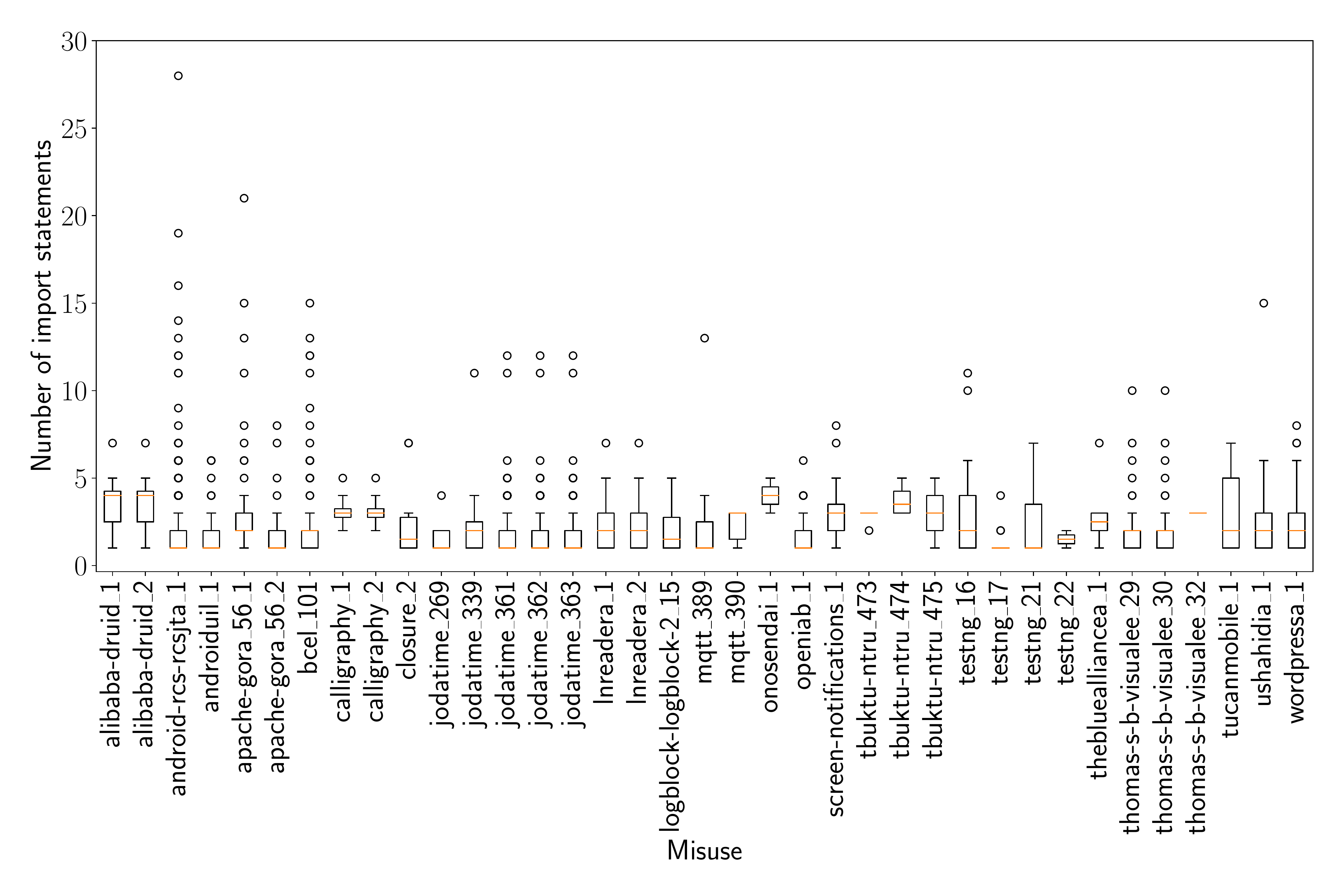}
	\caption{Distribution of the number import statements among the misuses for methods with at least one third-party import involved}
	\label{fig:importStmt}
\end{figure}

\begin{figure}
	\includegraphics[width=\textwidth]{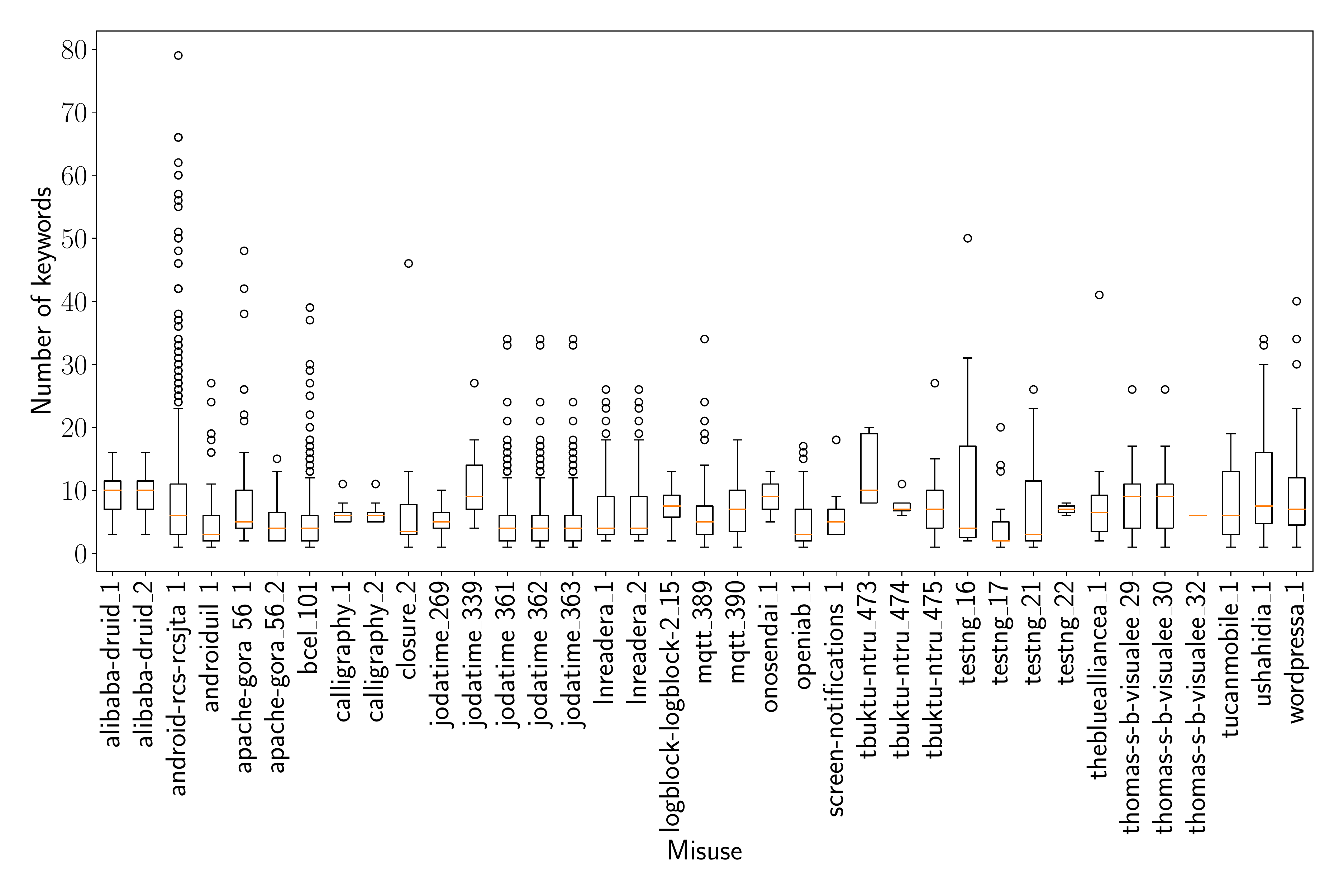}
	\caption{Distribution of the number extracted keywords among the misuses for methods with at least one third-party import involved}
	\label{fig:kwwords}
\end{figure}

\autoref{fig:importStmt} shows the distributions of the numbers of import statements among the changed methods with at least one external API for each misuse. We can observe that the majority of methods have at most 5 import statements estimated by the upper whiskers ($1.5$ of the interquartile range). On average, for methods with at least one imported third-party API, $1.8$ (median $1$) import statements were found. However, there are still outliers that refer up to 28 imports (e.g., \texttt{android-rcs-rcsjita\_1}).

Regarding the number of keywords (\autoref{fig:kwwords}), usually, at most 20 keywords are extracted - once again estimated by the upper whiskers. In extreme cases (e.g., \texttt{android-rcs-rcsjita\_1}), the number of keywords rises up to 79. On average, $6.3$ keywords are extracted (median $4$).

Moreover, we checked the ability of the code change analysis to extract the import statement(s) of the misused API(s) in the investigated method. This is important since our method uses these imports to find similar API usages. In case we did not extract the misused API, we hardly expect to find similar API usages for that misuse. We know the misused API(s) based on the meta-description in the MUBench dataset. For \emph{31} out of 37 misuses, our method successfully extracted the import statement of the misused API.

We did not precisely measure the execution time for a single change-based analysis. However, we can recall from the timestamps of the generated files, that a single execution takes from several seconds up to at most three minutes for large commits (e.g., \texttt{android-rcs-rcsjta\_1}). Note that this time excludes the downloading of the repository source files.

\paragraph{Implications}

Our results show that the change-based API analysis can effectively reduce the amount of source code that has to be analyzed.  At the same time, it is still able to determine the respective misused API in 31 out of 37 cases and on average does not extract too many import statements (mean $1.8$) and keywords (mean $6.3$). However, some extreme cases with 642 changed methods, 79 extracted keywords, and 28 different external import statements remain and require additional analyses to reduce the amount of data. For example, one can perform the API misuse detection approach only on the most suspicious methods (e.g., very complex methods, frequently changed methods), which are indicated by properties found in the change-based error detection domain (cf. Section~\ref{ssec:change-based}).

\newpage

\subsection{Filtering Analysis (\ref{rq-2})}
\label{ssec:results_rq2}
\paragraph{Methodology}

\begin{table}
	\centering
	\begin{tabular}{|p{2cm}|p{1cm}|p{1cm}|p{1cm}|p{1cm}|p{1cm}|p{1cm}|}
		\hline
		\textbf{$search_{loc}$} & \multicolumn{3}{p{3cm}|}{internal} & \multicolumn{3}{p{3cm}|}{external}\\
		\hline
		\textbf{$search_{imp}$} & \multicolumn{3}{p{3cm}|}{all imports} & \multicolumn{3}{p{3cm}|}{misused imports}\\		
		\hline
		\textbf{$filter_{file}$} & sr=0.0 & sr=0.25 & \multicolumn{2}{c|}{sr=0.5} & sr=0.75 & sr=1.0\\		
		\hline
		\textbf{$filter_{method}$} & \multicolumn{3}{p{3cm}|}{applied} & \multicolumn{3}{p{3cm}|}{not applied}\\\hline
	\end{tabular}
	\caption{Different configurations for the analysis of file searching (i.e., $search_{loc}$ and $search_{imp}$) and filtering (i.e., $filter_{file}$ and method $filter_{method}$)}
	\label{tab:strategies}
\end{table}

We conducted the analysis of all search and filter strategies (i.e., $search_{loc}$, $search_{imp}$, $filter_{file}$, and $filter_{method}$) as described in Section~\ref{ssec:approach-stratgies}. As illustrated in \autoref{tab:strategies}, the analyzed strategies comprise 40 different configurations, all of which were evaluated for each of the 37 misuses from MUBench. This sums up to 1,480 different configurations.
We implemented different scripts for conducting the strategies and obtained similar source files (i.e., $search_{loc}$) as described in Section~\ref{ssec:data}.
Regarding the $search_{imp}$-step, we obtained the misused import statements by the meta-description of the misuses in the MUBench benchmark.
With respect to the $filter_{file}$ strategy, we tested different values for the satisfaction ratio in the interval $[0,1]$ and including the extremes $sr=0.0$ (no keyword has to be matched) and $sr=1.0$ (all keywords have to be matched). Since we estimate the best value for the satisfaction ratio without having too many configurations, we split the interval into four quarters, i.e., $[0, 0.25, 0.5, 0.75, 1]$. 
Finally, we applied the $filter_{method}$-step by using the extracted sets of keywords.

For each configuration, we computed the \emph{relative pattern frequency}. This describes how often a fixing pattern was found in the set of methods obtained from the retrieved source files. We obtained this by conducting the following five steps:
\begin{enumerate}
	\item We manually distilled one or multiple variants of the fixing pattern using the known fix from the MUBench benchmark.
	\item We generated the AUG for each fix
	\item We generated the AUG of all methods obtained from the particular configuration.
	\item We count how often the fixing pattern AUG is a subgraph in the set of AUGs obtained by a particular configuration.
	\item We selected that pattern by the highest number of occurrences and divided that number by the number of AUGs.
\end{enumerate}

Since the subgraph isomorphism problem is NP-hard, we only checked a relaxed condition. In particular, we only checked whether the set of nodes and the set of edges of the fixing pattern AUG is a subset of the set of nodes and the set of edges of the candidate AUG. Consequently, this introduces an overestimation, i.e., the real number of pattern occurrences might be lower. Therefore, we investigate the subsequent pattern mining ability in \ref{rq-3}.

We then compared the different configurations based on the relative pattern frequency.
For that purpose, we applied the non-parametric Wilcoxon signed-rank test to determine whether the differences in the set of different configuration groups (if any exist) are significant. We chose this test instead of, for instance, the parametric t-test, since we cannot be sure that the relative pattern frequency follows a normal distribution. Particularly, the Wilcoxon signed-rank test has the null hypothesis that two paired groups originate from the same distribution. We reject the null hypothesis with $\alpha=0.05$. For our analysis, these groups represent the relative pattern frequencies obtained from the strategies. Thus, the elements of the groups are paired by their misuse. Since the tests assume the elements of the single groups to be independent, we cannot simply split all frequency values from all configurations into two sets. For example, the 740 configurations using $filter_{method}$ are not completely independent since 592 configurations used the same source files but with a different $filter_{file}$ strategy. Therefore, to determine the real effect of a single strategy, we only compared those configurations using a single search strategy. When we determined the effect of one strategy the respective other filter strategies were left out. The concrete conditions under which the groups for comparison of the single strategies were obtained are depicted in \autoref{tab:test_condition}.

\begin{table}
	\begin{tabular}{p{2.5cm}p{8cm}}
		\hline
		\textbf{Comparison of} & \textbf{Condition to obtain independent groups}\\\hline
		$search_{loc}$ & no $filter_{file}$ (i.e., $sr=0$); no $filter_{method}$; each single $search_{imp}$-strategy\\\hline
		$search_{imp}$ &no $filter_{file}$ (i.e., $sr=0$); no $filter_{method}$; each single $search_{loc}$-strategy\\\hline
		$filter_{file}$ & no $filter_{method}$; each single $search_{imp}$ and $search_{loc}$ strategy\\\hline
		$filter_{method}$ & no $filter_{file}$ (i.e., $sr=0$); each single $search_{imp}$ and $search_{loc}$ strategy\\\hline		
	\end{tabular}
	\caption{Overview under which conditions the groups for statistical comparison of the single strategies were obtained}
	\label{tab:test_condition}
\end{table}	

\paragraph{Results}

In 748 out of 1480 cases, we obtained at least one similar source file fitting the criteria of the respective configuration. In 383 of those configurations, we found at least one occurrence of the fixing pattern. Regarding the misuses, for 33 misuses we found similar files. As denoted before, only in 31 cases, we were able to correctly extract the import of the misused API. Only for one of these 31 cases, we were not able to obtain any similar source files. For the 3 misuses, for which we found similar source files but not based on the misused API, our approach used other imports of shared third-party APIs. Consequently, it only selected source files that did not contain an occurrence of the fixing pattern.
For 22 of those 30 misuses for which the misused API import was correctly extracted and for which we found similar source files, we found at least one fixing pattern with one of the 40 configurations.

Considering the strategies in detail, we were interested in which one has a significant positive impact in increasing the relative pattern frequency.
Therefore, we first checked for how many misuses a particular strategy found at least one fixing pattern and second, whether the differences between single independent groups (denoted by \autoref{tab:test_condition}) were significant w.r.t. the previously described test.

\begin{figure}
	\includegraphics[width=\textwidth]{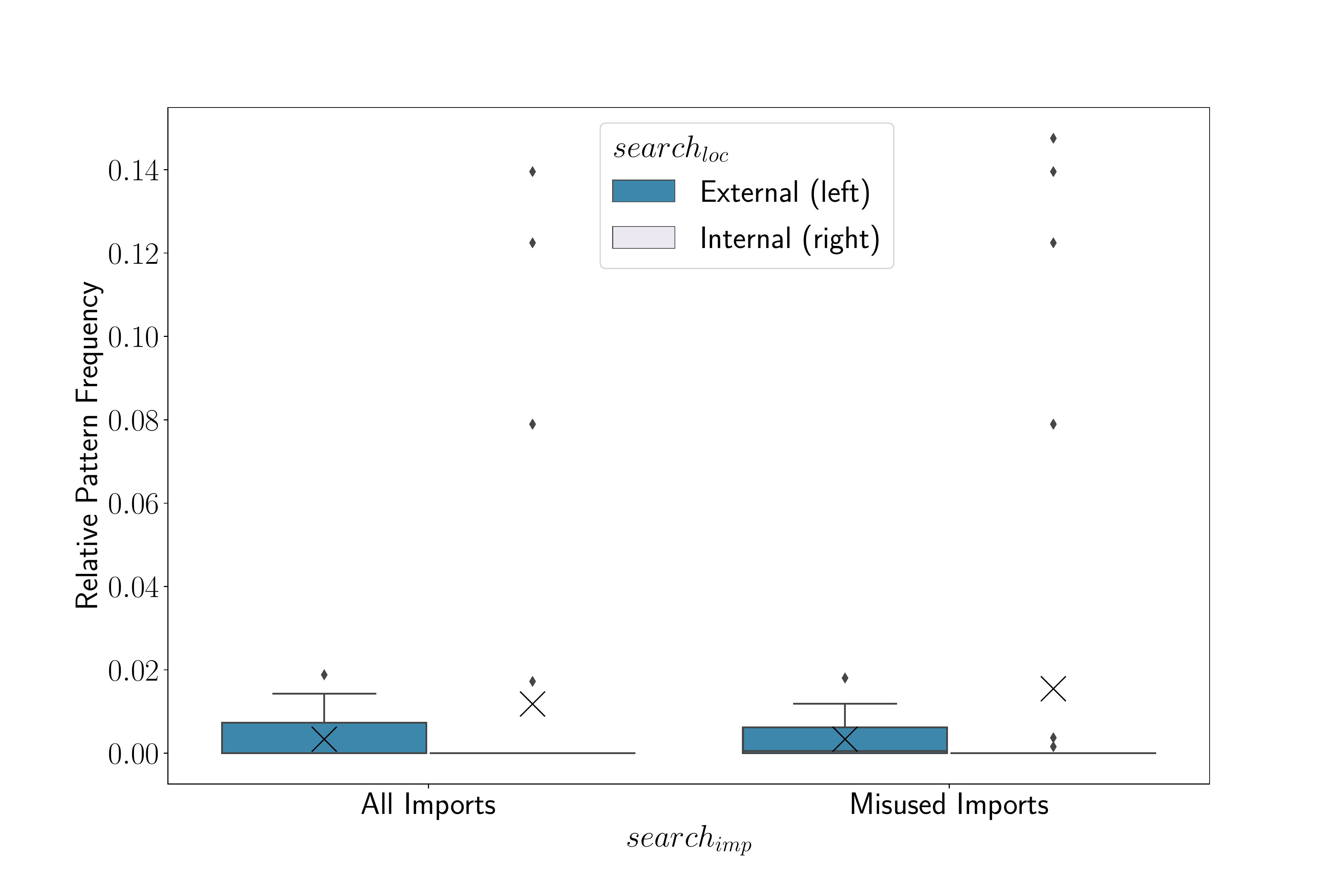}
	\caption{Distribution of the relative pattern frequency using different file search strategies grouped by API search strategy}
	\label{fig:internal-vs-external}
\end{figure}

The $search_{loc}$ strategy distinguish between \emph{internal} and \emph{external} search. The internal $search_{loc}$ found similar files containing at least one occurrence of the fixing pattern for seven misuses, while the external $search_{loc}$ found them for 22 misuses. Thus quantitatively, we found more fixing patterns externally. This matches the observations made in previous work~\citep{Amann2018a}. However, considering the distribution of relative frequencies in case a fixing pattern was found (\autoref{fig:internal-vs-external}), we can see that the mean relative frequency (indicated by the ``x"-mark) is higher for the internal $search_{loc}$. This is true across both API search strategies. Nevertheless, the differences in the means may arise only from outliers of internal $search_{loc}$, while most considered misuses result in a relative pattern frequency of zero. Using the Wilcoxon signed-rank test, we could also not determine a significant difference in the distributions of the two file search strategies. Therefore, it indicates that using both, internal and external $search_{loc}$ in a cascaded manner could be useful. For example, before searching externally it might be worth first searching within the project itself. This relates to the idea of the plastic surgery theorem from the automatic program repair domain~\cite{LeGoues2019}.

\begin{figure}
	\includegraphics[width=\textwidth]{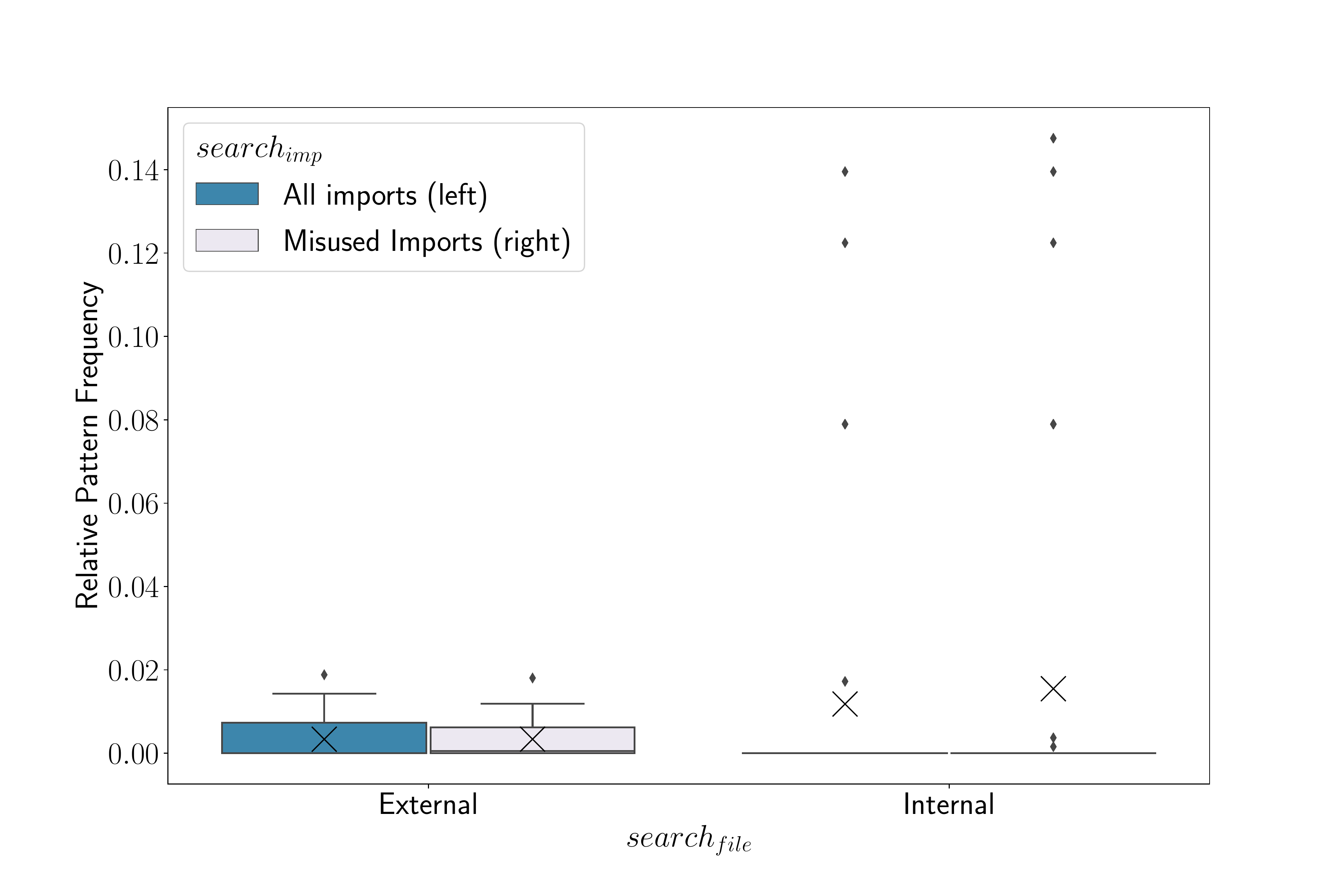}
	\caption{Distribution of the relative pattern frequency using different API search strategies grouped by each file search strategy}
	\label{fig:all-vs-misused}
\end{figure}

Next, we evaluated whether prior knowledge of the misused API imports has a significant positive impact on finding the fixing pattern. This is represented by the $search_{imp}$-strategy. As our results show, we found fixing patterns for 22 misuses when using only the misused-imports-$search_{imp}$, compared to 17 misuses when using all-imports-$search_{imp}$. When considering the relative pattern frequency in \autoref{fig:all-vs-misused}, we cannot observe a significant difference between the two strategies using either internal or external code search. The Wilcoxon test also supported this observation. This indicates that prior knowledge of the misused API has only a moderate impact on increasing the number of fixing patterns. Considering that likely there exits no perfect method for identifying the misused API, it is reassuring to see that the results without such a method are not that much worth it.

\begin{table}
	\centering
	\begin{tabular}{lr}
		\toprule
		\textbf{sr} &  \textbf{misuses with} \\
		& \textbf{fixing patterns} \\
		\midrule
		0.00 &      22 \\
		0.25 &      21 \\
		0.50 &      21 \\
		0.75 &      18 \\
		1.00 &       4 \\
		\bottomrule
	\end{tabular}
	\caption{Number of misuses per satisfaction ratio for which at least one fixing pattern was found}
	\label{tab:fixPerSrc}
\end{table}

$filter_{file}$ was applied for five different $sr$ values (cf. \autoref{tab:fixPerSrc}). We observed that the number of misuses, for which we found at least one fixing pattern, was relatively stable (slightly drops from 22 to 18 misuses) with increasing $sr$, however, it drastically drops to four misuses for $sr=1$.

\begin{figure}
	\includegraphics[width=\textwidth]{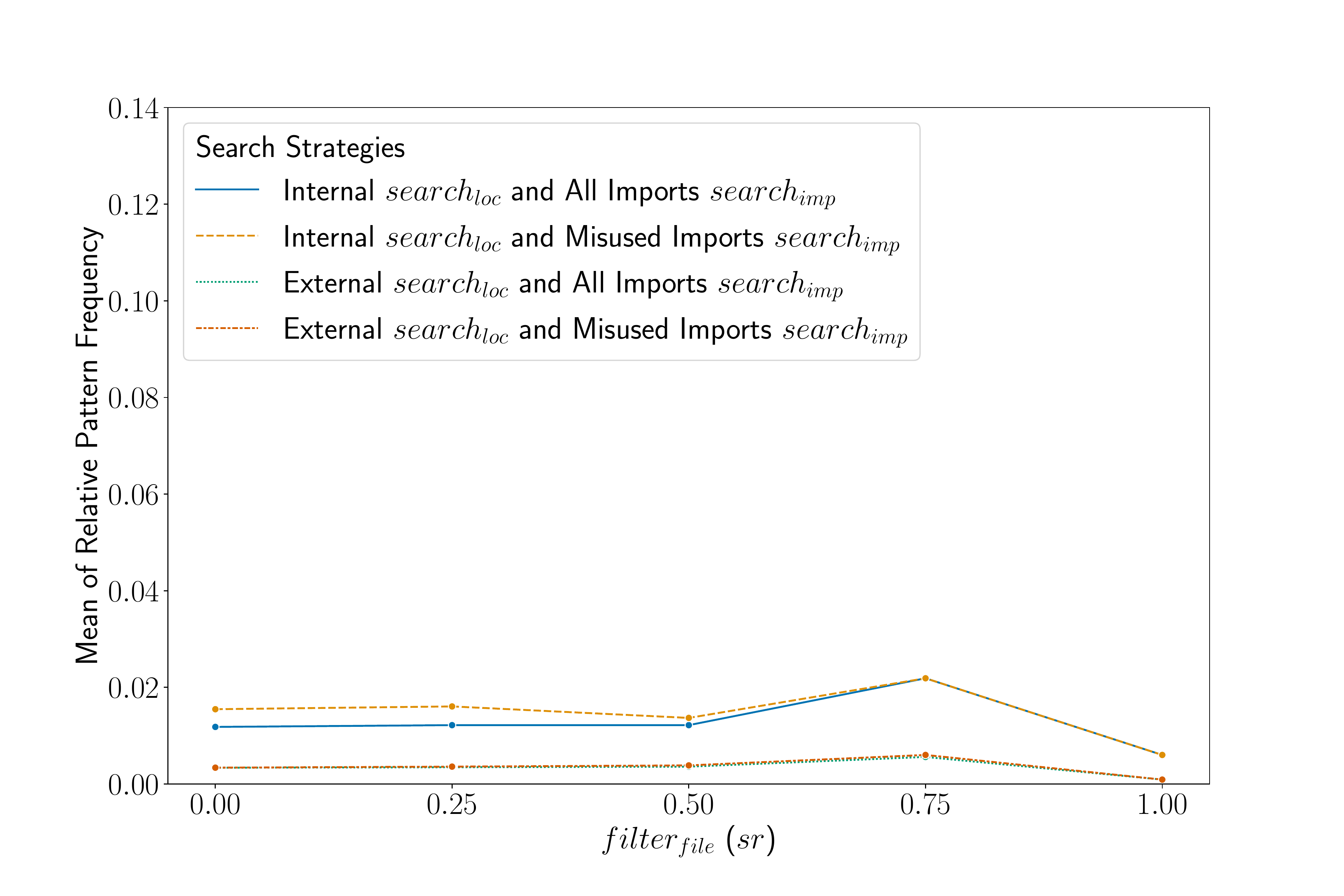}
	\caption{Mean values of the relative pattern frequency among different file filter strategies (satisfaction ratio) grouped by certain search strategies}
	\label{fig:srs_means}
\end{figure}

The relative pattern frequency is usually constant (w.r.t. to the means) while dropping for $sr=1$ (cf. \autoref{fig:srs_means}).
When comparing the distributions of $sr$s combined with different strategies for $search_{loc}$ and $search_{imp}$, we determined a significant difference between $sr=1$ and every other group of $search_{loc}$ and $search_{imp}$ except for applying internal $search_{loc}$ with all imports $search_{imp}$. Note that for the internal $search_{loc}$ there are very few results so that the statistical tests may not be as reliable as for the external $search_{loc}$. Regarding the aspect of its lower relative frequency (e.g., in means), we can conclude that $filter_{file}$ with $sr=1$ has a negative effect. While indicated by the mean values, the Wilcoxon test could not determine a significant difference in the distributions for $filter_{file}$ with $sr=0.75$ and all distributions with a lower $sr$. For configurations using external $search_{loc}$, we could determine a significant difference between $filter_{file}$ with $sr=0$ and $sr=0.25$ as well as $sr=0$ and $sr=0.5$. With respect to the mean values of these groups $filter_{file}$ has a slightly positive effect on the relative pattern frequency. Note that the respective median values are almost always zero for the different independent groups of $filter_{file}$. Therefore, we conclude that $filter_{file}$ usually has a moderate positive effect on the relative pattern frequency up to $sr=0.5$.

\begin{figure}
	\begin{subfigure}[c]{\textwidth}
		\includegraphics[width=\textwidth]{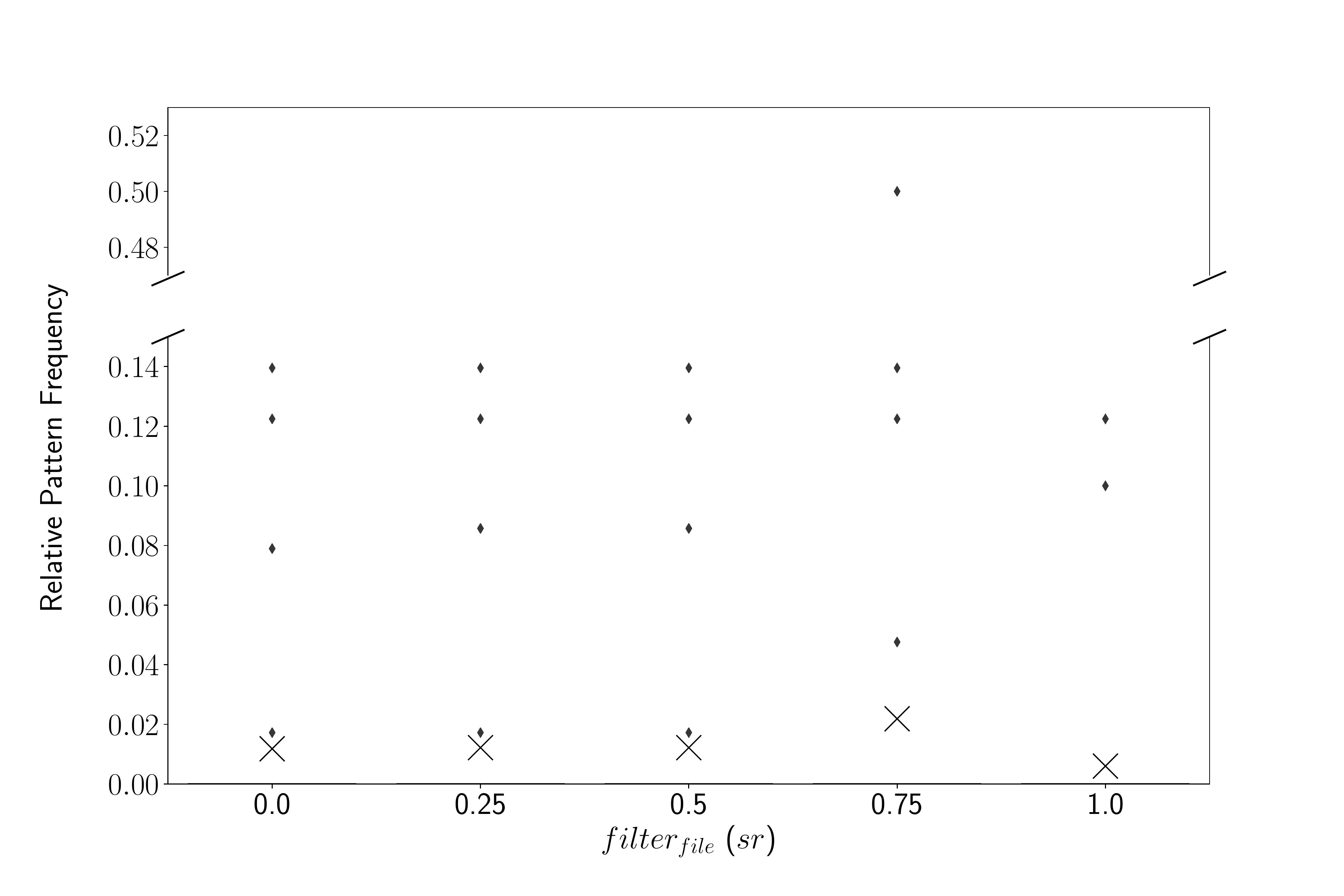}
		\subcaption{Internal/All Imports}
	\end{subfigure}
	\begin{subfigure}[c]{\textwidth}
		\includegraphics[width=\textwidth]{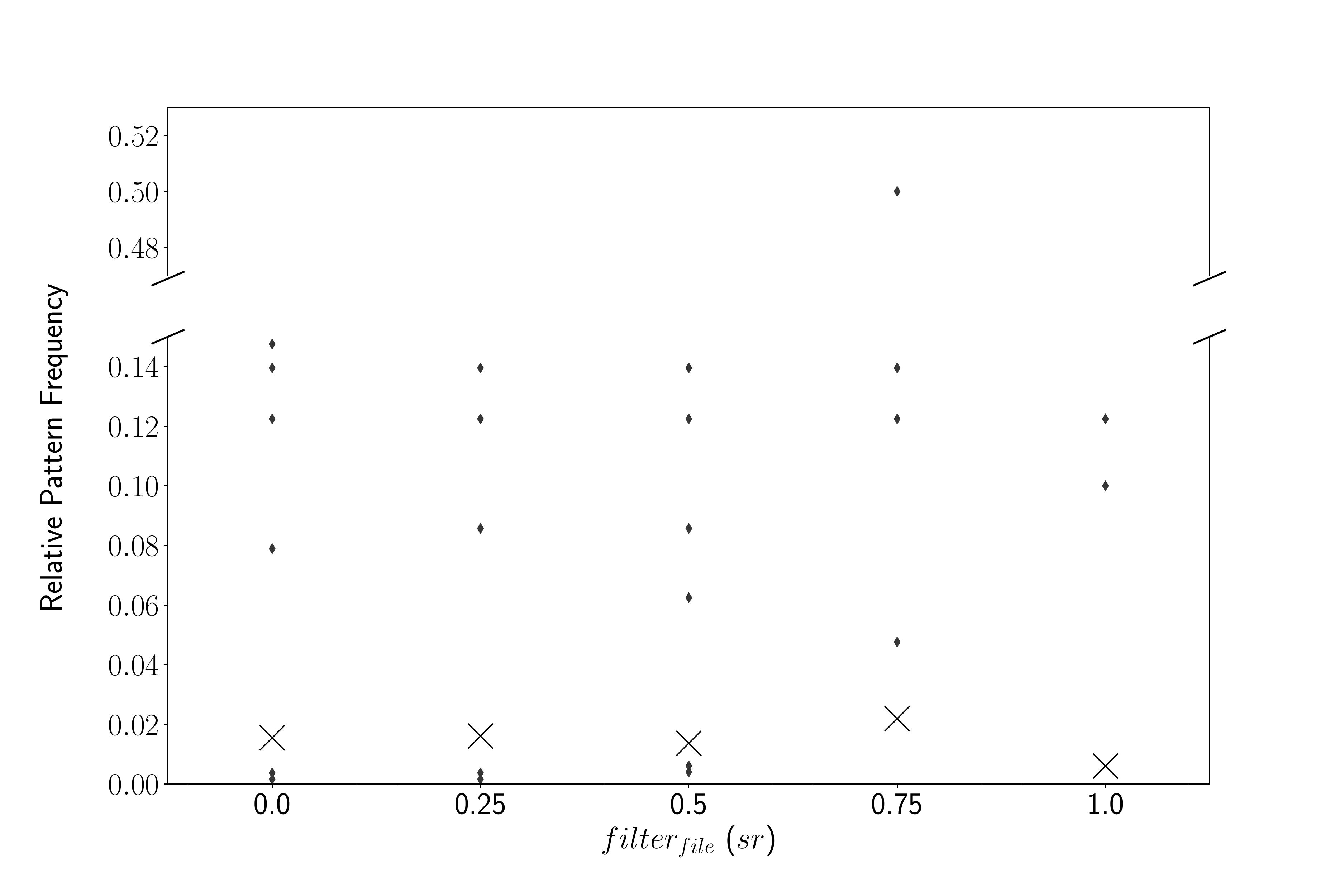}
		\subcaption{Internal/Misused Imports}
	\end{subfigure}
	\caption{Distribution of the relative pattern frequency using different file filter strategies grouped by each API search strategy of internally found source code}
	\label{fig:srs_internal}
\end{figure}

\begin{figure}
	\begin{subfigure}[c]{\textwidth}
	\includegraphics[width=\textwidth]{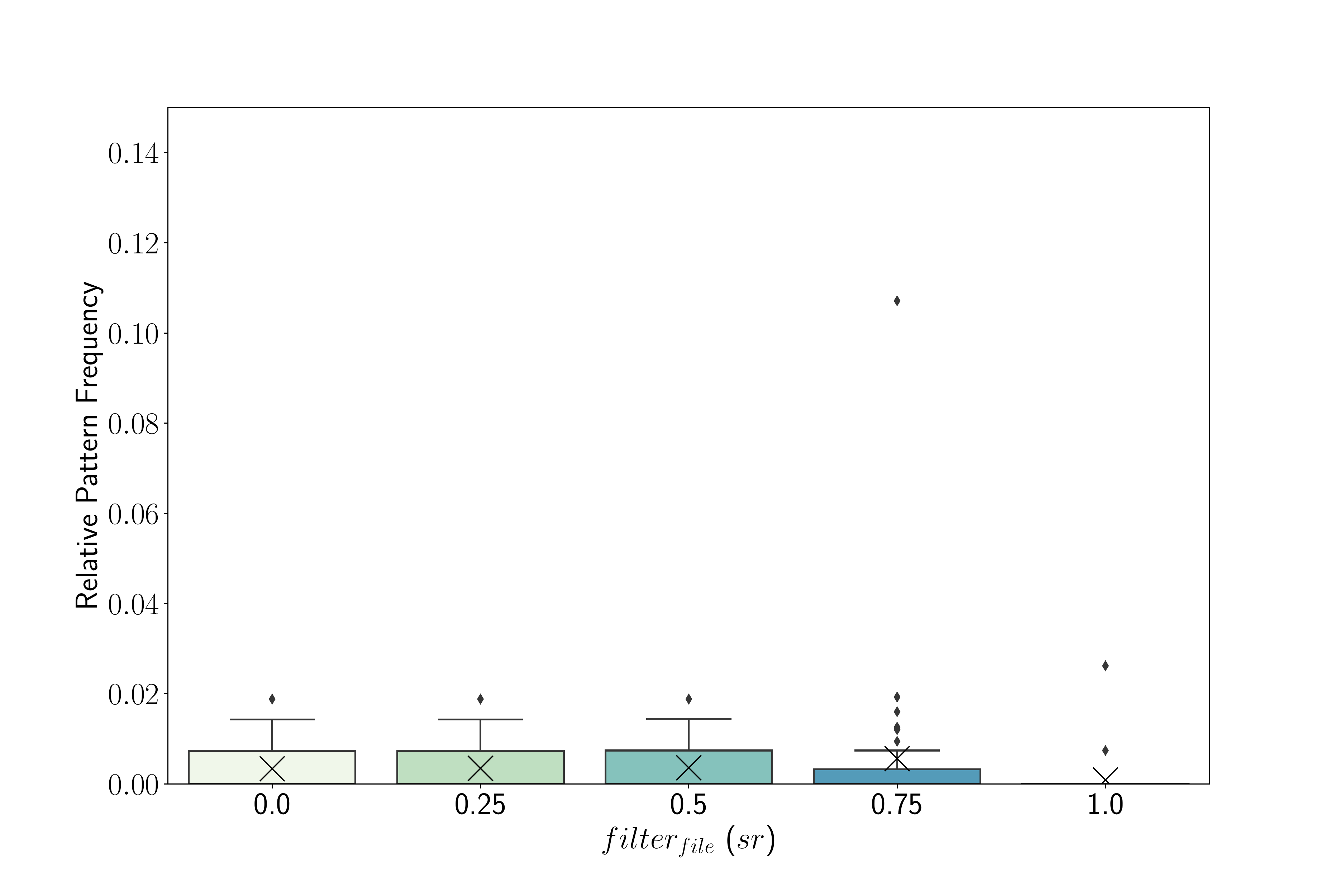}
	\subcaption{External/All Imports}
\end{subfigure}
\begin{subfigure}[c]{\textwidth}
	\includegraphics[width=\textwidth]{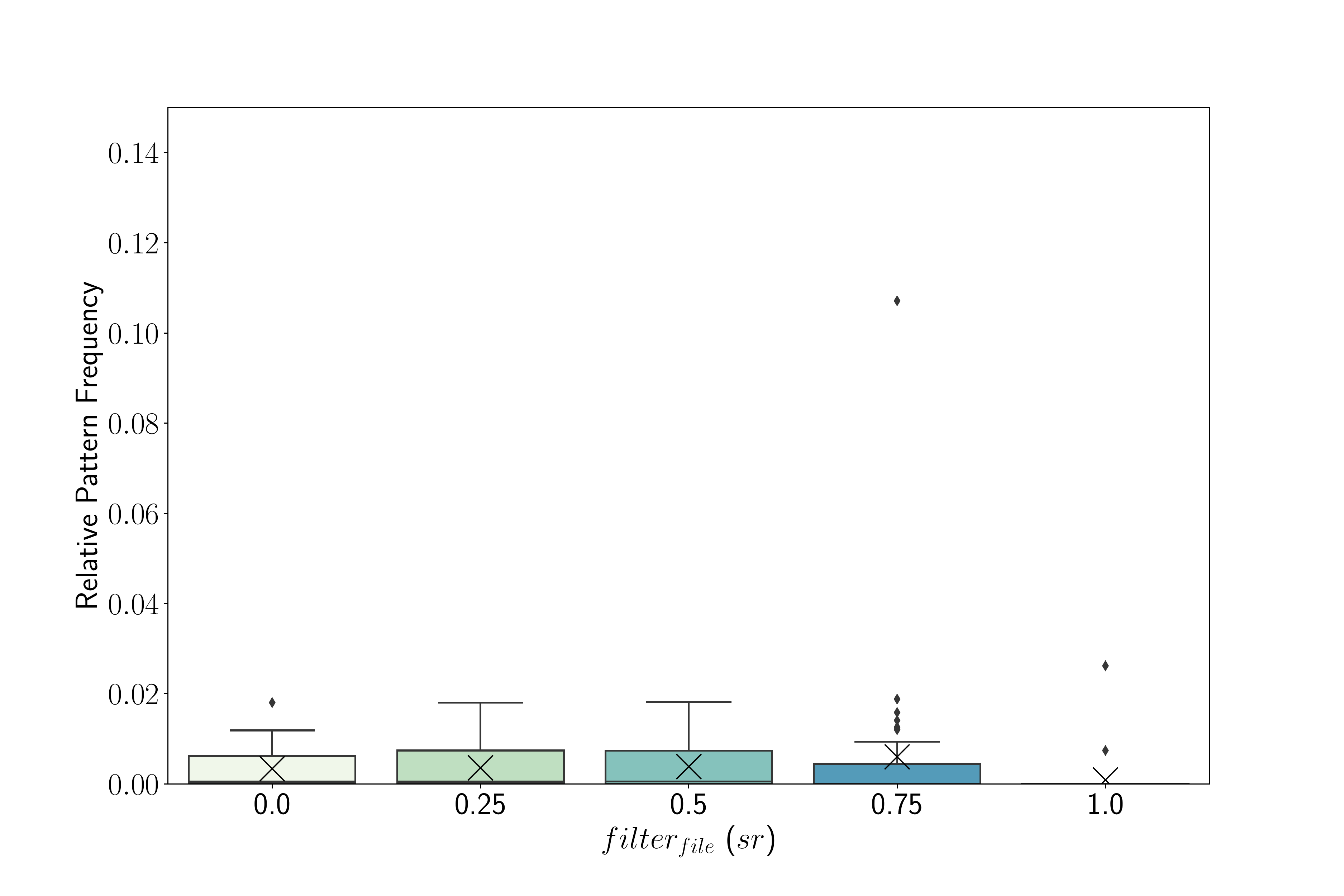}
	\subcaption{External/Misused Imports}
\end{subfigure}	
	\caption{Distribution of the relative pattern frequency using different file filter strategies grouped by each API search strategy of externally found source code}
\label{fig:srs_external}
\end{figure}

\begin{figure}
	\begin{subfigure}[c]{\textwidth}
		\includegraphics[width=\textwidth]{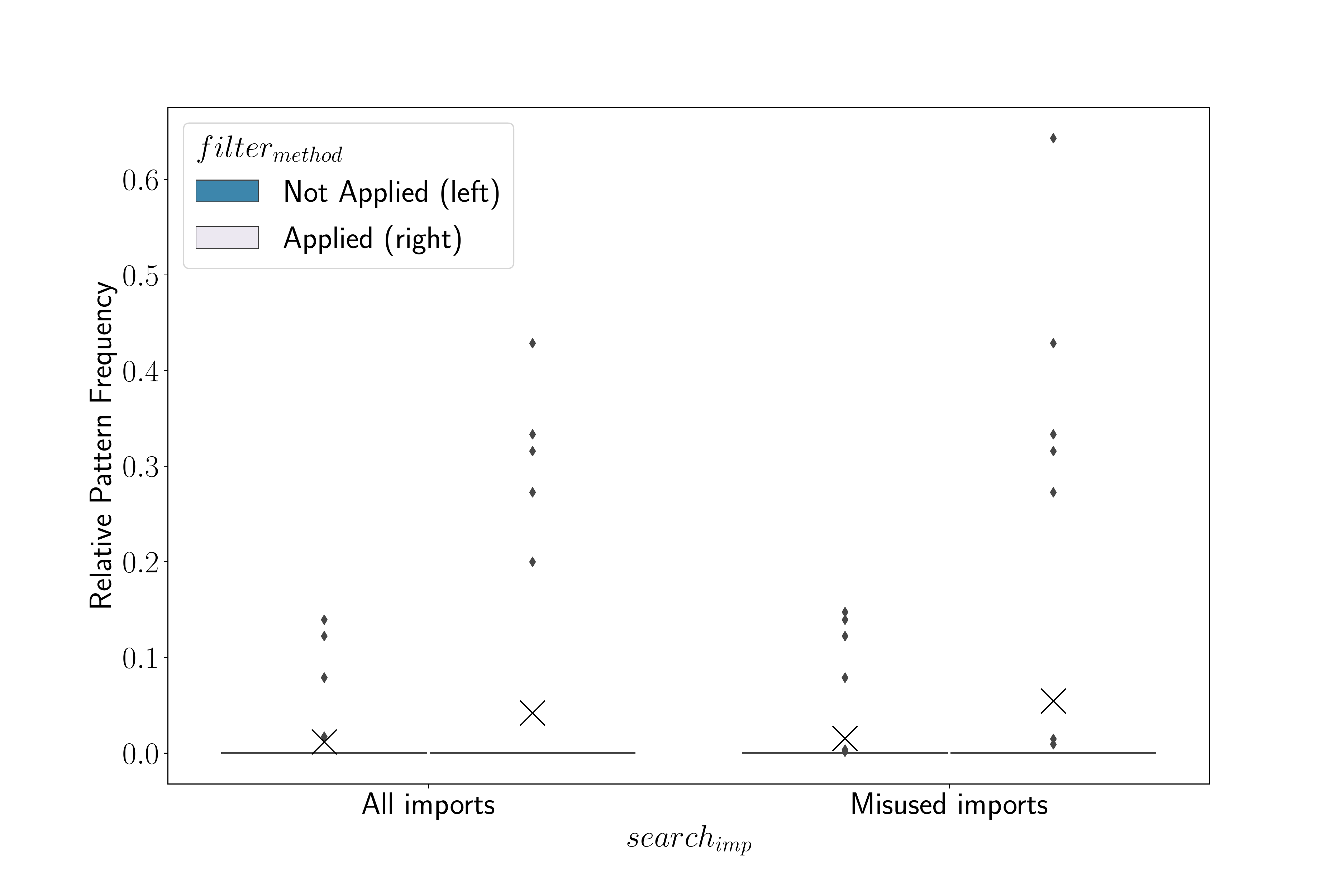}
		\subcaption{Internal File Search with an adapted y-axis for relative pattern frequency}
	\end{subfigure}
	\begin{subfigure}[c]{\textwidth}
		\includegraphics[width=\textwidth]{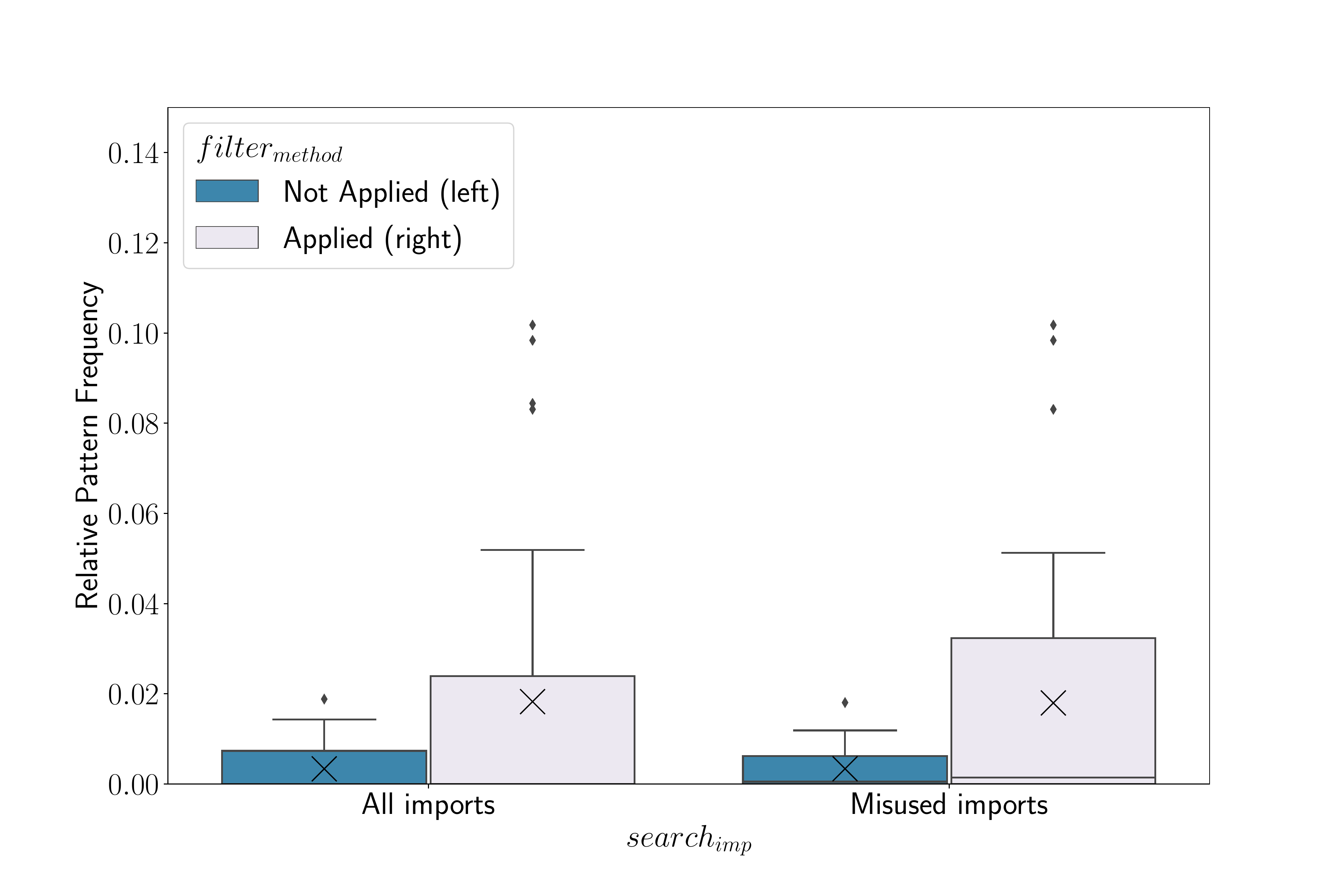}
		\subcaption{External File Search}
	\end{subfigure}
	\caption{Distribution of the relative pattern frequency applying the method filter strategy grouped by each File and API search strategy}
	\label{fig:methodFilter}
\end{figure}

Finally, we analyzed the $filter_{method}$ strategy. Our findings are that by using $filter_{method}$ we could find fixing patterns for 21 misuses while finding fixes for 22 misuses when applying no $filter_{method}$.
As depicted in \autoref{fig:methodFilter}, we observe a higher relative pattern frequency when applying the method filter strategy. Using the Wilcoxon signed-rank test, we could determine that the difference in distributions between applying and not applying $filter_{method}$ is significant. Note that due to the small number of unequal results for internal $search_{loc}$ the normal approximation used by the test might not hold and therefore should be taken with caution. However, this result indicates that $filter_{method}$ has a positive effect on increasing the relative pattern frequency.

We also analyzed the $sr$-values for each method that contains at least one keyword based on the raw source files (externally and internally collected as described in Section \ref{ssec:data}). The average $sr$ for internally found methods is $\approx0.11$ while for externally found ones it is slightly higher with $\approx0.19$. This supports our claim that the optimal $sr$ for method filtering is lower. Thus, assuming that the average number of extracted keywords is $6.3$  (cf. results of \ref{rq-1}), we consider a single keyword match is sufficient for method filtering since otherwise, we would remove too many methods (i.e., more than average). A more detailed analysis of the $sr$ in the methods can be found in our replication package\textsuperscript{\ref{fn:replpkg}}.

Regarding all 40 configurations, the overall best configuration (with mean $\approx0.058$) consisted of (1) using the internal $search_{loc}$, (2) using only the misused imports $search_{imp}$, (3) applying $filter_{file}$ with $sr=0.25$ and (4) applying the $filter_{method}$. The best configuration using the external $search_{loc}$ (mean $\approx0.026$)) used the misused imports $search_{imp}$, $filter_{file}$ with $sr=0.75$, and applying $filter_{method}$.

Once again, we did not precisely determine the execution times for the single filtering steps. Based on the timestamps of the generated files, we can recall that the filtering for a single misuse takes at most two minutes. Note that this time excludes the searching and downloading of similar source files on Searchcode.

\paragraph{Implications}

Our findings show that even though an internal $search_{loc}$ might not find a pattern as often as the external $search_{loc}$, their distributions do not significantly differ. Therefore, we suggest a cascaded approach, which first searches for a pattern within the project and then in foreign projects.

Moreover, we conclude that prior knowledge of the misused API is likely to have only little benefits compared to the effort associated with the required preliminary analysis (cf. the results when applying $search_{imp}$). A possible explanation of why multiple imports still find fixing patterns is that these imports describe a kind of context, i.e., APIs that are frequently used together. Including this context representation in the search increases the chance of finding fixing patterns that fit the actual misuse.

When filtering source code, we found that $filter_{file}$ up to $sr=0.5$ has a moderate effect on the relative pattern frequency, and thus may be applied to further reduce the number of source files. 

In contrast, the $filter_{method}$ proved to be far more effective at increasing the relative pattern frequency. A possible explanation is that searching keywords in the method scope yields a  more accurate representation of the misuse context than searching on the file level.

\subsection{Mining Analysis (\ref{rq-3})}
\label{ssec:results_rq3}
\paragraph{Methodology}

As versatile as the intermediate data representations are, there are also numerous ways of mining patterns from them. Usually, frequent pattern mining applies a variant of the well-known Apriori algorithm~\citep{Agrawal1993} with the extension of using closed pattern~\citep{Pasquier1999}. A closed pattern is a pattern for which all super-patterns have lower support values. 
Mining algorithms differently rank the patterns according to some metrics, e.g. by support (how frequently does the pattern occur in the data set) or by confidence (how frequently do the elements of the pattern co-occur with the pattern). It was also found out that other metrics could be more effective in pattern mining~\citep{Le2015}.

In our experiments, we applied the API usage pattern miner by Amann, since it works with the data structure of API usage graphs~\citep{Amann2018a}. This algorithm applies the Apriori algorithm with closed patterns on subgraphs by starting with individual AUG nodes and successively extending these depending on their neighbors in the AUG. For the extension process, the algorithm clusters isomorphic extensions, i.e., subgraphs. Note that, to cope with the graph isomorphism problem, they are using a graph vectorization heuristic. If the hash values of two graph vectors are equal, the graphs are considered isomorphic. Then a support threshold is used to identify which recurring sub-graphs should be reported as patterns. Further details can be found in the work by Amann et al.~\citep{Amann2018a,Amann2019}.

Further, we used the miner's \emph{cross-method} support definition which counts only in how many different methods a pattern occurs.

In a first experiment, we only considered those 22 misuses from MUBench for which we could find at least one fixing pattern in the previous evaluation. Essentially, we wanted to assess the effect of our selected filter strategies compared to a mining process that does no pre-processing of the source files at all.
For that purpose, we used the results of the source file collection as described in Section~\ref{ssec:data} as input for the un-filtered mining. Note that we conducted individual mining processes for internally and externally collected source files.
For the filtered case, based on our previous analysis, we selected the filter strategy by searching all API imports, setting $sr=0.5$, and applying the method filtering for both internally and externally collected source files.

Since we rank patterns regarding their support value, we have to set a minimum threshold. Based on our previous observation, we set the relative minimum support value (i.e.,  the ratio of absolute support to the number of all methods) for internal mining in both configurations (i.e., non-filtered vs. filtered) to $0.08$. This value is selected based on the lower quartile of the distribution of configurations using internal file filtering which has a relative pattern frequency greater than zero and rounded this value down.
For the external mining, we applied the same strategy based on the external filtering results and set the minimum relative support value to $0.004$ for both mining configurations (i.e., non-filtered vs. filtered). 

After mining, we sorted the patterns based on their absolute support (i.e., number of occurrences) and selected all patterns up to rank 20. Note that if multiple patterns had the same support, they share the same rank, while the next pattern with lower support has the next rank value increased by the number of patterns sharing the previous rank. For example, if two patterns share rank 1 then the next pattern has rank 3.

Then, the first two authors independently reviewed the patterns and validated whether they depicted fixing ones. Particularly, we compared them with the fixing patterns that we already used in Section~\ref{ssec:results_rq2} and reviewed - if necessary - the respective documentation of the API to check for equivalent patterns.
We further distinguished between the fixing type of \emph{is pattern} and \emph{is sub-pattern or equivalent pattern}. The rationale is that we rarely saw the pure form of the pattern by itself but rather the pattern being used as a sub-pattern in a bigger context. This is a result of closed pattern mining. Moreover, we found that some patterns to be semantically equivalent, which we did not consider as fixes in the first place.
Note that if we classified the pattern as \emph{is pattern}, we also marked it as \emph{is sub-pattern or equivalent pattern}. Hence the right-hand side of \autoref{tab:mining-result-no-filter} and \autoref{tab:mining-result-filter} can be considered as accumulated results.

We measured the agreement of the reviewers using Cohen's $\kappa$~\cite{Cohen1960}. Reviewing the filtered results of internal code, we found no disagreements (i.e., $\kappa=1$) regarding both cases, \emph{is pattern} and \emph{is sub-pattern or equivalent pattern}. For externally obtained patterns in the filtered case, we achieved a moderate agreement (i.e., $\kappa\approx0.43$) for the \emph{is pattern} review and substantial agreement (i.e., $\kappa\approx0.69$) for the \emph{is sub-pattern or equivalent pattern} review. 
Regarding the review of the unfiltered internal patterns, we could not compute $\kappa$ since we only obtained five negative results among which the reviewers agreed. For external patterns the reviewers had perfect agreement (i.e., $\kappa=1$) for the \emph{is pattern} and almost perfect agreement (i.e., $\kappa\approx0.84$) for \emph{is sub-pattern or equivalent pattern}. Note that the reviews of the filtered and non-filtered results were done by the same reviewers but the latter one several months later, which could constitute a bias. 
During the second review process, we noticed that we falsely classified one pattern type in the filtered results. Thus, we reevaluated these patterns individually in the respective misuses and corrected our results. In our replication package, we explicitly marked these corrected results and our Cohen's $\kappa$ computation also covers disagreements made in this reevaluation.
Finally, we then discussed the conflicting points to understand the reasons why the respective reviewer accepted or rejected a particular pattern as a fix. Based on our discussion, we then agreed on a final classification. 

In a second experiment, we analyze the effect of the filtering on each API usage in the AU500 dataset~\cite{Kang2021}. For that purpose, we conducted the internal and external code search for each method (including the exclusion steps as discussed before), applied the selected search and filter strategy (i.e., all imports, $sr=0.5$, and method filtering), and mined usage patterns for the filtered and non-filtered case. Since a manual validation of patterns for those 500 API usages was infeasible, we applied the automatic violation detection from MUDetect by Amann et al.~\cite{Amann2018a,Amann2019} to distinguish misuses from correct usages. While MUDetect employs many different violation techniques, we used a rather simple one that mines patterns with a certain minimum support and ranks the violations by the overlap between pattern and the usage. Particularly, this overlap between a pattern AUG $p$ and an arbitrary API usage AUG $u$ describes:

\noindent {$overlap(p,u) = \frac{|matchedNodes(p,u)|+|matchedEdges(p,u)|}{|nodes(p)|+|\{e | e \in edges(p) \land notConnectedToAny(e, nodes(p)\setminus nodes(u))\}|}$

where $matchedNodes$ and $matchedEdges$ denote the set of matched nodes and edges between the two AUGs, $nodes$ and $edges$ denote the set of nodes and edges of an AUG, and $notConnectedToAny$ is a predicate determining whether an edge $e$ is not connected to any node in the set represented by the second parameter (i.e., $nodes(p)\setminus nodes(u)$). Particularly, it checks whether the edge $e$ is not connected to any node from pattern $p$ missing in the API usage AUG $u$. This way, we obtain an \emph{overlap}\footnote{In the MUDetect implementation this value is denoted as confidence. Since this can be confused with the confidence definition from the pattern mining domain, we denote this value as \emph{overlap}.} ranging from 0 to 1, where 0 is associated with no overlap while 1 depicts a perfect overlap. In our analysis, we marked all results with an overlap of exactly 0 or 1 as correct usages while usages with an overlap between 0 and 1 are denoted as misuses. We further configured MUDetect to find patterns with a minimal absolute support of 2 (for internally found code) and 10 (for externally found code), respectively. Note that in contrast to the previous experiments, we do not use relative support values, since MUDetect only supports absolute values, and thus, we selected those values based on typical absolute support values, we have seen in the previous experiments. 
To keep our experiment in a reasonable time frame, we timeout the external mining for each entry after ten minutes and the internal mining after five minutes. 
Since the labels of AU500 represent a ground truth of correct usages and misuses, we report the difference between the filtered and non-filtered case by the precision and the recall. To this end, we mark a detected misuse (i.e., an overlap between 0 and 1) as true-positive if it was also labeled as misuse in the ground truth and as false-positive if not. In case the detection did not determine a misuse, we mark it as true-negative if it was labeled as a correct usage and as false-negative otherwise.

\paragraph{Results}

\begin{table}
	\centering
	\begin{tabular}{ll|cccc|cccc}
		\toprule
		\multicolumn{2}{l|}{\textbf{misuse}} & \multicolumn{4}{c|}{\textbf{pattern in Top}} & \multicolumn{4}{c}{\textbf{add. is sub-pattern or}} \\
		& & \multicolumn{4}{c|}{} & \multicolumn{4}{c}{\textbf{equiv. pattern in Top}} \\
		& & @1 & @5 & @10 & @20 & @1 & @5 & @10 & @20 \\	
		\toprule
		\multicolumn{2}{l|}{$\sum = 22$}& \textbf{3} & \textbf{3} & \textbf{4} & \textbf{8} & \textbf{3} & \textbf{6} & \textbf{7} & \textbf{8} \\
		\midrule
		1  &         alibaba\_druid\_1 &      - &      - &       - &       - &       - &       - &        - &        - \\
		2  &         alibaba\_druid\_2 &      - &      - &       - &       - &       - &       - &        - &        - \\
		3  &    android\_rcs\_rcsjta\_1 &      - &      - &       - &       - &       - &       - &        - &        - \\
		4  &        apache\_gora\_56\_1 &      - &      - &       - &       - &       - &       - &        - &        - \\
		5  &        apache\_gora\_56\_2 &      - &      - &       - &       - &       - &       - &        - &        - \\
		6  &                bcel\_101 &      - &      - &       - &       - &       - &       - &        - &        - \\
		7  &            jodatime\_269 &      - &      - &       E &       E &       - &       - &        E &        E \\
		8  &            jodatime\_361 &      E &      E &       E &       E &       E &       E &        E &        E \\
		9  &            jodatime\_362 &      E &      E &       E &       E &       E &       E &        E &        E \\
		10  &            jodatime\_363 &      E &      E &       E &       E &       E &       E &        E &        E \\
		11 &           logblock\_2\_15 &      - &      - &       - &       - &       - &       - &        - &        - \\
		12 &                mqtt\_389 &      - &      - &       - &       - &       - &       - &        - &        - \\
		13 &                mqtt\_390 &      - &      - &       - &       - &       - &       - &        - &        - \\
		14 &         tbuktu\_ntru\_473 &      - &      - &       - &       - &       - &       - &        - &        - \\
		15 &         tbuktu\_ntru\_474 &      - &      - &       - &       - &       - &       - &        - &        - \\
		16 &         tbuktu\_ntru\_475 &      - &      - &       - &       - &       - &       - &        - &        - \\
		17 &               testng\_16 &      - &      - &       - &       - &       - &       - &        - &        - \\
		18 &      thebluealliancea\_1 &      - &      - &       - &       - &       - &       - &        - &        - \\
		19 &  thomas\_s\_b\_visualee\_29 &      - &      - &       - &       E &       - &       E &        E &        E \\
		20 &  thomas\_s\_b\_visualee\_30 &      - &      - &       - &       E &       - &       E &        E &        E \\
		21 &  thomas\_s\_b\_visualee\_32 &      - &      - &       - &       E &       - &       E &        E &        E \\
		22 &             ushahidia\_1 &      - &      - &       - &       E &       - &       - &        - &        E \\
		\bottomrule
		\multicolumn{9}{c}{E: pattern found in external source files;}\\	\multicolumn{9}{c}{I: pattern found in internal source files;}\\
		\multicolumn{9}{c}{I/E: pattern found in internal and external source files}
	\end{tabular}
	\caption{Number of fixing patterns found in the Top@k patterns by mining without any filtering in the MUBench dataset}
	\label{tab:mining-result-no-filter}
\end{table}

\begin{table}
	\centering
	\begin{tabular}{ll|cccc|cccc}
		\toprule
		\multicolumn{2}{l|}{\textbf{misuse}} & \multicolumn{4}{c|}{\textbf{pattern in Top}} & \multicolumn{4}{c}{\textbf{add. is sub-pattern or}} \\
		& & \multicolumn{4}{c|}{} & \multicolumn{4}{c}{\textbf{equiv. pattern in Top}} \\
		& & @1 & @5 & @10 & @20 & @1 & @5 & @10 & @20 \\	
		\toprule
		\multicolumn{2}{l|}{$\sum = 22$}& \textbf{4} & \textbf{7} & \textbf{7} & \textbf{8} & \textbf{5} & \textbf{9} & \textbf{10} & \textbf{13} \\
		\midrule
		1  &         alibaba\_druid\_1 &      - &      - &       - &       - &       - &       - &        - &        - \\
		2  &         alibaba\_druid\_2 &      - &      - &       - &       - &       E &       E &        E &        E \\
		3  &    android\_rcs\_rcsjta\_1 &      - &      - &       - &       - &       - &       - &        E &        E \\
		4  &        apache\_gora\_56\_1 &      - &      - &       - &       - &       - &       - &        - &        - \\
		5  &        apache\_gora\_56\_2 &      - &      - &       - &       - &       - &       - &        - &        - \\
		6  &                bcel\_101 &      - &      - &       - &       - &       - &       - &        - &        - \\
		7  &            jodatime\_269 &      - &      - &       - &       E &       - &       - &        - &        E \\
		8  &            jodatime\_361 &      E &      E &       E &       E &       E &       E &        E &        E \\
		9  &            jodatime\_362 &      - &      E &       E &       E &       - &       E &        E &        E \\
		10  &            jodatime\_363 &      E &      E &       E &       E &       E &       E &        E &        E \\
		11 &           logblock\_2\_15 &      - &      - &       - &       - &       - &       - &        - &        - \\
		12 &                mqtt\_389 &      - &      I &       I &     I/E &       - &       I &        I &      I/E \\
		13 &                mqtt\_390 &      - &      I &       I &       I &       - &       I &        I &        I \\
		14 &         tbuktu\_ntru\_473 &      - &      - &       - &       - &       - &       - &        - &        E \\
		15 &         tbuktu\_ntru\_474 &      - &      - &       - &       - &       - &       - &        - &        E \\
		16 &         tbuktu\_ntru\_475 &      - &      - &       - &       - &       - &       - &        - &        - \\
		17 &               testng\_16 &      - &      - &       - &       - &       - &       - &        - &        - \\
		18 &      thebluealliancea\_1 &      - &      - &       - &       - &       - &       - &        - &        - \\
		19 &  thomas\_s\_b\_visualee\_29 &      I &      I &       I &       I &     I/E &     I/E &      I/E &      I/E \\
		20 &  thomas\_s\_b\_visualee\_30 &      I &      I &       I &       I &     I/E &     I/E &      I/E &      I/E \\
		21 &  thomas\_s\_b\_visualee\_32 &      - &      - &       - &       - &       - &       - &        - &        - \\
		22 &             ushahidia\_1 &      - &      - &       - &       - &       - &       E &        E &        E \\
		\bottomrule
		\multicolumn{9}{c}{E: pattern found in external source files;}\\	\multicolumn{9}{c}{I: pattern found in internal source files;}\\
		\multicolumn{9}{c}{I/E: pattern found in internal and external source files}
	\end{tabular}
	\caption{Number of fixing patterns found in the Top@k patterns by mining with our predefined configurations in the MUBench dataset}
	\label{tab:mining-result-filter}
\end{table}

\autoref{tab:mining-result-no-filter} and \autoref{tab:mining-result-filter} depict the results of our evaluation of the MUBench dataset regarding the API usage pattern mining for the non-filtered and the filtered mining configurations, respectively. 

In the unfiltered configuration, the miner found the fixing pattern in its `pure' form for four misuses in the Top@10 and eight misuses in the Top@20 most frequent patterns, respectively. If we relax the requirement so that the pattern can be a sub-pattern or an equivalent (sub-)pattern, this number increases to seven for Top@10. For the Top@20 case, it remains constant.

In the filtered configuration the `pure' form was found for seven misuses in the Top@10 and eight misuses in the Top@20. Concerning the relaxed case, the numbers increase to 10 and 13 misuses for the Top@10 and Top@20 frequent patterns, respectively.

Since our mining used the same original data and applied the same configuration (i.e., minimum support), we observed that our selected searching and filtering strategy had a positive effect on the number of detected fixing patterns for the analyzed misuses.
We analyzed whether the difference in the number of occurrences of fixing patterns in the Top@20 is significant. For that purpose, we used the $\chi^2$-test ($\alpha=0.05$), which due to the small sample size of 22 analyzed misuses, was corrected with the Yates correction ~\cite{Yates1934}. Based on this test the difference is not significant.

However qualitatively, we observed in the unfiltered case that fixing patterns were solely inferred from the externally found source code, while in the filtered case it found patterns from the internal code as well. We assume that this is due to the fact that the external code is being filtered to some degree (i.e., it was collected by searching with related imports on Searchcode), while the internal code was simply used as it is. 
Moreover, the unfiltered mining only found fixing patterns for misuses from three different projects, while the filtered mining does so for seven. Thus, this indicates that the positive effect of our strategy is more general in terms of different projects.

In more detail, the filtered case found more patterns, sub-patterns, or equivalent patterns in external code than in internal code, i.e., 12 compared to 4. For three misuses (i.e., \texttt{mqtt\_389}, \texttt{thomas\_s\_b\_visualee\_29}, and \\\texttt{thomas\_s\_b\_visualee\_30}) both strategies, i.e., internal and external, found a fixing pattern. This indicates that the combination of the two configurations, namely, internal and external code search, is beneficial.

Nevertheless, for nine misuses, the filtered case did not find a fixing pattern in the Top@20 highest-ranked patterns. Of these, the miner could not obtain any patterns for three misuses, namely, \texttt{apache\_gora\_56\_2}, \texttt{testng\_16}, and \texttt{thomas\-\_s\_b\_\-visualee\_32}. The miner failed to find patterns for the external configuration only in these three cases, while for the internal configuration, it did not find any patterns for 17 misuses. For seven misuses, we found patterns but could not retrieve a fixing pattern in the set of Top@20 ranked patterns. 

For the cases \texttt{jodatime\_269}, \texttt{thomas\-\_s\_b\_\-visualee\_29}-\texttt{32}, and \texttt{ushahidia\_1} the filtering actually removed too many valid occurrences so that either the fixing patterns were not ranked as highly as without filtering or did not find the patterns at all.

During the analysis of the filtered results, we made the observation that many patterns tend to be very close to the fixing pattern but miss some essential parts to be considered as a solution. Moreover, we observed many very similar patterns distributed among the highest-ranked patterns. By merging these sub-patterns one may re-build the fixing pattern.

Regarding the execution time of the miner step, we can roughly recall from the timestamps of the generated files that a single execution in most cases takes at most a minute regardless of whether the results were filtered or not. Only in a single case (filtered external source files of \texttt{thomas\_s\_b\_visualee\_29}), the mining took almost 18 minutes.

\begin{table}
	\centering
	\begin{tabular}{p{1.5cm}lrrrrrr}
		\hhline{--------}
		\textbf{Applied filtering} & \textbf{$search_{loc}$} & \textbf{\#tp} & \textbf{\#fp} & \textbf{\#tn} & \textbf{\#fn} &  \textbf{precision} &   \textbf{recall} \\
		\hhline{--------}
		\multirow{3}{*}{No} &   external &   8 &  16 & 369 & 107 &    33.33\% &  6.96\% \\
		& internal &   1 &  10 & 375 & 114 &     9.09\% &  0.87\% \\
		\hhline{~-------}
		 & \cellcolor{Lightgray}both &  \cellcolor{Lightgray}9 &  \cellcolor{Lightgray}26 & \cellcolor{Lightgray}359 & \cellcolor{Lightgray}106 &    \cellcolor{Lightgray}25.71\% &  \cellcolor{Lightgray}7.83\% \\		
		\hhline{--------}
		\multirow{3}{*}{Yes} &   external &   8 &  12 & 373 & 107 &     40.0\% &  6.96\% \\
		& internal &   9 &  27 & 358 & 106 &     25.0\% &  7.83\% \\
		\hhline{~-------}
		&	\cellcolor{Lightgray} both &\cellcolor{Lightgray} 13 &  \cellcolor{Lightgray}31 & \cellcolor{Lightgray}354 & \cellcolor{Lightgray}102 &    \cellcolor{Lightgray}29.55\% &  \cellcolor{Lightgray}11.3\% \\
		\hhline{--------}
	\end{tabular}
	\caption{Results of the Misuse Detection on the AU500 dataset using the violation detection technique from MUDetect with number of true positives (\#tp), false positives (\#fp), true negatives (\#tn), false negatives (\#fn), precision, and recall}
	\label{tab:au500_results}
\end{table}

Finally, we analyzed the effect of the previous searching and filtering on the violation detection by MUDetect~\cite{Amann2018a,Amann2019} applied on the AU500 dataset by Kang et al.~\cite{Kang2021}. \autoref{tab:au500_results} summarizes our results, by individually depicting the results of externally and internally searched code as well as a cascaded approach by applying both search strategies. Please note, that the row showing the values for both strategies does not necessarily represent the sum of the external and internal row since the sets of detected misuses from external and internal search can overlap.
Moreover, we observed that we could apply our approach only to 480 out of those 500 API usages in the dataset. In our computation, we considered these 20 entries as if the approach did not detect a misuse (i.e., they add either to \#tn or \#fn).

In general, we can observe a positive effect of the filtering. Particularly, the number of true positives for the internal search increased, and thus, their precision and their recall. When analyzing the reasons, we found out that this was an effect of the decreased number of AUGs to generate and mine from. In case we did not use the filtering strategy, the number of methods was too large, and thus, only 20 API usages could be analyzed using the internal,  non-filtered case, while the others were interrupted by the timeout. In contrast, for the filtered case, we were enabled to analyze 453 API usages. Regarding the external case, we could slightly improve the precision due to a lower number of false positives, while the recall was consistent. Interestingly, even though the number of true positives is equal for both external strategies, the detected API misuses differ. In detail, both strategies, the filtered and non-filtered external search, independently enabled the violation detection to correctly detect a misuse for a common set of five misuses. Every individual strategy, however, enabled the violation detection to find misuses for three API usages, which were not detected using the respective other strategy. The cause of this is that the filtering decreases the support values for those three patterns too much pushing it below the minimum support. In the other, non-filtered case, we observed that the mining was interrupted by a timeout. 
For the cascaded approach, we observed a lower precision than for the external setup, while the recall is higher. When applying the filtering, we observe an increase for both precision (i.e., $+3.84\%$) and recall (i.e., $+3.47\%$) compared to the non-filtered case. A subsequent $\chi^2$-test ($\alpha=0.05$) found the difference of the true positive values to be non-significant. While we found the precision to be comparable with previous results~\cite{Amann2019, Kang2021} the recall is rather low. We will discuss the implications of this observation in the following paragraph.

\paragraph{Implications}

In both experiments on the MUBench~\cite{Amann2016} and the AU500 dataset~\cite{Kang2021}, we could not determine a significant effect on the difference in the number of found fixing patterns and true positives when applying filtering compared to no filtering. However, from a qualitative perspective, we found that filtering still has a positive effect, for instance, we found fixing patterns for more distinct projects and based on internally collected source files. Its main effect, however, is the reduction of code samples, allowing faster mining on fewer examples without negatively influencing the misuse detection capability. 

Further, we found that our search strategy is not perfect. For MUBench, we found patterns for around 59\% of the considered misuses and around 35\% of all 37 misuses. For the AU500 dataset, the maximum achieved recall was 11.3\%. Therefore, we will consider in future work whether other code search strategies perform better. For instance in the work by Kang et al.~\cite{Kang2021}, the authors applied the recently published artifact \texttt{AUSearch}~\cite{Asyrofi2020} and achieved higher recall values. 
Moreover, we applied very simple ranking strategies (support and violation overlap). As it has been shown by previous work, other metrics could further improve the results~\citep{Le2015,Amann2018a}. Additionally, recent research has come up with potential further datasets for API misuses~\cite{Nielebock2021, Kechagia2021}, which provide a more diverse set of validation data.

We could confirm that the combination of both code search strategies contributed to retrieving more fixing patterns, while the external strategy found patterns for more misuses than the internal one in the MUBench dataset. In the AU500 dataset, both had an almost equal number of true positives while the external search achieved usually a higher precision.

Additionally, we observed many very similar patterns as well as incomplete patterns in the ranking of the filtered results of MUBench. By clustering these patterns, similar to previous work~\citep{zhong2009mapo}, we could decrease the number of similar patterns to a lower number of clusters. The patterns within these clusters would then represent a set of different possible options for applying an API in a particular context. Depending on the strategy, one could simply pick the most frequent pattern from each cluster or merge patterns regarding some heuristic to re-combine the missing parts. This will be part of future work.

\section{Threats to Validity}
\label{sec:threats}

The validity of our evaluation could be subject to different threats. With respect to related literature \citep{Siegmund2015}, we consider threats to \emph{internal} and \emph{external validity}.

\paragraph{Internal Validity}

Internal validity describes to which degree we can trust our results. Particularly, errors made in our process could harm the robustness of our results.

In our concept, we rely on identifying similar source code which is likely to contain fixes for a particular misuse. For external search, we used \emph{SearchCode}, which leverages data from different code repositories. Depending on the concrete misuse, significant time may have passed between the misuse introduction and our similar code search. Therefore, the discovered code may not yet have existed at the time the misuse was first introduced.
This may imply that the changed-based approach might find fewer fixes when executed just-in-time.

Moreover, the search results are biased through \emph{Searchcode}'s search algorithm, as a different external search might perform differently. However, we found the subsequent search and filter strategies to perform similarly among externally and internally found code sets. In fact, we found the same strategies for both sets as best fitting for the mining process of \ref{rq-3}. Additionally, we kept a consistent data set for all search strategies so that we assume this effect to be almost equal for all strategies. Based on this assumption, we still expect that our results express which strategy works best. 

Even though we filtered the similar code to exclude any files originating from the source project of the misuse, our process cannot guarantee that we do not find code originating from forks of that project.
However, these threats are to a large degree mitigated when, in particular for frequently used APIs, we find many similar usages.

Our process is capable of inferring fixing patterns. However, we can guarantee neither the patterns' correctness nor their completeness. Regarding the former, we did not check whether the fixing pattern would introduce new errors. The latter is due to the fact that we can never be sure to have found all possible variations of a fix. To some extent, this threat is again mitigated when the fixing patterns exhibit higher support values, as this would likely favor the relevant and more general fix variants.

We applied the relative pattern frequency to compare different search and filter strategies in their ability to find fixing patterns. However, this metric could be biased in case we retrieved a very low number of source files. For example, the number of internally found files is usually lower than those of external ones. Thus, comparative statistics such as the differences in means might be only by chance.

Finally, the first two authors independently validated whether the mined patterns represented the respective fix or not. However, as manual validation always has subjective aspects, this may introduce bias or noise into our evaluation. Moreover, the two separate review phases were separated by a gap of several months, which may bias the agreement metrics since the reviewers had a learning experience from the first review.
For this reason, we published all our results, data, and scripts as a replication package\textsuperscript{\ref{fn:replpkg}} and invite other researchers to re-validate our findings.

\paragraph{External Validity}

External Validity describes to which degree our results generalize to unknown data, i.e. other misuses from different projects.

Our methodology only considers static and intra-procedural API usage patterns, i.e., patterns within the scope of a single method declaration. However, API usage patterns may also be scattered among several methods, i.e., inter-procedural or may only be detected when the code is explicitly executed (i.e., dynamic pattern inference). Since our methodology does not detect such inter-procedural or dynamically inferred patterns, our results currently only refer to static, intra-procedural API usage pattern miners. 

Our MUBench-based case study features many similar misuses. Regarding this point, Sven Amann, one of the authors of MUBench, noted ``\emph{The benchmark dataset may not be representative for API misuses in the wild}''\citep[p. 75]{Amann2018a}. Therefore, if MUBench is subject to this limitation, the same is likely true for our case study.

Moreover, most of our evaluation considers only a small set of 37 real API misuses from MUBench. While specifically preprocessed to reduce potential bias (e.g., removing duplicates), future work may validate our results on larger datasets of API misuses such similar to ours on the AU500 dataset~\cite{Kang2021} as well as recently published ones~\cite{Nielebock2021, Kechagia2021}.

Our method and analysis refer only to API misuses in Java. We did not study whether this method would perform similarly for other programming languages, however, if the keyword extraction is adapted, this is arguably the case for other procedural and object-oriented languages. Regarding other programming paradigms, an adaptation would likely also require conceptual adaptations.

\section{Related Work}
\label{sec:related}

Our work relates to four further software engineering fields, which we consider in the following.

\subsection{Change-Based Error Detection}
\label{ssec:change-based}

The idea of detecting bugs based on commits is not new. Originally, these works investigated metrics indicating suspicious commits that introduced bugs. This is also known as just-in-time bug detection. Mockus and Weiss used change properties such as size or diffusion (e.g., number of distinct files that have to be changed) and build a model based on a logistic regression to estimate the probability of an error~\cite{Mockus2000}. Sliweski et al. introduced the SZZ-algorithm by which they could identify bug-introducing commits by using the version control system~\cite{Sliwerski2005}. We also used this algorithm to determine the API misuse-introducing commits.
Kim et al. trained a support vector machine (SVM) based on information from the source code metadata and achieved an accuracy of 78\% to detect bug-inducing commits~\cite{Kim2008}. However, they found the model to be too project-specific to be globally usable.
Instead Kamei et al.~\cite{Kamei2013}, similarly to Mockus and Weiss' idea, used a logistic regression and found a generic model achieving an average accuracy of 68\%. Their model indicates that commits with files that get large and frequent changes are associated with introducing bugs.
An and Khomh~\cite{An2015} confirmed this observation regarding the changes' sizes and found further correlations with low developers-experience, longer commit messages, and changes distributed across multiple files.
Augmenting these general characteristics of suspicious commits, we have begun to investigate the influence of API-specific information as shown in our preliminary work~\cite{Nielebock2018}.
Other tools like ChangeLocator~\cite{Wu2017} and ChangeRanker~\cite{Guo2020} improve the accuracy by applying information from automatically collected crash reports. Since we aim to detect API misuses at the time of the commit, this data is usually not available.
Other approaches detect bugs by modeling code changes as logic rules and try to detect exceptions from these rules within the code change~\cite{Kim2009}. Similarly, some automated bug detection and repair techniques use previous or historical bug fixes~\cite{Kim2006, Sun2010, Le2016, Long2016}. In contrast, our approach considers API usages at the time of a commit. However, we also worked on re-using previous fixes to detect misuses in other repositories~\cite{Nielebock2020}. Whether using historical fixes is beneficial for our approach will be the subject of our further work.

\subsection{API Selection and Usage Recommendation}
\label{ssec:api-recommendation}
API recommendation aims to assist developers in selecting suitable APIs for their use-case and also correctly applying those APIs.
While in our case the API selection is fixed by the given source code, some of the search strategies from this field inspired our approach.\\
Saul et al. implemented the FRAN (Find with RANdom walks) algorithm, which, given an API and a particular function, finds closely related functions of the same API~\cite{Saul2007}. The \texttt{Prospector} assists developers in creating objects by recommending code based on the desired type of the object and likely-relevant parameters~\cite{Mandelin2005}. Similarly, Chan et al. use a subgraph-based algorithm to find relevant code regarding a textual query~\cite{chan2012searching}. Other approaches use textual input from feature requests~\cite{thung2013automatic} or search for textual queries using additional sources such as code documentations or Q\&A webpages~\cite{Lv2015, rahman2016rack}. Thung et al. also developed an approach that suggests complete libraries based on the respective APIs currently used in a client project~\cite{Thung2013}. The \texttt{MUSE} approach finds usage examples for individual specified API methods by means of static slicing for simplification and various heuristics for ranking~\cite{moreno2015can}. 
In our approach, we include data types, called methods, and import statements into the context of the API usage and use these to find and filter similar API usages. Note that usually, we do not use all information at the same time. Moreover, we do not expect to have access to the declaration of those functions that are called by the query function.
A very recent approach uses API embeddings~\cite{Nguyen2017, Chen2019}, or joint natural text and API embeddings~\cite{Huang2018} to find similar API usages. These techniques enable finding semantically equivalent APIs even if they do not share syntactical similarities. This, however, requires the training of a neural network, which would not scale for our envisioned use case. Moreover, applying semantically equivalent APIs would require significant changes to the code (i.e., substituting complete libraries) and would thus potentially introduce further sources of bugs.

\subsection{Code Search and Code Recommendation}
\label{ssec:code search}

Code search is usually referred to as the task performed by developers to retrieve code of varying size from code snippets up to complete packages~\cite{gallardo2009internet}. The main motivations for code search are code reuse, -repair, -understanding, -location, impact analysis, or finding suitable third-party libraries~\cite{gallardo2009internet, sadowski2015developers, Xia2017}. This indicates that the code search is heavily domain and use-case-specific.
However, still, developers tend to prefer general-purpose search engines for code search~\cite{sadowski2015developers, Rahman2018}.
In contrast, our goal was to find similar code examples without requiring human interaction.

Several automatic approaches aim to facilitate the search process and the quality of the results. These approaches range from using a domain-specific languages~\cite{paul1994framework}, the code context~\cite{Holmes2005, Sahavechaphan2006}, test cases~\cite{lemos2007codegenie}, static and dynamic specifications in the form of method signatures and test cases~\cite{reiss2009semantics}, the documentation and slicing techniques~\cite{kim2010towards}, textual matching based on different ranking and natural language processing mechanism ~\cite{McMillan2011}, input/output code examples using SMT-solver (Satisfiability modulo theories solver)~\cite{Stolee2014}, or learned neural code embeddings~\cite{Gu2018}. All these approaches usually aim to find accurate search results. In contrast, our approach can cope with a certain degree of `noise' in the search results due to subsequent filter and mining steps. This allows us to keep the search algorithm as lightweight as possible. 

During the development of this work, the \texttt{AUSearch} tool was published using user queries to find similar API usages~\cite{Asyrofi2020}. Their technique uses type resolutions to retrieve better matching samples. It was successfully applied by Kang et al.~\cite{Kang2021} when applying their API misuse detector. However, they applied the user queries manually, while our approach is intended to run fully automatically. In future work, we incorporate \texttt{AUSearch} in the overall process.

Note that the keyword search using the context is very similar to the notions presented in Strathcona~\cite{Holmes2005} and the XSnippet tool~\cite{Sahavechaphan2006}. These tools, like our approach, used inheritance, type- and method-call information from the method declaration to retrieve similar code. Unfortunately, both tools were not available at the time we conducted our experiments. 

Some recent work on automated program repair by Xin and Reiss used code search to reduce the typical huge search space for patches~\cite{Xin2017, Xin2019}. 
In their work, they made the interesting observation that it is worth using different search strategies for internal and external code. We will consider this, as a potential extension of our work.

\subsection{Code Clone Detection}
\label{ssec:clone-detection}

Finding similar source code is related to retrieving syntactically and semantically equivalent source code, namely, code clones. The motivation for detecting code clones is manifold and includes reducing maintenance effort, detecting plagiarism, code compaction, analyzing software evolution, and bug detection.
A number of tools were developed to support clone detection ~\cite{Koschke2007,roy2009comparison}.
\texttt{Deckard} is well-known and uses parse tree vectorization to find clones~\cite{jiang2007deckard}. In the recent research other tools aim to find different levels of clones ranging from Type-3 (near-miss) to Type-4 (same functionality)~\cite{sajnani2016sourcerercc,white2016deep,Saini2018}.
For our use-case, however, clone detection is less relevant, since the fixing code for a target API misuse by definition cannot be a clone of the former.
\section{Conclusion and Further Work}
\label{sec:conclusion}

Recent research came up with a variety of automatic API misuse detectors that rely on the idea of inferring correct API usages from existing code samples.
However, we discovered that most approaches do not consider how these code samples can be collected in practice.
Therefore, this paper introduces a new approach to collect and improve the input source code files for API usage pattern mining to more effectively find patterns for API misuses detection. This approach uses a program-change analysis combined with similar code search and a filtering strategy. We also introduce a concept with which this approach can easily be integrated into an ordinary continuous integration process. This concept has two advantages:
First, it applies API misuse mining in a just-in-time manner whenever developers commit changes in their code to the version control system. Second, to improve the results of the miner, based on the concrete change, it applies different search and filter strategies to increase the relative frequencies of true positive patterns in a set of code samples before mining.

In our experiments, we determined the overall best search and filter strategy by analyzing 37 well-known API misuses from the MUBench dataset~\cite{Amann2016} and selecting that strategy that achieved the highest relative frequency. Using this strategy, we analyzed the effect on the pattern mining and misuse detection using the tooling by Amann et al.~\cite{Amann2018a,Amann2019} and two different API misuse datasets~\cite{Amann2016,Kang2021}. 

Our main findings are:
\begin{enumerate}
	\item Considering only changed methods that modified the usage of third-party libraries can effectively reduce the number of methods to investigate (i.e., on average reduction of 86.4\% of methods to be analyzed). 
	\item Both, internal (i.e., within the project) and external (i.e., in other projects) code search contribute to more fixing patterns being found.
	\item Including knowledge of which API was misused into the search has only a negligible effect and therefore it is generally sufficient to search for similar API usages without exactly knowing the misused API.
	\item File filtering with keywords has only a moderate effect on increasing the relative frequency of patterns, while method filtering with keywords has a significant positive effect without removing too many real patterns.
	\item In comparison to non-filtered results, with our strategy we retrieved more patterns that can fix a misuse. Even though the difference is not significant in a quantitative manner, we qualitatively observed that we found more patterns regarding different projects and for internally collected source code. Nevertheless, our search strategy is improvable, for instance, by recent work on similar API code search~\cite{Asyrofi2020}.
\end{enumerate}

Based on our results and existing related work, we plan the following additional steps to further integrate this knowledge into a full-fledged API misuse detection and correction tool.

\paragraph{Detecting API-Misuse Commits}
While we found that analyzing API usage on the commit level reduces the effort in terms of the number of methods to analyze, we still have some huge outliers. Moreover, related work~\cite{Mockus2000,Kim2008,Kamei2013,An2015} also suggests techniques for discriminating bug-containing from bug-free commits. 
We plan to include these techniques as a further pre-processing step so that API change analysis is only conducted when the commit is determined as suspicious, such as indicated by our prior work~\cite{Nielebock2018}. 
Additionally, to cope with huge changes consisting of several hundreds of potentially misuse-containing methods, we could also transfer these techniques to the method scope, i.e., detecting particularly suspicious methods (e.g., by the frequency of method's changes). 

\paragraph{Code Search and Filtering}
Up to now, we have aimed for a lightweight search and filter process. That means, we avoided costly static and dynamic code analyses. Assuming that we can further reduce the absolute number of methods to analyze, it could be worth introducing further analysis as those presented in Section~\ref{ssec:code search}, for instance, by applying \texttt{AUSearch}~\cite{Asyrofi2020}. 
We also made the observation that for some APIs there exist only a few code examples. To solve this problem a recent idea is to represent API usages in the form of a learned vector embedding, such as API2Vec~\cite{Nguyen2017}. This embedding depicts semantic relations between API usages as vector operations, such as $v(ListIterator.hasNext) \approx v(StringTokenizer.hasMoreTokens)\ -\ v(StringTokenizer.nextTokens)\ +\ v(ListIterator.next)$. A first notion of how this can be leveraged to map verified usage patterns from well-known APIs to equivalent, less-known APIs is published in~\cite {Nielebock2020A}.

\paragraph{API Misuse Detection}
Our analysis shows that in case we have a misuse introducing commit and we specifically investigate the misuse-containing method, we can improve the relative frequency of true positive patterns in retrieved similar API usages. However, we also have to consider to what degree our approach may falsely classify correct API usages as misuses. Our analysis on the AU500 dataset indicates that further steps for increasing the process precision are necessary. 
Finally, we need a human-based study on the usability of our approach, since there are a number of factors that may hamper developer's acceptance.

\paragraph{API Misuse Repair}
We envision a full-fledged process that not only detects API misuses automatically but also suggests patches. For that purpose, we have to incorporate the fixing pattern (i.e., that detected the misuse) back into the original code. This requires further post-processing such as mapping variables. Moreover, the patches need to be validated, which other automated program repair approaches typically do by using test suites. However, we do not have such a test suite or cannot ensure whether these tests check the API misuse behavior~\cite{LeGoues2019}. Thus, we require human intervention. To minimize that effort, we have to ensure that we have only a few patch candidates with a high rate of true positives.

\bibliographystyle{spbasic}      % basic style, author-year citations
%\bibliography{MYshort,paper-guided_pattern_mining_for_misuse_detection}
\bibliography{MYabrv,paper-guided_pattern_mining_for_misuse_detection}

\begin{thebibliography}{93}
\providecommand{\natexlab}[1]{#1}
\providecommand{\url}[1]{{#1}}
\providecommand{\urlprefix}{URL }
\expandafter\ifx\csname urlstyle\endcsname\relax
  \providecommand{\doi}[1]{DOI~\discretionary{}{}{}#1}\else
  \providecommand{\doi}{DOI~\discretionary{}{}{}\begingroup
  \urlstyle{rm}\Url}\fi
\providecommand{\eprint}[2][]{\url{#2}}

\bibitem[{Agrawal and Menzies(2018)}]{Agrawal2018}
Agrawal A, Menzies T (2018) {Is "Better Data" Better Than "Better Data
  Miners"?: On the Benefits of Tuning SMOTE for Defect Prediction}. In: {Proc.}
  40th {Int. Conf. Softw. Eng. (ICSE)}, ACM, New York, NY, USA, pp 1050--1061,
  \doi{10.1145/3180155.3180197}

\bibitem[{Agrawal et~al.(1993)Agrawal, Imieli\'{n}ski, and Swami}]{Agrawal1993}
Agrawal R, Imieli\'{n}ski T, Swami A (1993) {Mining Association Rules between
  Sets of Items in Large Databases}. { SIGMOD Rec} 22(2):207–216,
  \doi{10.1145/170036.170072}

\bibitem[{Allamanis and Sutton(2014)}]{Allamanis2014}
Allamanis M, Sutton C (2014) {Mining Idioms from Source Code}. In: {Proc.} 22nd
  {Int. Symp. Found. Softw. Eng. (FSE)}, ACM, New York, NY, USA, pp 472--483,
  \doi{10.1145/2635868.2635901}

\bibitem[{Amann(2018)}]{Amann2018a}
Amann S (2018) {A Systematic Approach to Benchmark and Improve Automated Static
  Detection of Java-API Misuses}. PhD thesis, Technische Universit{\"a}t
  Darmstadt, \urlprefix\url{http://tubiblio.ulb.tu-darmstadt.de/106302/}

\bibitem[{Amann et~al.(2016)Amann, Nadi, Nguyen, Nguyen, and
  Mezini}]{Amann2016}
Amann S, Nadi S, Nguyen HA, Nguyen TN, Mezini M (2016) {MUBench: A Benchmark
  for API-misuse Detectors}. In: {Proc.} 13th {Int. Work. Min. Softw. Repos.
  (MSR)}, ACM, New York, NY, USA, pp 464--467, \doi{10.1145/2901739.2903506}

\bibitem[{Amann et~al.(2018)Amann, Nguyen, Nadi, Nguyen, and
  Mezini}]{Amann2018}
Amann S, Nguyen HA, Nadi S, Nguyen TN, Mezini M (2018) {A Systematic Evaluation
  of Static API-Misuse Detectors}. {IEEE Trans Softw Eng}
  \doi{10.1109/TSE.2018.2827384}

\bibitem[{Amann et~al.(2019)Amann, Nguyen, Nadi, Nguyen, and
  Mezini}]{Amann2019}
Amann S, Nguyen HA, Nadi S, Nguyen TN, Mezini M (2019) {Investigating next
  Steps in Static API-Misuse Detection}. In: {Proc.} 16th {Int. Work. Min.
  Softw. Repos. (MSR)}, IEEE Press, p 265–275, \doi{10.1109/MSR.2019.00053}

\bibitem[{Ammons et~al.(2002)Ammons, Bod\'{\i}k, and Larus}]{Ammons2002}
Ammons G, Bod\'{\i}k R, Larus JR (2002) Mining specifications. In: {Proc.} 29th
  {Symp. Princ. Program. Lang. (POPL)}, ACM, New York, NY, USA, pp 4--16,
  \doi{10.1145/503272.503275}

\bibitem[{An and Khomh(2015)}]{An2015}
An L, Khomh F (2015) {An Empirical Study of Crash-inducing Commits in Mozilla
  Firefox}. In: {Proc.} 11th {Int. Conf. Predict. Models Softw. Eng.
  (Promise)}, ACM, New York, NY, USA, pp 5:1--5:10,
  \doi{10.1145/2810146.2810152}

\bibitem[{Asyrofi et~al.(2020)Asyrofi, Thung, Lo, and Jiang}]{Asyrofi2020}
Asyrofi MH, Thung F, Lo D, Jiang L (2020) {AUSearch: Accurate API Usage Search
  in GitHub Repositories with Type Resolution}. In: {Proc.} 27th {Int. Conf.
  Softw. Anal., Evol., Reeng. (SANER)}, pp 637--641,
  \doi{10.1109/SANER48275.2020.9054809}

\bibitem[{Chan et~al.(2012)Chan, Cheng, and Lo}]{chan2012searching}
Chan WK, Cheng H, Lo D (2012) {Searching Connected API Subgraph via Text
  Phrases}. In: {Proc.} 20th {Int. Symp. Found. Softw. Eng. (FSE)}, ACM, p~10,
  \doi{10.1145/2393596.2393606}

\bibitem[{Chen et~al.(2019)Chen, Xing, Liu, and Ong}]{Chen2019}
Chen C, Xing Z, Liu Y, Ong KLX (2019) {Mining Likely Analogical APIs across
  Third-Party Libraries via Large-Scale Unsupervised API Semantics Embedding}.
  {IEEE Trans Softw Eng} pp 1--1, \doi{10.1109/TSE.2019.2896123}

\bibitem[{Cohen(1960)}]{Cohen1960}
Cohen J (1960) {A Coefficient of Agreement for Nominal Scales}. {Educ and
  Psychol Meas} 20(1):37--46, \doi{10.1177/001316446002000104}

\bibitem[{Dyer et~al.(2013)Dyer, Nguyen, Rajan, and Nguyen}]{Dyer2013}
Dyer R, Nguyen HA, Rajan H, Nguyen TN (2013) {Boa: A Language and
  Infrastructure for Analyzing Ultra-large-scale Software Repositories}. In:
  {Proc.} 35th {Int. Conf. Softw. Eng. (ICSE)}, IEEE Press, Piscataway, NJ,
  USA, pp 422--431, \doi{10.1109/ICSE.2013.6606588}

\bibitem[{Ernst et~al.(2001)Ernst, Cockrell, Griswold, and Notkin}]{Ernst2001}
Ernst MD, Cockrell J, Griswold WG, Notkin D (2001) {Dynamically Discovering
  Likely Program Invariants to Support Program Evolution}. {IEEE Trans Softw
  Eng} 27(2):99--123, \doi{10.1109/32.908957}

\bibitem[{Ernst et~al.(2007)Ernst, Perkins, Guo, McCamant, Pacheco, Tschantz,
  and Xiao}]{Ernst2007}
Ernst MD, Perkins JH, Guo PJ, McCamant S, Pacheco C, Tschantz MS, Xiao C (2007)
  {The Daikon System for Dynamic Detection of Likely Invariants}. {Sci Comput
  Program} 69(1):35 -- 45, \doi{10.1016/j.scico.2007.01.015}, {Special Issue on
  Experimental Software and Toolkits}

\bibitem[{Frolin S.~Ocariza et~al.(2013)Frolin S.~Ocariza, Bajaj, Pattabiraman,
  and Mesbah}]{FrolinS.Ocariza2013}
Frolin S~Ocariza J, Bajaj K, Pattabiraman K, Mesbah A (2013) {An Empirical
  Study of Client-Side JavaScript Bugs}. In: {Proc.} 7th {Int. Symp. Empir.
  Softw. Eng. Meas. (ESEM)}, pp 55--64, \doi{10.1109/ESEM.2013.18}

\bibitem[{Gabel and Su(2008)}]{gabel2008javert}
Gabel M, Su Z (2008) {Javert: Fully Automatic Mining of General Temporal
  Properties from Dynamic Traces}. In: {Proc.} 16th {Int. Symp. Found. Softw.
  Eng. (FSE)}, ACM, pp 339--349, \doi{10.1145/1453101.1453150}

\bibitem[{Gallardo-Valencia and Sim(2009)}]{gallardo2009internet}
Gallardo-Valencia RE, Sim SE (2009) {Internet-Scale Code Search}. In: {Proc.}
  1st {Work. Search-Driven Dev.-Users, Infrastruct., Tools Eval. (SUITE)},
  IEEE, pp 49--52, \doi{10.1109/SUITE.2009.5070022}

\bibitem[{Gousios(2013)}]{Gousios2013}
Gousios G (2013) {The GHTorrent Dataset and Tool Suite}. In: {Proc.} 10th {Int.
  Work. Min. Softw. Repos. (MSR)}, pp 233--236, \doi{10.1109/MSR.2013.6624034}

\bibitem[{Gu et~al.(2018)Gu, Zhang, and Kim}]{Gu2018}
Gu X, Zhang H, Kim S (2018) {Deep Code Search}. In: {Proc.} 40th {Int. Conf.
  Softw. Eng. (ICSE)}, IEEE, pp 933--944, \doi{10.1145/3180155.3180167}

\bibitem[{Guo et~al.(2020)Guo, Li, Ma, Zhou, Lu, Chen, and Xu}]{Guo2020}
Guo Z, Li Y, Ma W, Zhou Y, Lu H, Chen L, Xu B (2020) {Boosting Crash-inducing
  Change Localization with Rank-performance-based Feature Subset Selection}.
  {Empir Softw Eng} pp 1--46, \doi{10.1007/s10664-020-09802-1}

\bibitem[{Holmes and Murphy(2005)}]{Holmes2005}
Holmes R, Murphy GC (2005) {Using Structural Context to Recommend Source Code
  Examples}. In: {Proc.} 27th {Int. Conf. Softw. Eng. (ICSE)}, ACM, New York,
  NY, USA, pp 117--125, \doi{10.1145/1062455.1062491}

\bibitem[{{Hou} and {Li}(2011)}]{Hou2011}
{Hou} D, {Li} L (2011) {Obstacles in Using Frameworks and APIs: An Exploratory
  Study of Programmers' Newsgroup Discussions}. In: {Proc.} 19th {Int. Conf.
  Program Compr. (ICPC)}, IEEE, pp 91--100, \doi{10.1109/ICPC.2011.21}

\bibitem[{Huang et~al.(2018)Huang, Xia, Xing, Lo, and Wang}]{Huang2018}
Huang Q, Xia X, Xing Z, Lo D, Wang X (2018) {API Method Recommendation without
  Worrying about the Task-API Knowledge Gap}. In: {Proc.} 33rd {Int. Conf.
  Autom. Softw. Eng. (ASE)}, ACM, New York, NY, USA, p 293–304,
  \doi{10.1145/3238147.3238191}

\bibitem[{Jiang et~al.(2007)Jiang, Misherghi, Su, and
  Glondu}]{jiang2007deckard}
Jiang L, Misherghi G, Su Z, Glondu S (2007) {DECKARD: Scalable and Accurate
  Tree-Based Detection of Code Clones}. In: {Proc.} 29th {Int. Conf. Softw.
  Eng. (ICSE)}, IEEE Computer Society, pp 96--105, \doi{10.1109/ICSE.2007.30}

\bibitem[{Kamei et~al.(2013)Kamei, Shihab, Adams, Hassan, Mockus, Sinha, and
  Ubayashi}]{Kamei2013}
Kamei Y, Shihab E, Adams B, Hassan AE, Mockus A, Sinha A, Ubayashi N (2013) {A
  Large-Scale Empirical Study of Just-in-Time Quality Assurance}. {IEEE Trans
  Softw Eng} 39(6):757--773, \doi{10.1109/TSE.2012.70}

\bibitem[{Kang and Lo(2021)}]{Kang2021}
Kang HJ, Lo D (2021) {Active Learning of Discriminative Subgraph Patterns for
  API Misuse Detection}. {IEEE Trans Softw Eng} pp 1--1,
  \doi{10.1109/TSE.2021.3069978}

\bibitem[{Kechagia et~al.(2021)Kechagia, Mechtaev, Sarro, and
  Harman}]{Kechagia2021}
Kechagia M, Mechtaev S, Sarro F, Harman M (2021) {Evaluating Automatic Program
  Repair Capabilities to Repair API Misuses}. {IEEE Trans Softw Eng} pp 1--1,
  \doi{10.1109/TSE.2021.3067156}

\bibitem[{Kim et~al.(2010)Kim, Lee, Hwang, and Kim}]{kim2010towards}
Kim J, Lee S, Hwang Sw, Kim S (2010) {Towards an Intelligent Code Search
  Engine}. In: {Proc.} 24th {AAAI Conf. Artif. Intell. (AAAI)},
  \urlprefix\url{https://www.aaai.org/ocs/index.php/AAAI/AAAI10/paper/download/1691/2212}

\bibitem[{Kim and Notkin(2009)}]{Kim2009}
Kim M, Notkin D (2009) {Discovering and Representing Systematic Code Changes}.
  In: {Proc.} 31st {Int. Conf. Softw. Eng. (ICSE)}, IEEE Computer Society, USA,
  p 309–319, \doi{10.1109/ICSE.2009.5070531}

\bibitem[{Kim et~al.(2006)Kim, Pan, and Whitehead}]{Kim2006}
Kim S, Pan K, Whitehead EEJ (2006) {Memories of Bug Fixes}. In: {Proc.} 14th
  {Int. Symp. Found. Softw. Eng. (FSE)}, ACM, New York, NY, USA, p 35–45,
  \doi{10.1145/1181775.1181781}

\bibitem[{Kim et~al.(2008)Kim, E.~James~Whitehead, and Zhang}]{Kim2008}
Kim S, E~James~Whitehead J, Zhang Y (2008) {Classifying Software Changes: Clean
  or Buggy?} {IEEE Trans Softw Eng} 34(2):181--196,
  \doi{10.1109/TSE.2007.70773}

\bibitem[{Koschke(2007)}]{Koschke2007}
Koschke R (2007) {Survey of Research on Software Clones}. In: Koschke R, Merlo
  E, Walenstein A (eds) { Dagstuhl Semin. Proc.: Duplic., Redundancy, and
  Similarity Softw.}, Internationales Begegnungs- und Forschungszentrum f{\"u}r
  Informatik (IBFI), Schloss Dagstuhl, Germany, Dagstuhl, Germany, no. 06301 in
  Dagstuhl Seminar Proceedings,
  \urlprefix\url{http://drops.dagstuhl.de/opus/volltexte/2007/962}

\bibitem[{Le and Lo(2015)}]{Le2015}
Le TDB, Lo D (2015) {Beyond Support and Confidence: Exploring Interestingness
  Measures for Rule-Based Specification Mining}. In: {Proc.} 22nd {Int. Conf.
  Softw. Anal., Evol., Reeng. (SANER)}, pp 331--340,
  \doi{10.1109/SANER.2015.7081843}

\bibitem[{Le et~al.(2016)Le, Lo, and {Le~Goues}}]{Le2016}
Le XBD, Lo D, {Le~Goues} C (2016) {History Driven Program Repair}. In: {Proc.}
  23rd {Int. Conf. Softw. Anal., Evol., Reeng. (SANER)}, vol~1, pp 213--224,
  \doi{10.1109/SANER.2016.76}

\bibitem[{{Le Goues} and Weimer(2012)}]{Goues2012}
{Le Goues} C, Weimer W (2012) {Measuring Code Quality to Improve Specification
  Mining}. {IEEE Trans Softw Eng} 38(1):175--190, \doi{10.1109/TSE.2011.5}

\bibitem[{Le~Goues et~al.(2019)Le~Goues, Pradel, and
  Roychoudhury}]{LeGoues2019}
Le~Goues C, Pradel M, Roychoudhury A (2019) {Automated Program Repair}. {Commun
  ACM} 62(12):56–65, \doi{10.1145/3318162}

\bibitem[{Lemos et~al.(2007)Lemos, Bajracharya, Ossher, Morla, Masiero, Baldi,
  and Lopes}]{lemos2007codegenie}
Lemos OAL, Bajracharya SK, Ossher J, Morla RS, Masiero PC, Baldi P, Lopes CV
  (2007) {CodeGenie: Using Test-Cases to Search and Reuse Source Code}. In:
  {Proc.} 22nd {Int. Conf. Autom. Softw. Eng. (ASE)}, ACM, pp 525--526,
  \doi{10.1145/1321631.1321726}

\bibitem[{Li and Zhou(2005)}]{li2005pr}
Li Z, Zhou Y (2005) {PR-Miner: Automatically Extracting Implicit Programming
  Rules and Detecting Violations in Large Software Code}. In: {Proc.} 5th {Eur.
  Softw. Eng. Conf./Found. Softw. Eng. (ESEC/FSE)}, ACM, pp 306--315,
  \doi{10.1145/1095430.1081755}

\bibitem[{Livshits and Zimmermann(2005)}]{livshits2005dynamine}
Livshits B, Zimmermann T (2005) {DynaMine: Finding Common Error Patterns by
  Mining Software Revision Histories}. In: {Proc.} 5th {Eur. Softw. Eng.
  Conf./Found. Softw. Eng. (ESEC/FSE)}, ACM, vol~30, pp 296--305,
  \doi{10.1145/1095430.1081754}

\bibitem[{Long and Rinard(2016)}]{Long2016}
Long F, Rinard M (2016) {Automatic Patch Generation by Learning Correct Code}.
  In: {Proc.} 43rd {Symp. Princ. Program. Lang. (POPL)}, ACM, New York, NY,
  USA, p 298–312, \doi{10.1145/2837614.2837617}

\bibitem[{Lv et~al.(2015)Lv, Zhang, Lou, Wang, Zhang, and Zhao}]{Lv2015}
Lv F, Zhang H, Lou Jg, Wang S, Zhang D, Zhao J (2015) {CodeHow: Effective Code
  Search Based on API Understanding and Extended Boolean Model}. In: {Proc.}
  30th {Int. Conf. Autom. Softw. Eng. (ASE)}, IEEE Press, p 260–270,
  \doi{10.1109/ASE.2015.42}

\bibitem[{Mandelin et~al.(2005)Mandelin, Xu, Bod\'{\i}k, and
  Kimelman}]{Mandelin2005}
Mandelin D, Xu L, Bod\'{\i}k R, Kimelman D (2005) {Jungloid Mining: Helping to
  Navigate the API Jungle}. In: {Proc.} 26th {ACM SIGPLAN Conf. Program. Lang.
  Des. Implement. (PLDI)}, ACM, New York, NY, USA, pp 48--61,
  \doi{10.1145/1065010.1065018}

\bibitem[{McMillan et~al.(2011)McMillan, Grechanik, Poshyvanyk, Xie, and
  Fu}]{McMillan2011}
McMillan C, Grechanik M, Poshyvanyk D, Xie Q, Fu C (2011) {Portfolio: Finding
  Relevant Functions and Their Usage}. In: {Proc.} 33rd {Int. Conf. Softw. Eng.
  (ICSE)}, ACM, New York, NY, USA, p 111–120, \doi{10.1145/1985793.1985809}

\bibitem[{Mockus and Weiss(2000)}]{Mockus2000}
Mockus A, Weiss DM (2000) {Predicting Risk of Software Changes}. { Bell Labs
  Tech J} 5(2):169--180, \doi{10.1002/bltj.2229}

\bibitem[{Moreno et~al.(2015)Moreno, Bavota, Di~Penta, Oliveto, and
  Marcus}]{moreno2015can}
Moreno L, Bavota G, Di~Penta M, Oliveto R, Marcus A (2015) {How Can I Use This
  Method?} In: {Proc.} 37th {Int. Conf. Softw. Eng. (ICSE)}, IEEE, vol~1, pp
  880--890, \doi{10.1109/ICSE.2015.98}

\bibitem[{Murali et~al.(2017)Murali, Chaudhuri, and
  Jermaine}]{murali2017bayesian}
Murali V, Chaudhuri S, Jermaine C (2017) {Bayesian Specification Learning for
  Finding API Usage Errors}. In: {Proc.}11th {Eur. Softw. Eng. Conf./Found.
  Softw. Eng. (ESEC/FSE)}, {ACM}, pp 151--162, \doi{10.1145/3106237.3106284}

\bibitem[{Nadi et~al.(2016)Nadi, Kr\"{u}ger, Mezini, and Bodden}]{Nadi2016}
Nadi S, Kr\"{u}ger S, Mezini M, Bodden E (2016) {Jumping through Hoops: Why Do
  Java Developers Struggle with Cryptography APIs?} In: {Proc.} 38th {Int.
  Conf. Softw. Eng. (ICSE)}, ACM, New York, NY, USA, p 935–946,
  \doi{10.1145/2884781.2884790}

\bibitem[{Nguyen et~al.(2017)Nguyen, Nguyen, Phan, and Nguyen}]{Nguyen2017}
Nguyen TD, Nguyen AT, Phan HD, Nguyen TN (2017) {Exploring API Embedding for
  API Usages and Applications}. In: {Proc.} 39th {Int. Conf. Softw. Eng.
  (ICSE)}, IEEE, p 438–449, \doi{10.1109/ICSE.2017.47}

\bibitem[{Nguyen et~al.(2009)Nguyen, Nguyen, Pham, Al-Kofahi, and
  Nguyen}]{Nguyen2009}
Nguyen TT, Nguyen HA, Pham NH, Al-Kofahi JM, Nguyen TN (2009) Graph-based
  mining of multiple object usage patterns. In: {Proc.} 7th {Eur. Softw. Eng.
  Conf./Found. Softw. Eng. (ESEC/FSE)}, ACM, New York, NY, USA, pp 383--392,
  \doi{10.1145/1595696.1595767}

\bibitem[{Nielebock et~al.(2018)Nielebock, Heum\"{u}ller, and
  Ortmeier}]{Nielebock2018}
Nielebock S, Heum\"{u}ller R, Ortmeier F (2018) {Commits As a Basis for API
  Misuse Detection}. In: {Proc.} 7th { Int. Work. on Softw. Min.}, ACM, New
  York, NY, USA, pp 20--23, \doi{10.1145/3242887.3242890}

\bibitem[{Nielebock et~al.(2020{\natexlab{a}})Nielebock, Heum\"{u}ller,
  Kr\"{u}ger, and Ortmeier}]{Nielebock2020}
Nielebock S, Heum\"{u}ller R, Kr\"{u}ger J, Ortmeier F (2020{\natexlab{a}})
  {Cooperative API Misuse Detection Using Correction Rules}. In: ACM (ed)
  {Proc.} 42nd {Int. Conf. Softw. Eng. (ICSE)}, ACM,
  \doi{10.1145/3377816.3381735}, {ICSE-NIER} track - forthcoming

\bibitem[{Nielebock et~al.(2020{\natexlab{b}})Nielebock, Heum\"{u}ller,
  Kr\"{u}ger, and Ortmeier}]{Nielebock2020A}
Nielebock S, Heum\"{u}ller R, Kr\"{u}ger J, Ortmeier F (2020{\natexlab{b}})
  {Using API-Embedding for API-Misuse Repair}. In: ACM (ed) {Proc.} 1st { Int.
  Work. Autom. Program Repair (APR)}, \doi{10.1145/3387940.3392171},
  forthcoming

\bibitem[{Nielebock et~al.(2021)Nielebock, Blockhaus, Krüger, and
  Ortmeier}]{Nielebock2021}
Nielebock S, Blockhaus P, Krüger J, Ortmeier F (2021) {AndroidCompass: A
  Dataset of Android Compatibility Checks in Code Repositories}. In: {Proc.}
  18th {Int. Work. Min. Softw. Repos. (MSR)}, pp 535--539,
  \doi{10.1109/MSR52588.2021.00069}

\bibitem[{Oliveira et~al.(2018)Oliveira, Lin, Rahman, Akefirad, Ellis, Perez,
  Bobhate, DeLong, Cappos, and Brun}]{Oliveira2018}
Oliveira DS, Lin T, Rahman MS, Akefirad R, Ellis D, Perez E, Bobhate R, DeLong
  LA, Cappos J, Brun Y (2018) {API} blindspots: Why experienced developers
  write vulnerable code. In: {Proc.} 14th { Symp. on Usable Priv. and Secur.
  (SOUPS)}, {USENIX} Association, Baltimore, MD, pp 315--328

\bibitem[{Pasquier et~al.(1999)Pasquier, Bastide, Taouil, and
  Lakhal}]{Pasquier1999}
Pasquier N, Bastide Y, Taouil R, Lakhal L (1999) {Discovering Frequent Closed
  Itemsets for Association Rules}. In: Beeri C, Buneman P (eds) {Proc.} 7th {
  Int. Conf. Database Theory (ICDT)}, Springer Berlin Heidelberg, Berlin,
  Heidelberg, pp 398--416, \doi{10.1007/3-540-49257-7_25}

\bibitem[{Paul and Prakash(1994)}]{paul1994framework}
Paul S, Prakash A (1994) {A Framework for Source Code Search using Program
  Patterns}. {IEEE Trans Softw Eng} 20(6):463--475, \doi{10.1109/32.295894}

\bibitem[{Pradel and Gross(2009)}]{pradel2009automatic}
Pradel M, Gross TR (2009) {Automatic Generation of Object Usage Specifications
  from Large Method Traces}. In: {Proc.} 24th {Int. Conf. Autom. Softw. Eng.
  (ASE)}, IEEE Computer Society, pp 371--382, \doi{10.1109/ASE.2009.60}

\bibitem[{Rahman et~al.(2016)Rahman, Roy, and Lo}]{rahman2016rack}
Rahman MM, Roy CK, Lo D (2016) {RACK: Automatic API Recommendation Using
  Crowdsourced Knowledge}. In: {Proc.} 23rd {Int. Conf. Softw. Anal., Evol.,
  Reeng. (SANER)}, IEEE, vol~1, pp 349--359, \doi{10.1109/SANER.2016.80}

\bibitem[{Rahman et~al.(2018)Rahman, Barson, Paul, Kayani, Lois, Quezada,
  Parnin, Stolee, and Ray}]{Rahman2018}
Rahman MM, Barson J, Paul S, Kayani J, Lois FA, Quezada SF, Parnin C, Stolee
  KT, Ray B (2018) {Evaluating How Developers Use General-Purpose Web-Search
  for Code Retrieval}. In: {Proc.} 15th {Int. Work. Min. Softw. Repos. (MSR)},
  ACM, New York, NY, USA, p 465–475, \doi{10.1145/3196398.3196425}

\bibitem[{Reiss(2009)}]{reiss2009semantics}
Reiss SP (2009) {Semantics-Based Code Search}. In: {Proc.} 31st {Int. Conf.
  Softw. Eng. (ICSE)}, IEEE Computer Society, pp 243--253,
  \doi{10.1109/ICSE.2009.5070525}

\bibitem[{Robillard and Deline(2011)}]{Robillard2011}
Robillard MP, Deline R (2011) {A Field Study of API Learning Obstacles}. {Empir
  Softw Eng} 16(6):703--732, \doi{10.1007/s10664-010-9150-8}

\bibitem[{Robillard et~al.(2013)Robillard, Bodden, Kawrykow, Mezini, and
  Ratchford}]{robillard2013automated}
Robillard MP, Bodden E, Kawrykow D, Mezini M, Ratchford T (2013) {Automated API
  Property Inference Techniques}. {IEEE Trans Softw Eng} 39(5):613--637,
  \doi{10.1109/TSE.2012.63}

\bibitem[{Roy et~al.(2009)Roy, Cordy, and Koschke}]{roy2009comparison}
Roy CK, Cordy JR, Koschke R (2009) {Comparison and Evaluation of Code Clone
  Detection Techniques and Tools: A Qualitative Approach}. {Sci Comput Program}
  74(7):470--495, \doi{10.1016/j.scico.2009.02.007}

\bibitem[{Sadowski et~al.(2015)Sadowski, Stolee, and
  Elbaum}]{sadowski2015developers}
Sadowski C, Stolee KT, Elbaum S (2015) {How Developers Search for Code: A Case
  Study}. In: {Proc.} 10th {Eur. Softw. Eng. Conf./Found. Softw. Eng.
  (ESEC/FSE)}, ACM, pp 191--201, \doi{10.1145/2786805.2786855}

\bibitem[{Sahavechaphan and Claypool(2006)}]{Sahavechaphan2006}
Sahavechaphan N, Claypool K (2006) {XSnippet: Mining For Sample Code}. In:
  {Proc.} 21st {Conf. Object-Oriented Program., Syst., Lang. Appl. (OOPSLA)},
  ACM, New York, NY, USA, pp 413--430, \doi{10.1145/1167473.1167508}

\bibitem[{Saied et~al.(2020)Saied, Raelijohn, Batot, Famelis, and
  Sahraoui}]{Saied2020}
Saied MA, Raelijohn E, Batot E, Famelis M, Sahraoui H (2020) {Towards assisting
  developers in API usage by automated recovery of complex temporal patterns}
  119:106213, \doi{https://doi.org/10.1016/j.infsof.2019.106213},
  \urlprefix\url{https://www.sciencedirect.com/science/article/pii/S0950584918301642}

\bibitem[{Saini et~al.(2018)Saini, Farmahinifarahani, Lu, Baldi, and
  Lopes}]{Saini2018}
Saini V, Farmahinifarahani F, Lu Y, Baldi P, Lopes CV (2018) {Oreo: Detection
  of Clones in the Twilight Zone}. In: {Proc.} 26th {Eur. Softw. Eng.
  Conf./Found. Softw. Eng. (ESEC/FSE)}, ACM, New York, NY, USA, pp 354--365,
  \doi{10.1145/3236024.3236026}

\bibitem[{Sajnani et~al.(2016)Sajnani, Saini, Svajlenko, Roy, and
  Lopes}]{sajnani2016sourcerercc}
Sajnani H, Saini V, Svajlenko J, Roy CK, Lopes CV (2016) {SourcererCC: Scaling
  Code Clone Detection to Big-Code}. In: {Proc.} 38th {Int. Conf. Softw. Eng.
  (ICSE)}, IEEE, pp 1157--1168, \doi{10.1145/2884781.2884877}

\bibitem[{Saul et~al.(2007)Saul, Filkov, Devanbu, and Bird}]{Saul2007}
Saul ZM, Filkov V, Devanbu P, Bird C (2007) {Recommending Random Walks}. In:
  {Proc.} 6th {Eur. Softw. Eng. Conf./Found. Softw. Eng. (ESEC/FSE)}, ACM, New
  York, NY, USA, pp 15--24, \doi{10.1145/1287624.1287629}

\bibitem[{Siegmund et~al.(2015)Siegmund, Siegmund, and Apel}]{Siegmund2015}
Siegmund J, Siegmund N, Apel S (2015) {Views on Internal and External Validity
  in Empirical Software Engineering}. In: {Proc.} 37th {Int. Conf. Softw. Eng.
  (ICSE)}, vol~1, pp 9--19, \doi{10.1109/ICSE.2015.24}

\bibitem[{\'{S}liwerski et~al.(2005)\'{S}liwerski, Zimmermann, and
  Zeller}]{Sliwerski2005}
\'{S}liwerski J, Zimmermann T, Zeller A (2005) {When Do Changes Induce Fixes?}
  In: {Proc.} 2nd {Int. Work. Min. Softw. Repos. (MSR)}, ACM, New York, NY,
  USA, pp 1--5, \doi{10.1145/1082983.1083147}

\bibitem[{Stolee et~al.(2014)Stolee, Elbaum, and Dobos}]{Stolee2014}
Stolee KT, Elbaum S, Dobos D (2014) {Solving the Search for Source Code}. {ACM
  Trans Softw Eng Methodology} 23(3), \doi{10.1145/2581377}

\bibitem[{Sun et~al.(2010)Sun, Shu, Podgurski, Li, Zhang, and Yang}]{Sun2010}
Sun B, Shu G, Podgurski A, Li S, Zhang S, Yang J (2010) {Propagating Bug Fixes
  with Fast Subgraph Matching}. In: {Proc.} 21st {Int. Symp. Softw. Reliab.
  Eng. (ISSRE)}, pp 21--30, \doi{10.1109/ISSRE.2010.36}

\bibitem[{Thummalapenta and Xie(2007)}]{Thummalapenta2007}
Thummalapenta S, Xie T (2007) {Parseweb: A Programmer Assistant for Reusing
  Open Source Code on the Web}. In: {Proc.} 22nd {Int. Conf. Autom. Softw. Eng.
  (ASE)}, ACM, New York, NY, USA, pp 204--213, \doi{10.1145/1321631.1321663}

\bibitem[{Thung et~al.(2013{\natexlab{a}})Thung, Lo, and Lawall}]{Thung2013}
Thung F, Lo D, Lawall J (2013{\natexlab{a}}) {Automated Library
  Recommendation}. In: {Proc.} 20th {Work. Conf. Reverse Eng. (WCRE)}, pp
  182--191, \doi{10.1109/WCRE.2013.6671293}

\bibitem[{Thung et~al.(2013{\natexlab{b}})Thung, Wang, Lo, and
  Lawall}]{thung2013automatic}
Thung F, Wang S, Lo D, Lawall J (2013{\natexlab{b}}) {Automatic Recommendation
  of API Methods from Feature Requests}. In: {Proc.} 28th {Int. Conf. Autom.
  Softw. Eng. (ASE)}, IEEE Press, pp 290--300, \doi{10.1109/ASE.2013.6693088}

\bibitem[{Wasylkowski and Zeller(2011)}]{wasylkowski2011mining}
Wasylkowski A, Zeller A (2011) {Mining Temporal Specifications from Object
  Usage}. {Autom Softw Eng} 18(3-4):263--292, \doi{10.1007/s10515-011-0084-1}

\bibitem[{Wasylkowski et~al.(2007)Wasylkowski, Zeller, and
  Lindig}]{wasylkowski2007detecting}
Wasylkowski A, Zeller A, Lindig C (2007) {Detecting Object Usage Anomalies}.
  In: {Proc.} 6th {Eur. Softw. Eng. Conf./Found. Softw. Eng. (ESEC/FSE)}, ACM,
  pp 35--44, \doi{10.1145/1287624.1287632}

\bibitem[{Weimer and Necula(2005)}]{Weimer2005}
Weimer W, Necula GC (2005) {Mining Temporal Specifications for Error
  Detection}. In: Halbwachs N, Zuck LD (eds) {Proc.} 11th {Int. Conf. Tools
  Algorithms Constr. Anal. Syst. (TACAS)}, Springer Berlin Heidelberg, Berlin,
  Heidelberg, pp 461--476, \doi{10.1007/978-3-540-31980-1_30}

\bibitem[{White et~al.(2016)White, Tufano, Vendome, and
  Poshyvanyk}]{white2016deep}
White M, Tufano M, Vendome C, Poshyvanyk D (2016) {Deep Learning Code Fragments
  for Code Clone Detection}. In: {Proc.} 31st {Int. Conf. Autom. Softw. Eng.
  (ASE)}, ACM, pp 87--98, \doi{10.1145/2970276.2970326}

\bibitem[{Wu et~al.(2017)Wu, Wen, Cheung, and Zhang}]{Wu2017}
Wu R, Wen M, Cheung SC, Zhang H (2017) {ChangeLocator: Locate Crash-Inducing
  Changes Based on Crash Reports}. {Empir Softw Eng}
  \doi{10.1007/s10664-017-9567-4}

\bibitem[{Xia et~al.(2017)Xia, Bao, Lo, Kochhar, Hassan, and Xing}]{Xia2017}
Xia X, Bao L, Lo D, Kochhar PS, Hassan AE, Xing Z (2017) {What do Developers
  Search for on the Web?} {Empir Softw Eng} 22(6):3149--3185,
  \doi{10.1007/s10664-017-9514-4}

\bibitem[{Xin and Reiss(2017)}]{Xin2017}
Xin Q, Reiss SP (2017) {Leveraging Syntax-Related Code for Automated Program
  Repair}. In: {Proc.} 32nd {Int. Conf. Autom. Softw. Eng. (ASE)}, IEEE Press,
  p 660–670, \doi{10.1109/ASE.2017.8115676}

\bibitem[{Xin and Reiss(2019)}]{Xin2019}
Xin Q, Reiss SP (2019) {Better Code Search and Reuse for Better Program
  Repair}. In: {Proc.} 6th, IEEE Press, p 10–17, \doi{10.1109/GI.2019.00012}

\bibitem[{Yang and Evans(2004)}]{yang2004automatically}
Yang J, Evans D (2004) {Automatically Inferring Temporal Properties for Program
  Evolution}. In: {Proc.} 15th {Int. Symp. Softw. Reliab. Eng. (ISSRE)}, IEEE,
  pp 340--351, \doi{10.1109/ISSRE.2004.11}

\bibitem[{Yang et~al.(2006)Yang, Evans, Bhardwaj, Bhat, and
  Das}]{yang2006perracotta}
Yang J, Evans D, Bhardwaj D, Bhat T, Das M (2006) {Perracotta: Mining Temporal
  API Rules from Imperfect Traces}. In: {Proc.} 28th {Int. Conf. Softw. Eng.
  (ICSE)}, ACM, pp 282--291, \doi{10.1145/1134285.1134325}

\bibitem[{Yates(1934)}]{Yates1934}
Yates F (1934) {Contingency Tables Involving Small Numbers and the chi2 Test}.
  { Suppl J R Stat Soc} 1(2):217--235, \doi{10.2307/2983604}

\bibitem[{Zhong and Su(2015)}]{zhong2015empirical}
Zhong H, Su Z (2015) {An Empirical Study on Real Bug Fixes}. In: {Proc.} 37th
  {Int. Conf. Softw. Eng. (ICSE)}, IEEE/ACM, IEEE Press, pp 913--923,
  \doi{10.1109/ICSE.2015.101}

\bibitem[{Zhong et~al.(2009)Zhong, Xie, Zhang, Pei, and Mei}]{zhong2009mapo}
Zhong H, Xie T, Zhang L, Pei J, Mei H (2009) {MAPO: Mining and Recommending API
  Usage Patterns}. {Proc}23rd{Eur Conf Object-Oriented Program (ECOOP)} pp
  318--343, \doi{10.1007/978-3-642-03013-0_15}

\bibitem[{{Zhou} et~al.(2017){Zhou}, {Gu}, {Chen}, {Huang}, {Panichella}, and
  {Gall}}]{Zhou2017}
{Zhou} Y, {Gu} R, {Chen} T, {Huang} Z, {Panichella} S, {Gall} H (2017)
  {Analyzing APIs Documentation and Code to Detect Directive Defects}. In:
  {Proc.} 39th {Int. Conf. Softw. Eng. (ICSE)}, pp 27--37,
  \doi{10.1109/ICSE.2017.11}

\bibitem[{{Zibran} et~al.(2011){Zibran}, {Eishita}, and {Roy}}]{Zibran2011}
{Zibran} MF, {Eishita} FZ, {Roy} CK (2011) {Useful, But Usable? Factors
  Affecting the Usability of APIs}. In: {Proc.} 18th {Work. Conf. Reverse Eng.
  (WCRE)}, IEEE, pp 151--155, \doi{10.1109/WCRE.2011.26}

\end{thebibliography}
%\bibliography{MYfull,paper-guided_pattern_mining_for_misuse_detection}
\noindent\includegraphics[width=.3\textwidth]{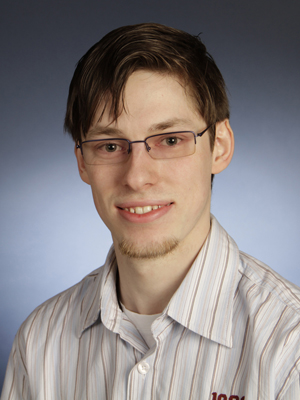}\\
\noindent\textbf{Sebastian Nielebock} is a Ph.D. student at the Otto-von-Guericke University of Magdeburg.
He received his B.Sc. and M.Sc. degrees in Computer Systems in Engineering from there as well.
Since October 2013 he is a member of the Chair of Software Engineering.
His research focuses on empirical and automated software engineering, i.e., programming language analysis, automated error detection, and automated program repair.

\noindent\includegraphics[width=.3\textwidth]{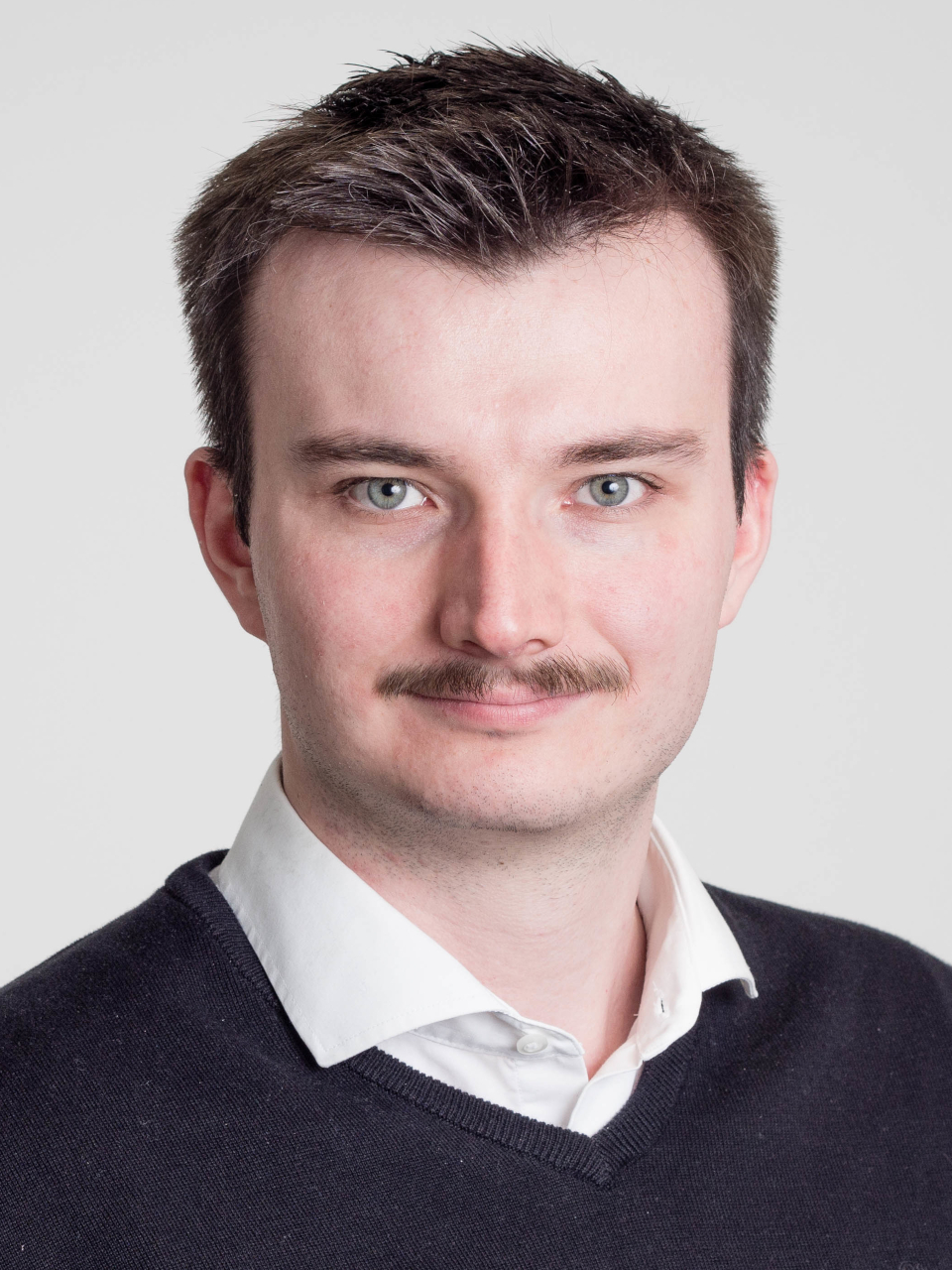}\\
\noindent\textbf{Robert Heum\"uller} is a Ph.D. student at Otto-von-Guericke University of Magdeburg, where he also received his B.Sc. and M.Sc. degrees in Computer Science.
In January 2016 he joined the Chair of Software Engineering, to research on automated software engineering, model-driven software development, and domain-specific languages.

\newpage

\noindent\includegraphics[width=.3\textwidth]{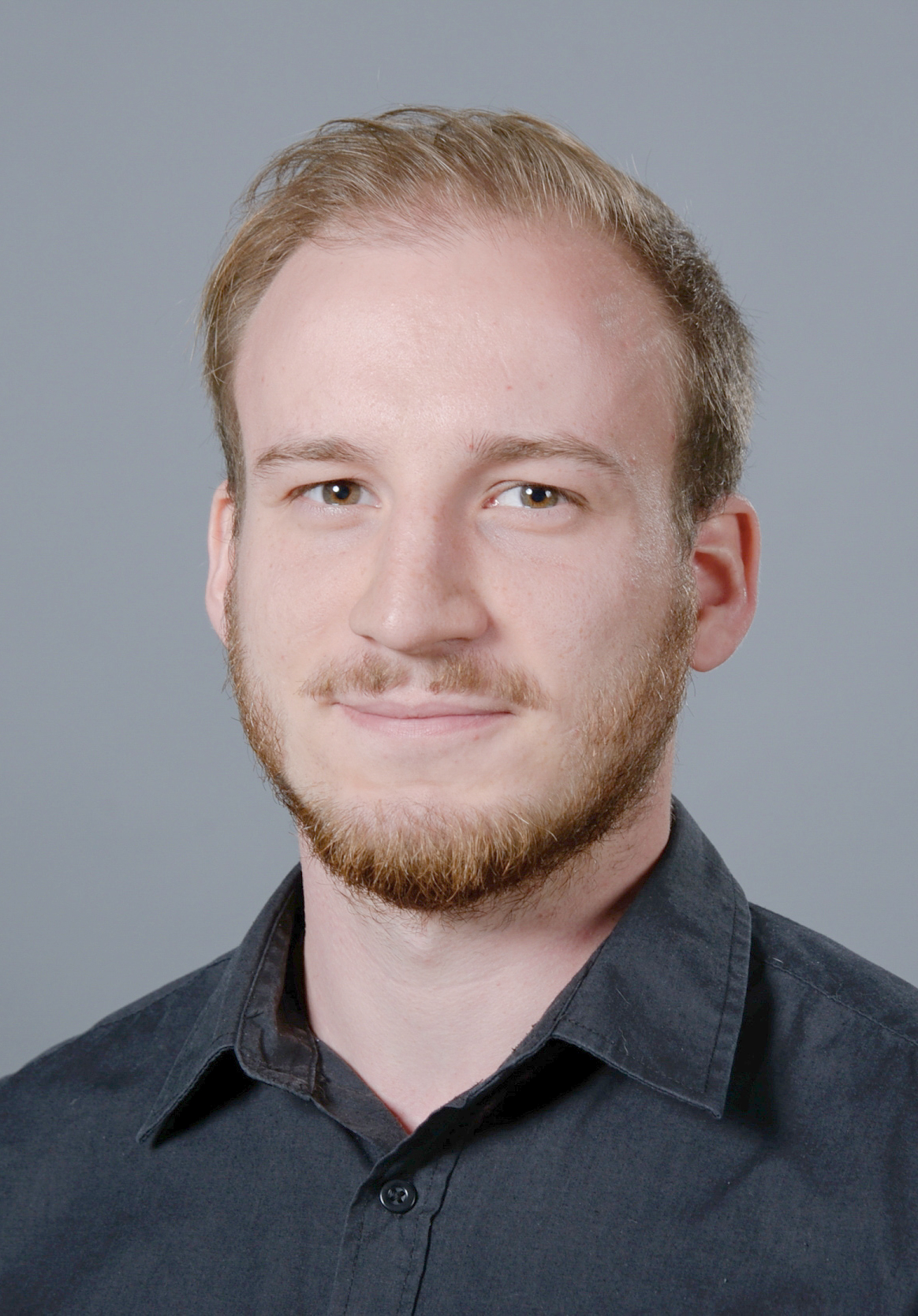}\\
\noindent\textbf{Kevin Michael Schott} is currently studying for his M.Sc. degree in Computer Science at the Otto-von-Guericke University of Magdeburg. He received the B.Sc. degree in Computer Science, where his thesis dealt with the topic API Misuse Detection. During his recent master studies he was engaged at the Chair of Software Engineering and continued his research in the field of automated program repair.

\noindent\includegraphics[width=.3\textwidth]{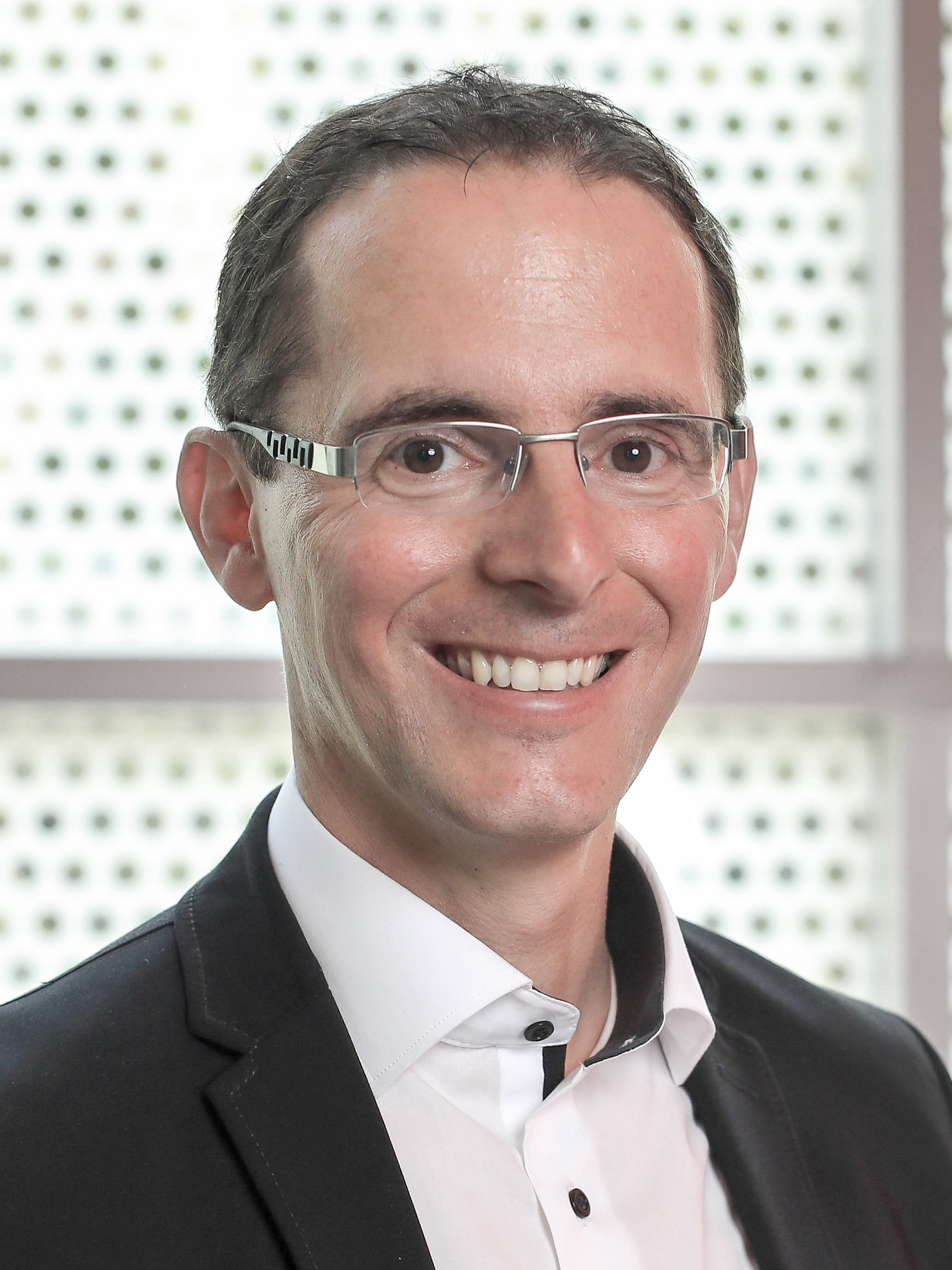}\\
\noindent\textbf{Frank Ortmeier} is a full professor and head of the ``Chair of Software Engineering (CSE)'' at the Otto-von-Guericke 
University of Magdeburg, Germany.
He received his Ph.D. degree from the University of Augsburg in 2005. After three years employed as a Post-Doc in Augsburg, he became an associate 
professor for ``Computer Systems in Engineering'' in Magdeburg in 2008. Since 2013 he is holding the Chair fo Software Engineering at OvGU.
Currently, he is leading several research projects, coordinating the Bachelor' degree program ``Computer Systems in Engineering'' as well as the Master's degree  
program ``Digital Engineering''. He is a founding member of the university's Center for Digital Engineering, Management and Operations (CeDEMO).
His research is driven by the idea of improving engineering tasks with methods from computer science -- with a special focus on methods 
from Software Engineering, formal specification techniques, mobile assistance, and robotics.

\balance

\end{document}